\documentclass[reprint,superscriptaddress,amsmath,amssymb,aps,pra,floatfix,]{revtex4-2} 
\usepackage{float}
\makeatletter
\let\newfloat\newfloat@ltx
\makeatother
\usepackage{subcaption} 
\usepackage{bm}
\usepackage{dcolumn}
\usepackage{bm}
\usepackage[english]{babel}
\usepackage[utf8]{inputenc}
\usepackage{amsthm}
\usepackage{enumitem}
\usepackage{mathtools}
\usepackage{physics}
\usepackage{algorithm}
\usepackage{algpseudocode}
\usepackage{amsmath}
\usepackage{xcolor}
\usepackage{graphicx}
\usepackage[left=23mm,right=13mm,top=35mm,columnsep=15pt]{geometry} 
\usepackage{adjustbox}
\usepackage{placeins}
\usepackage[T1]{fontenc}
\usepackage{lipsum}
\usepackage{csquotes}
\usepackage{float}
\usepackage{babel,blindtext}
\usepackage{hyperref}
\usepackage[normalem]{ulem}
\usepackage{footnote}
\usepackage{physics}
\usepackage{braket}
\usepackage{esvect}
\usepackage{caption}
\usepackage[color=yellow]{todonotes}
\begin{document}

\preprint{APS/123-QED} 

\title{Quantum annealing eigensolver as a NISQ era tool for probing strong correlation effects in quantum chemistry} 

\author{Aashna Anil Zade}
\email{aashnazade@gmail.com}
\affiliation{Centre for Quantum Engineering, Research and Education, TCG CREST, Kolkata 700091, India} 
\affiliation{Department of Physics, IIT Tirupati, Chindepalle, Andhra Pradesh 517619, India}
\author{Kenji Sugisaki}
\affiliation{Centre for Quantum Engineering, Research and Education, TCG CREST, Kolkata 700091, India}
\affiliation{Graduate School of Science and Technology, Keio University, 7-1 Shinkawasaki, Saiwai-ku, Kawasaki, Kanagawa 212-0032, Japan}
\affiliation{Quantum Computing Center, Keio University, 3-14-1 Hiyoshi, Kohoku-ku, Yokohama, Kanagawa 223-8522, Japan} 
\affiliation{Keio University Sustainable Quantum Artificial Intelligence Center (KSQAIC), Keio University, 2-15-45 Mita, Minato-ku, Tokyo 108-8345, Japan}
\author{Matthias Werner}
\affiliation{Qilimanjaro Quantum Tech, Carrer de Veneçuela 74, 08019 Barcelona, Spain} 
\affiliation{Departament de F\'{i}sica Qu\`{a}ntica i Astrof\'{i}sica, Facultat de F\'{i}sica,
Universitat de Barcelona, 08028 Barcelona, Spain}
\affiliation{Institut de Ci\`{e}ncies del Cosmos, Universitat de Barcelona,
ICCUB, Mart\'{i} i Franqu\`{e}s 1, 08028 Barcelona, Spain}
\author{Ana Palacios} 
\affiliation{Qilimanjaro Quantum Tech, Carrer de Veneçuela 74, 08019 Barcelona, Spain} 
\affiliation{Departament de F\'{i}sica Qu\`{a}ntica i Astrof\'{i}sica, Facultat de F\'{i}sica,
Universitat de Barcelona, 08028 Barcelona, Spain}
\affiliation{Institut de Ci\`{e}ncies del Cosmos, Universitat de Barcelona,
ICCUB, Mart\'{i} i Franqu\`{e}s 1, 08028 Barcelona, Spain}
\author{Jordi Riu}
\affiliation{Qilimanjaro Quantum Tech, Carrer de Veneçuela 74, 08019 Barcelona, Spain} 
\affiliation{Universitat Politècnica de Catalunya, Carrer de Jordi Girona, 3, 08034 Barcelona, Spain}
\author{Jan Nogué}
\affiliation{Qilimanjaro Quantum Tech, Carrer de Veneçuela 74, 08019 Barcelona, Spain} 
\affiliation{Universitat Politècnica de Catalunya, Carrer de Jordi Girona, 3, 08034 Barcelona, Spain}
\author{Artur Garcia-Saez}
\affiliation{Qilimanjaro Quantum Tech, Carrer de Veneçuela 74, 08019 Barcelona, Spain} 
\affiliation{Barcelona Supercomputing Center, Pl. Eusebi G\"uell 1, 08032 Barcelona, Spain} 
\author{Arnau Riera}
\affiliation{Qilimanjaro Quantum Tech, Carrer de Veneçuela 74, 08019 Barcelona, Spain} 
\author{V. S. Prasannaa}
\email{srinivasaprasannaa@gmail.com}
\affiliation{Centre for Quantum Engineering, Research and Education, TCG CREST, Kolkata 700091, India} 
\affiliation{Academy of Scientific and Innovative Research (AcSIR), Ghaziabad 201002, India} 

\date{\today}

\begin{abstract} 
The quantum-classical hybrid variational quantum eigensolver (VQE) algorithm is arguably the most popular noisy intermediate-scale quantum (NISQ) era approach to quantum chemistry. We consider the underexplored quantum annealing eigensolver (QAE) algorithm as a worthy alternative. We use a combination of numerical calculations for a system where strong correlation effects dominate, and conclusions drawn from our preliminary scaling analysis for QAE and VQE to make the case for QAE as a NISQ era contender to VQE for quantum chemistry. For the former, we pick the representative example of computing avoided crossings in the H$_4$ molecule in a rectangular geometry, and demonstrate that we obtain results to within about 1.2\% of the full configuration interaction value on the D-Wave Advantage system 4.1 hardware. We carry out analyses on the effect of the number of shots, anneal time, and the choice of Lagrange multiplier on our obtained results. Following our numerical results, we carry out a detailed yet preliminary analysis of the scaling behaviours of both the QAE and the VQE algorithms. We analyze the non-recurring and recurring costs involved in both the algorithms and arrive at their net scaling behaviours. 
\end{abstract} 

\maketitle 

\tableofcontents

\section{Introduction} \label{sec:introduction}

The Variational Quantum Eigensolver (VQE) algorithm is the most well-known and arguably the most widely used NISQ era approach to quantum chemistry. Several works in literature have employed the algorithm, both on quantum hardware (for example, see Refs.~\cite{Peruzzo2014VQE,kandala2017hardware,google2020hartree,nam2020ground,Pan2023vqe12q,C2vpathway,guo2024experimental} and simulation fronts (Refs.~\cite{Ostaszewski2021optimisation_qc, Tang2021AdaptVqe,Villela2022sim_ionizationenergy,sugisaki2021quantum, Sumeet2022precision_vqe,Bharti2022NISQalgo,Mario2021QCalgos_chemistry,Palak2024VQE}). However, in gate-based NISQ hardware, noise poses significant problems, and thus obtaining results of high quality using VQE (or for that matter, any other gate-based algorithm applied to chemistry) is extremely challenging. On the other hand, on the quantum annealing front, the Quantum Annealer Eigensolver (QAE)~\cite{Teplukhin2019CalculationAnnealer}, which is also a NISQ era quantum-classical hybrid approach and which is also built on the variational principle, has proven to yield highly accurate results~\cite{Vikrant2024}. Furthermore, due to its general purpose nature, QAE has been widely applied to a variety of problems. They  include molecular vibrational spectra~\cite{Teplukhin2019CalculationAnnealer}, complex eigenvalue problems~\cite{Teplukhin2020complex}, energy calculations for molecular electronic states \cite{Teplukhin2020ElectronicAnnealer, Teplukhin2021ComputingAnnealer}, relativistic calculations of fine structure splitting in highly charged atomic ions~\cite{Vikrant2024}, simulating particle physics~\cite{Illaqae}, and lattice gauge theories \cite{Rahmanqae}. However, despite its general scope and better resilience to noise, the QAE algorithm has been vastly underexplored relative to VQE. 

The versatility and robustness of QAE to quantum hardware errors as well as dearth in the range of applications in the context of quantum chemistry naturally prompts further exploration of obtaining molecular properties using the algorithm. In particular, it is interesting and timely to study the performance of QAE in domains where predicting strong correlation effects, one of the cornerstones of quantum many-body theoretic calculations applied to quantum chemistry, are at play. This would be a step towards mirroring the tremendous efforts that have been dedicated towards capturing strong correlation effects in the VQE framework~\cite{scorrel000,scorrel00,scorrel01,scorrel02,scorrel0,scorrel1}. 

While other quantum annealing algorithms for quantum chemistry do exist, such as the Xia--Bian--Kais (XBK) transformation ~\cite{Xia2018ElectronicHamiltonian,XBKhardware} and the Qubit Coupled Cluster (QCC)~\cite{QCC,QuantChemonQA} method, they have practical limitations. For instance, the XBK approach incurs a significant qubit overhead due to the need to encode a k-local electronic structure Hamiltonian (initially represented as a weighted sum of tensor product of all Pauli operators) into a form involving only Pauli $Z$ operators. The qubit count increases further due to quadratization required to transform it into a 2-local Ising Hamiltonian. In contrast, QAE minimizes an energy functional expressed as a double sum over Hamiltonian matrix elements. Post encoding, this functional naturally results in a 2-local form compatible with the quantum annealer (see Eq. \eqref{qubo} below) thus avoiding the need for quadratization, the tradeoff here being that the encoded problem always exhibits a full connectivity among qubits. The QCC method on the other hand, although requiring fewer qubits than XBK~\cite{QCC2}, relies heavily on a classical optimizer to minimize the energy functional which can involve parameters scaling exponentially with the number of entanglers used to capture electron correlation~\cite{QuantChemonQA}. Moreover, it was highlighted in Ref.~\cite{QuantChemonQA} that the QCC method tends to exhibit the issue of convergence to different local minima. This problem is particularly severe for strongly correlated systems.  

\begin{figure*}[t] 
\centering
\begin{tabular}{ccc}
\includegraphics[width=.33\textwidth]{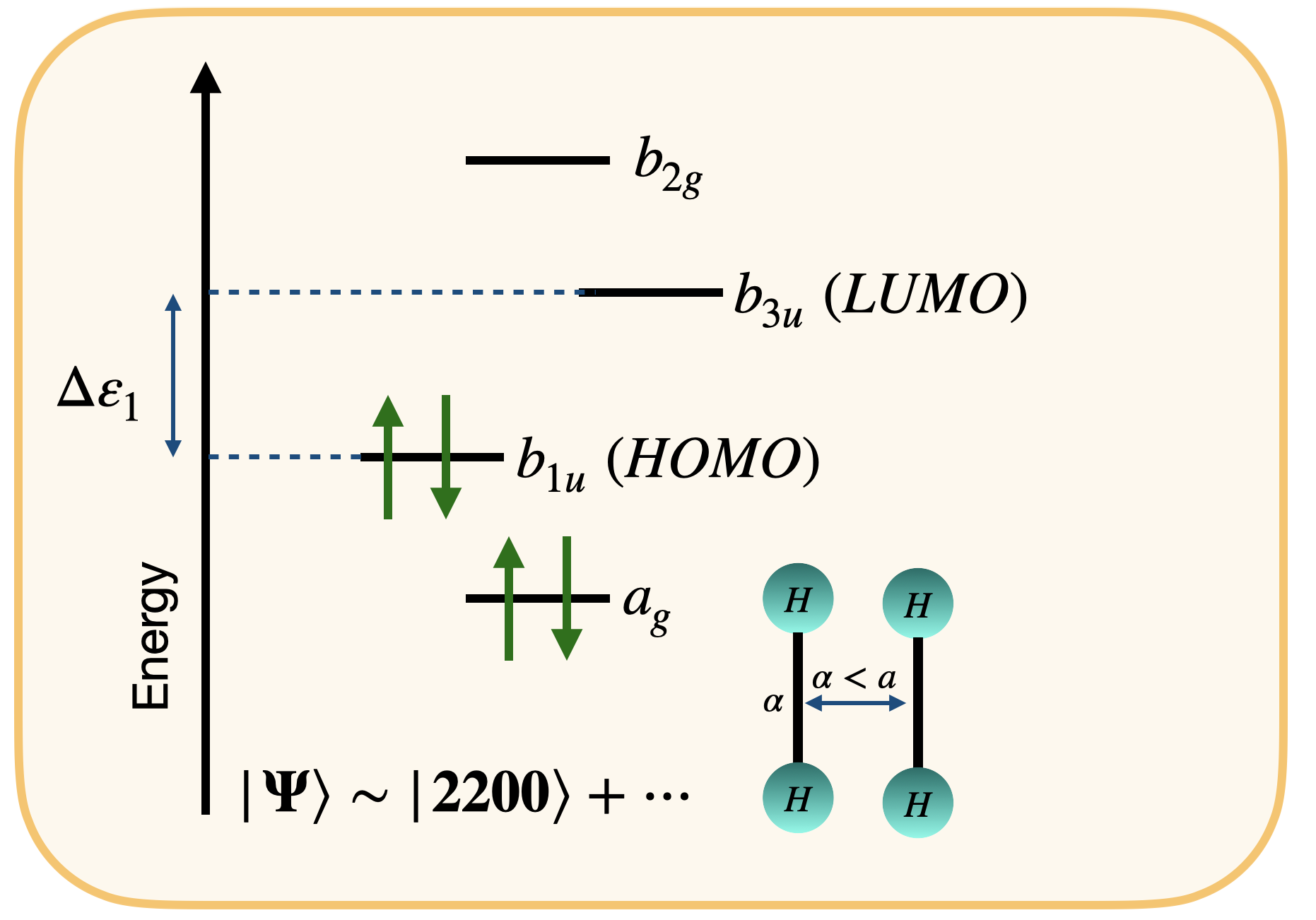}&
\includegraphics[width=.33\textwidth]{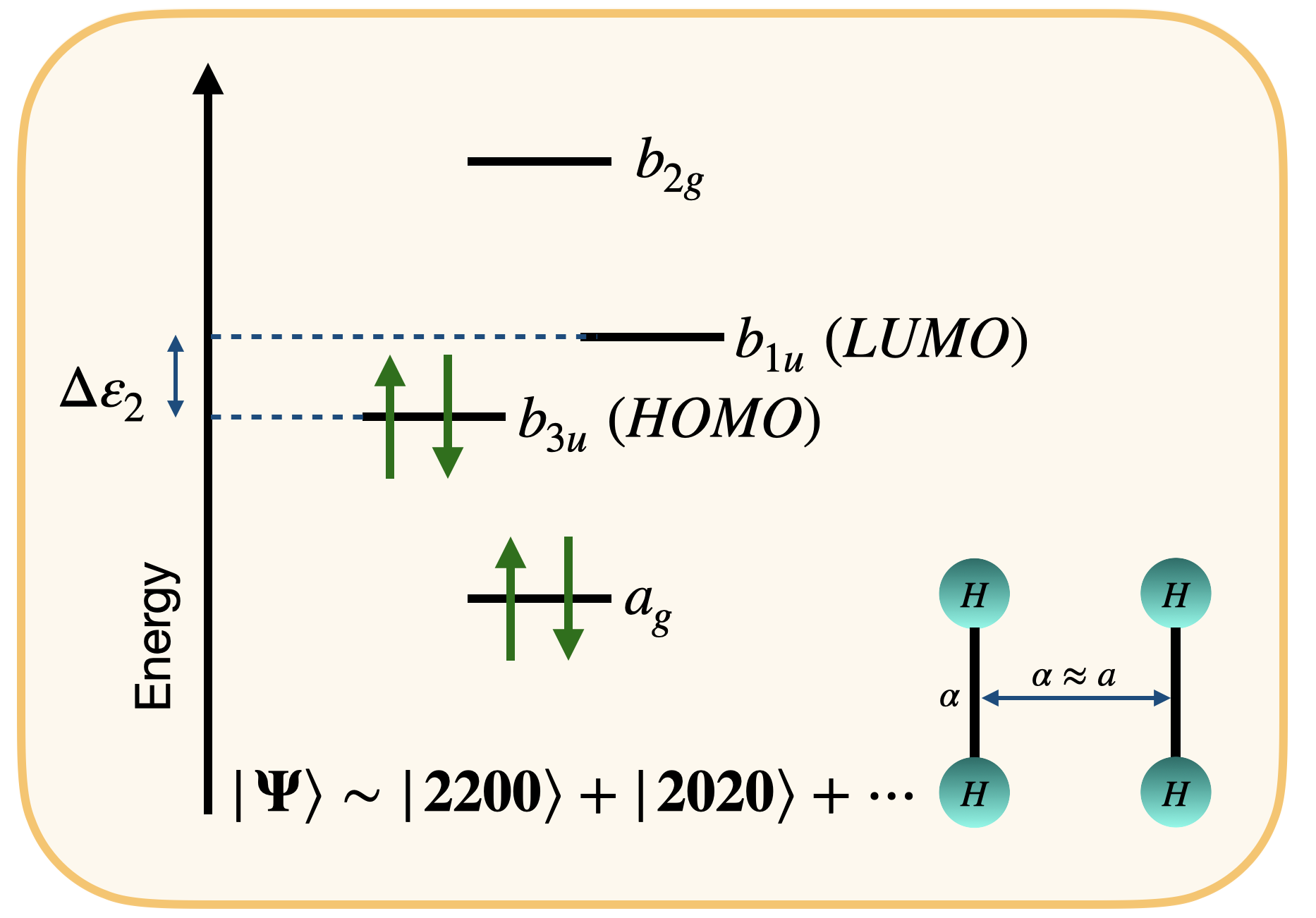}&
\includegraphics[width=.33\textwidth]{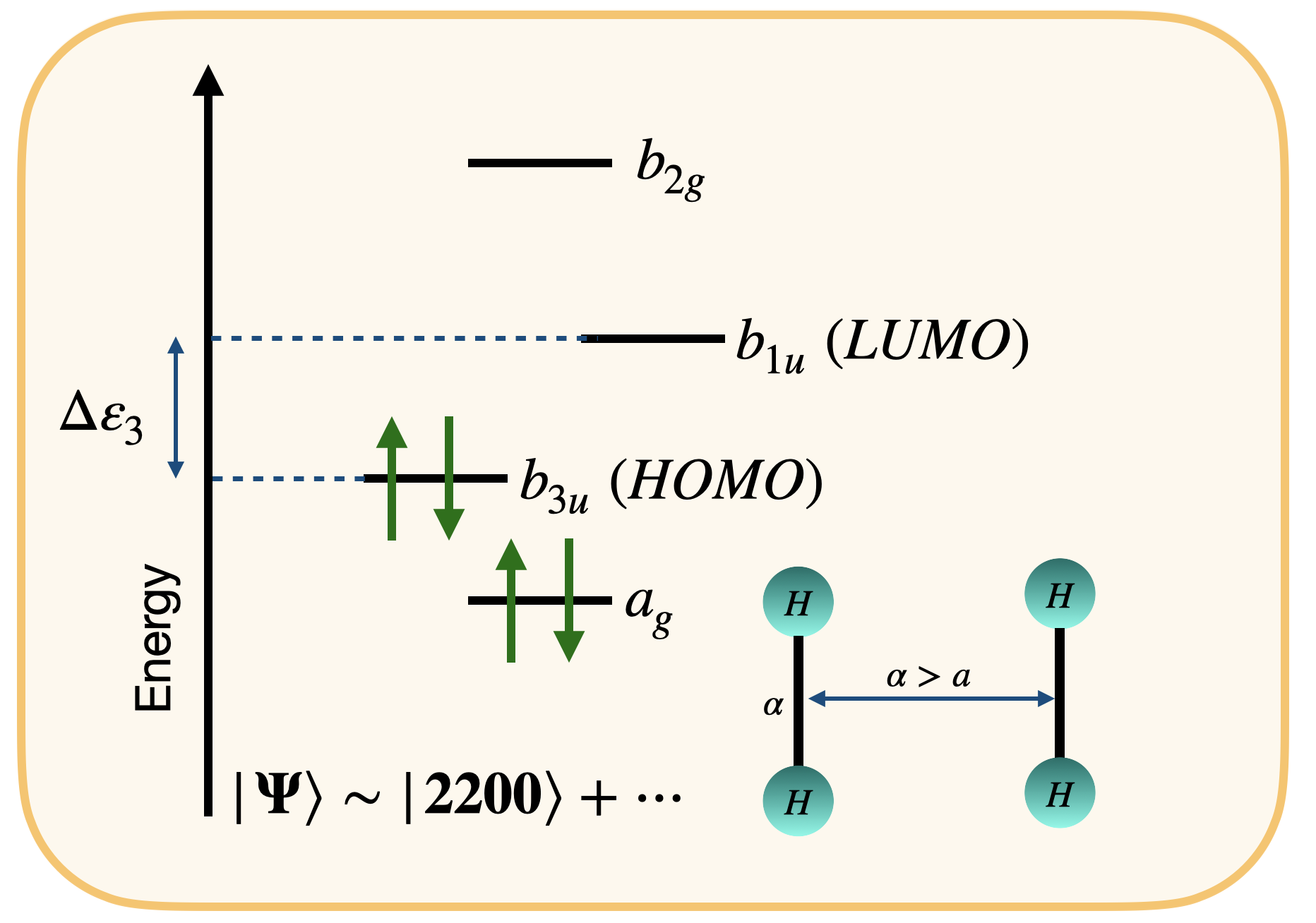}\\
(a)&(b)&(c)
\end{tabular}
\caption{Molecular orbital diagram for H$_4$ across its potential energy curve, for the cases (a) $\alpha < a$, (b) $\alpha \approx a$, and (c) $\alpha >a$, where $\alpha$ denotes the distance between two H$_2$ sub-systems that constitute H$_4$, whereas $a$ is the bond length of the H$_2$ molecule. In the sub-figures, $a_{g}$, $b_{1u}$, etc refer to the irreducible representations of the $D_{2h}$ point group, while $\Delta \varepsilon_i$ is the difference in energy between the HOMO and LUMO across the PEC. HOMO and LUMO are the highest occupied and the lowest unoccupied molecular orbitals respectively.} 
\label{MOH4}
\end{figure*} 

In this work, we begin by carrying out avoided crossings (AC) calculations on the well-known prototypical system: H$_4$ in a rectangular geometry~\cite{H4paper}, to showcase the strength of QAE in predicting physical effects in the strong correlation regime. We expand the many-body wave function in a basis of configuration state functions (CSFs) and recast the energy functional as an Ising Hamiltonian. We then determine the ground state energy and the excited state energy of interest to us using three approaches: quantum annealing (for which we employ the QAE algorithm; on the D-Wave Advantage machine), simulated annealing (for which we use simulated annealing eigensolver (SAE) approach), and the graphical unitary group approach full configuration interaction (GUGA-FCI; which we shall hereafter shorten to FCI for brevity) provided by GAMESS-US~\cite{GAMESS} for benchmarking our results. We also carry out an analysis of the `knobs' of the QAE algorithm, such as the effect of the number of shots, the anneal time, and the choice of the Lagrange parameter on our AC results. For completeness, we also carry out VQE calculation for ground state energy estimation on the IBM superconducting devices.  

Our numerical computations are followed by a preliminary analysis on the scaling behaviours of both the QAE and the VQE approaches (classical as well as quantum resources), with the goal of comparing the two. To that end, we quantify the cost of both the algorithms by categorizing the algorithmic steps into recurring and non-recurring and identify the highest scaling term among these to be the net scaling for the algorithm.

The rest of the sections are organized as follows: Section \ref{sec:Theory} presents the relevant theory and methodology, while Section \ref{sec:Results} discusses our findings from numerical calculations. In Section \ref{sec:Comparison_with_vqe}, we study the scaling behaviours of QAE and VQE, and conclude with Section \ref{sec:Conclusion}. \\ 

\section{Theory and Methodology} \label{sec:Theory} 

\subsection{Quantum annealing}

Quantum annealing (QA) is a meta-heuristic that leverages quantum fluctuations, and is designed to solve computationally hard optimization problems, typically by encoding them into the ground state of a Hamiltonian ~\cite{Kadowaki1998quantum_annealing, Kadowaki2002quantum_annealing, Farhi2000adiabatic_computation}. 
QA involves preparing the ground state of an easily constructible Hamiltonian, $H_I$, which is typically chosen to be the transverse field Hamiltonian, and then gradually transforming it into another Hamiltonian, $H_F$, generally chosen to be the Ising Hamiltonian whose ground state encodes the solution to the problem. As one gradually goes from $H_I$ (at time 
$t=0, s=0, A(0)\gg B(0)$)
to $H_F$ (at time 
$t=t_a, s=1, A(1)\ll B(1)$) according to $H(s)= A(s)H_I + B(s)H_F;\ s\in[0,1]$, the system transitions smoothly from the ground state of $H_I$ to the ground state of $H_F$. We add that that while the adiabatic model described above is the theoretical model underlying QA, the latter, in practice, operates far from the adiabatic limit. In this work we use QAE ~\cite{Teplukhin2019CalculationAnnealer} to map the electronic structure Hamiltonian, typically a k-local operator expressed as a weighted sum of Pauli strings, into a related Ising form. We subsequently use QA to find its ground state.

\subsection{The QAE algorithm: a primer} 

The QAE algorithm solves the eigenvalue equation $H\ket{\Psi} = E\ket{\Psi}$ by transforming it into an energy minimization problem. The energy functional considered is the expectation value of the Hamiltonian $H$ with respect to an unknown state $\ket{\Psi}$. To avoid trivial solutions, the normalization constraint $\langle\Psi|\Psi\rangle-1=0$ is enforced by the inclusion of a Lagrange multiplier $\lambda$ into the energy functional.  The modified formulation after dropping the irrelevant constant is given by

\begin{align}\label{efunc2}
	\epsilon &= \langle \Psi |H |\Psi\rangle  - \lambda\langle\Psi| \Psi\rangle . 
\end{align} 
Assuming an ansatz of the form $\ket{\Psi} = \sum^{d_{CI}}_{i=1} c_i \ket{\Phi_i}$ where $\{\ket{\Phi_i}\}$ represent a set of $d_{CI}$ known basis functions, the QAE algorithm aims to determine the unknown coefficients $\{c_i\}$ with $c_i \in [-1,1]$ in the energy functional 
\begin{eqnarray}\label{energy_funcc}
	\epsilon(\vec{c}, \lambda)=\sum_{i,j=1}^{d_{CI}}c_{i}c_{j}H_{ij}-\lambda\sum_{i=1}^{d_{CI}}c_{i}^2. 
\end{eqnarray} 
We subsequently use the fixed point encoding scheme where each coefficient $c_{i}$ is encoded using K qubits $q^{i}_{K} \in \{-1,1\}$ in the following manner
\begin{equation}\label{encode_q_c}
    c_{i} = \sum^{K-1}_{\alpha=1}2^{\alpha-K}q^{i}_{\alpha} - q^{i}_{K} 
\end{equation}
This approach converts the problem from a  continuous optimization over the variables $\{c_i\}$ to a discrete optimization problem over the binary variables $\{q_i\}$ i.e.
\begin{equation}
\begin{aligned}
   \epsilon(\vec{q}, \lambda)&=\sum_{i,j=1}^{d_{CI}}\Big(\sum_{\alpha,\beta=1}^{K-1}2^{\alpha+\beta-2K} q^{i}_{\alpha}q^{j}_{\beta}+\\ &2^{\alpha+1-K} q^{i}_{\alpha} \pm 
   1 \Big)\Big(H_{ij} - \lambda\delta_{ij}\Big)
\end{aligned}\label{qubo}
\end{equation}
making it suitable for implementation on current quantum annealers. In the above equation, $H_{ij} \equiv \langle \Phi_i|H|\Phi_j\rangle $ and $\delta_{ij}$ refers to the Kronecker delta function. The $\pm$ sign arises from the product $q^{i}_{K}q^{j}_K$ which can be 1 or $-1$. The pseudocode (see Algorithm \ref{algcap}) summarizes the QAE implementation employed. 

\begin{algorithm}
\caption{QAE}\label{algcap}
\begin{algorithmic}
		\State Get $H$, Guess $\lambda_{\pm}$
		\For {$\lambda \in [\lambda_{-}, \lambda_{+}]$}
		\State $\epsilon(\vec{q}, \lambda) \gets \epsilon(\vec{c}, \lambda)$  \Comment{Encoding}
		\State $\vec{q}_{\text{optimum}} \gets [\epsilon(\vec{q}, \lambda)]_{min}$        \Comment{Annealing}
		\State $\vec{c}_{\text{optimum}} \gets \vec{q}_{\text{optimum}}$  \Comment{Reverse encoding}
		\State $E_{\lambda} \gets \epsilon(\vec{c}_{\text{optimum}}, \lambda)$
		\EndFor
	\State Get $min\{E_{\lambda}\}$
\end{algorithmic}
\end{algorithm} 

In the succeeding paragraphs, we describe the theory behind avoided crossings, followed by details on obtaining them in the QAE framework. 

\subsection{Avoided crossings} 

An accurate description of ground and excited state potential energy curves (PECs) in regions where the electronic states interact strongly poses a challenge for well-known electronic structure methods such as the single reference coupled cluster approach~\cite{Piecuch,Hubbard,Dukelsky,Henderson}. These strongly interacting regions, termed as avoided crossings, involve geometries far from equilibrium where an accurate description of the electronic structure requires accounting for both strong and weak correlation effects, of which the former are predominant. In fact, ACs are a key indicator of strong correlation effects wherein the Hartree--Fock (HF) state alone no longer acts as a good reference for methods such as coupled cluster to compute accurate wave functions. 

In the case of H$_4$, this behaviour originates from the quasi-degenerate nature of the molecular orbitals, which can be continuously varied by changing the parameter that defines the geometry, $\alpha$. The parameter corresponds to the distance between the two H$_2$ sub-systems, each with bond distance separation given by $a$. Specifically, when $\alpha \neq a$, the energy gap between the highest occupied molecular orbital (HOMO) and the lowest unoccupied molecular orbital (LUMO) is substantial. This makes the HF configuration, given by $\ket{2200}$ in the occupancy number representation (specification of the occupancy of each orbital), dominant such that 
$\ket{\Psi}_{\alpha \neq a} \sim \ket{2200}$, thus enabling single reference methods to accurately describe the wave function in these regions. Conversely, when 
$\alpha \approx a$, the HOMO and LUMO become nearly degenerate
and an additional configuration $\ket{2020}$ begins to contribute significantly alongside HF resulting in $\ket{\Psi}_{\alpha \approx a} \sim \ket{2200} - \ket{2020}$. Figure~\ref{MOH4} depicts the orbital degeneracies along varying parameter values. The `$\sim$' symbol is to indicate that we are ignoring the normalization constants for brevity, whereas the `$\cdots$' symbol indicates that the rest of the states besides those mentioned on the right hand side do not contribute significantly. 

The AC in H$_4$ can also be understood from a group theoretic point of view. The non crossing rule put forth by Neumann and Wigner in 1929~\cite{Neumann2000ONTB} states that PECs corresponding to electronic states of same point group symmetry do not cross. Formally symmetry of an electronic state can be determined by evaluating the direct product of the irreducible representation of each of the electrons involved in that state. For H$_4$, this corresponds to the A$_g$ symmetry of the D$_{2h}$ point group. 

ACs are significant in quantum chemistry for studying reaction dynamics. They often correspond to the energy of transition states in chemical reactions. By accurately determining the energy at an AC, we can identify the major reaction pathway among multiple possible transition states. In this work, we use QAE to predict ACs in the H$_4$ molecule. The primary source of avoided crossing arises from the symmetry of the electronic states. Additionally, this model system allows us to vary the degree of orbital quasi-degeneracies by simply changing the geometry defining parameter.

\section{Results and Discussions} 
\label{sec:Results} 

\subsection{QAE results for avoided crossings} \label{QAE-results} 

\subsubsection{Input parameters} 
\begin{figure}[h]
\hspace{-1cm}
\includegraphics[width=8cm]{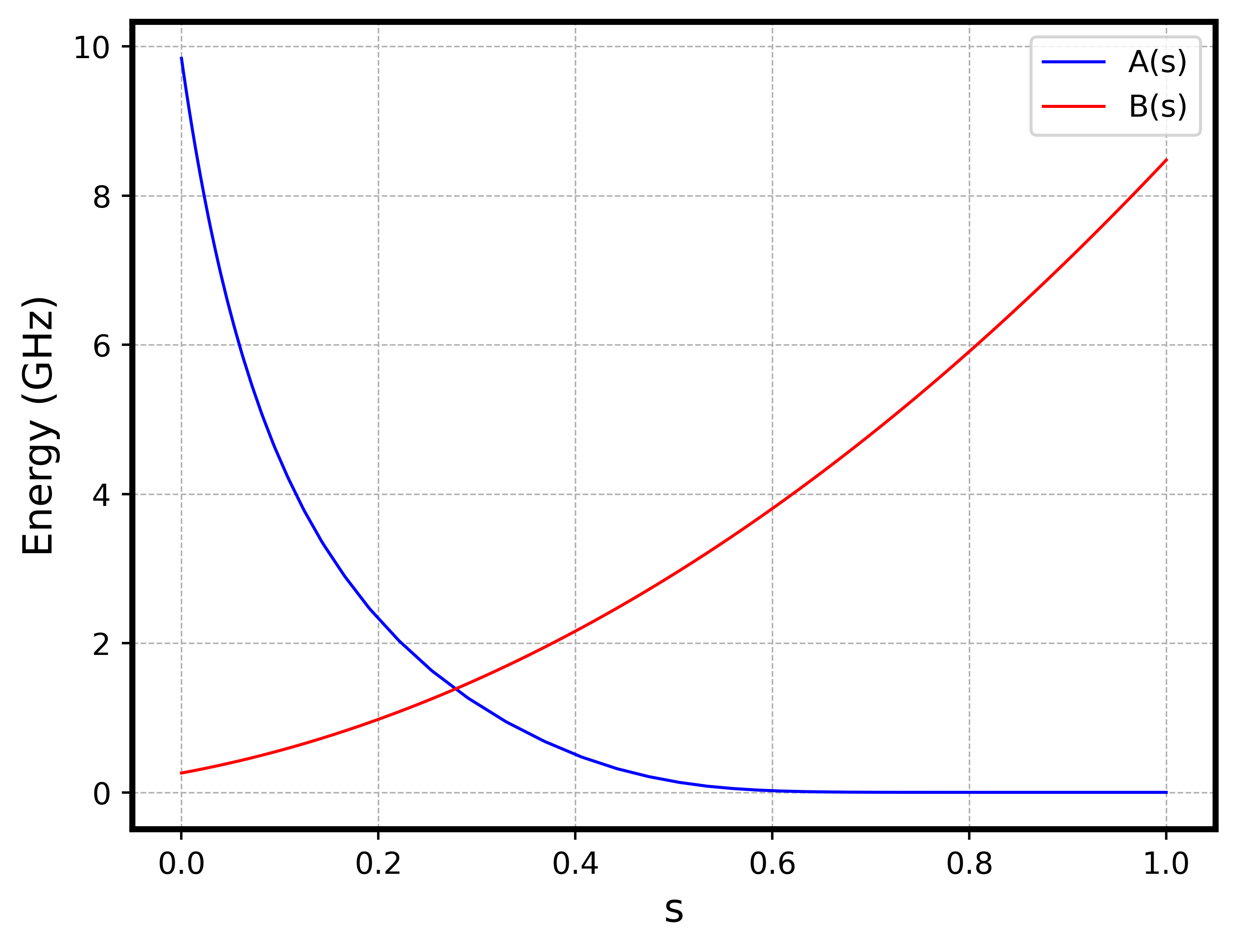}
 \caption{Anneal schedule used in our computation. }\label{anneal_schedule}
\centering
\end{figure} 

We begin by detailing the input parameters associated with our chosen molecule. H$_4$ is a planar four-electron model consisting of two interacting hydrogen molecules. The geometry is defined by intermolecular and intramolecular H$\dots$H distances $\alpha$ and $a$ respectively, where the latter is fixed to 2 Bohr. We calculate the ground and excited state energies at five geometries ($\alpha$ = 1.8, 1.9, 2.0, 2.1, and 2.2 Bohr; for the purposes of this work, we fix the resolution to 0.1 Bohr in view of the cost of quantum computational resources). We work with the STO-3G minimal basis set and employ the $D_{2h}$ point group symmetry in our calculations. 

To generate the CI matrix elements, $H_{ij}$, we use the GUGA-FCI approach available in the GAMESS-US program~\cite{GAMESS}. There are 8 CSFs for our 4 electron 4 orbital active space. As it will be detailed later, this corresponds to an $80 \times 80$ all-to-all connected QUBO problem with $K=10$. 

We now comment on the computational details and machine parameters in the QAE part of our workflow. We use the D-Wave Advantage system 4.1 for all the QAE computations. We need to choose a range of $\lambda$ values across which we scan. For this purpose, we first carry out several SA computations, each with a different choice for the $\lambda$ range. A given $\lambda$ range is subdivided into 1000 parts, and for each of those $\lambda$ values, we carry out SAE (the approach involves all of the steps from QAE, except that the quantum annealing is replaced by simulated annealing) with 1000 shots. We pick that $\lambda$ range for our QAE calculation that gives the best energy value. All of our SAE calculations are performed using the D-Wave Ocean Toolkit~\cite{Ocean}. For our QAE calculations, we subdivide the chosen $\lambda$ range into 100 values, and perform QA with 1000 shots for each of those $\lambda$ values. For each anneal, we pick the default anneal time of 20 microseconds. Once we find the lowest energy from the procedure, we repeat the process 10 times (we term this as 10 repetitions hereafter) and choose the lowest energy among the set of values for our final result for the ground state energy. We note that for an excited state calculation, we follow the same procedure, except that the Hamiltonian, $H_{e}$, is constructed from the ground state Hamiltonian, $H_{g}$, and the ground state wave function, $\ket{\Psi_g}$, by invoking the Brauer's theorem as 
\begin{align*}
	H_{e} = H_{g} + S_0 \ket{\Psi_g}\bra{\Psi_g}. 
\end{align*} 
This theorem states that the lowest eigenvalue of $H_{e}$ corresponds to the second-lowest eigenvalue of $H_{g}$, which represents the first excited state energy we are looking for. Here, $S_0 > E_e - E_g$. We choose $S_0$ to be 1 in our work.

The quantum annealing process takes place over the interval $t=0$ to $t_a$, where $t_a$ is the anneal time that is set to be $20\mu s$ for our work. To implement the problem into the Advantage 4.1 system, which does not have all-to-all connectivity, we must first embed our fully connected problem into its Pegasus topology by means of a qubit overhead as we shall see later. This is known as the minor embedding process, which we perform with the EmbeddingComposite() class from the Ocean SDK. Our implementation follows the device's default anneal schedule as shown in Figure \ref{anneal_schedule}, which depicts the anneal schedule in terms of the change in energy as a function of scaled time (s)~\cite{Ocean}. 

\begin{figure}[t]
	\includegraphics[width=9cm]{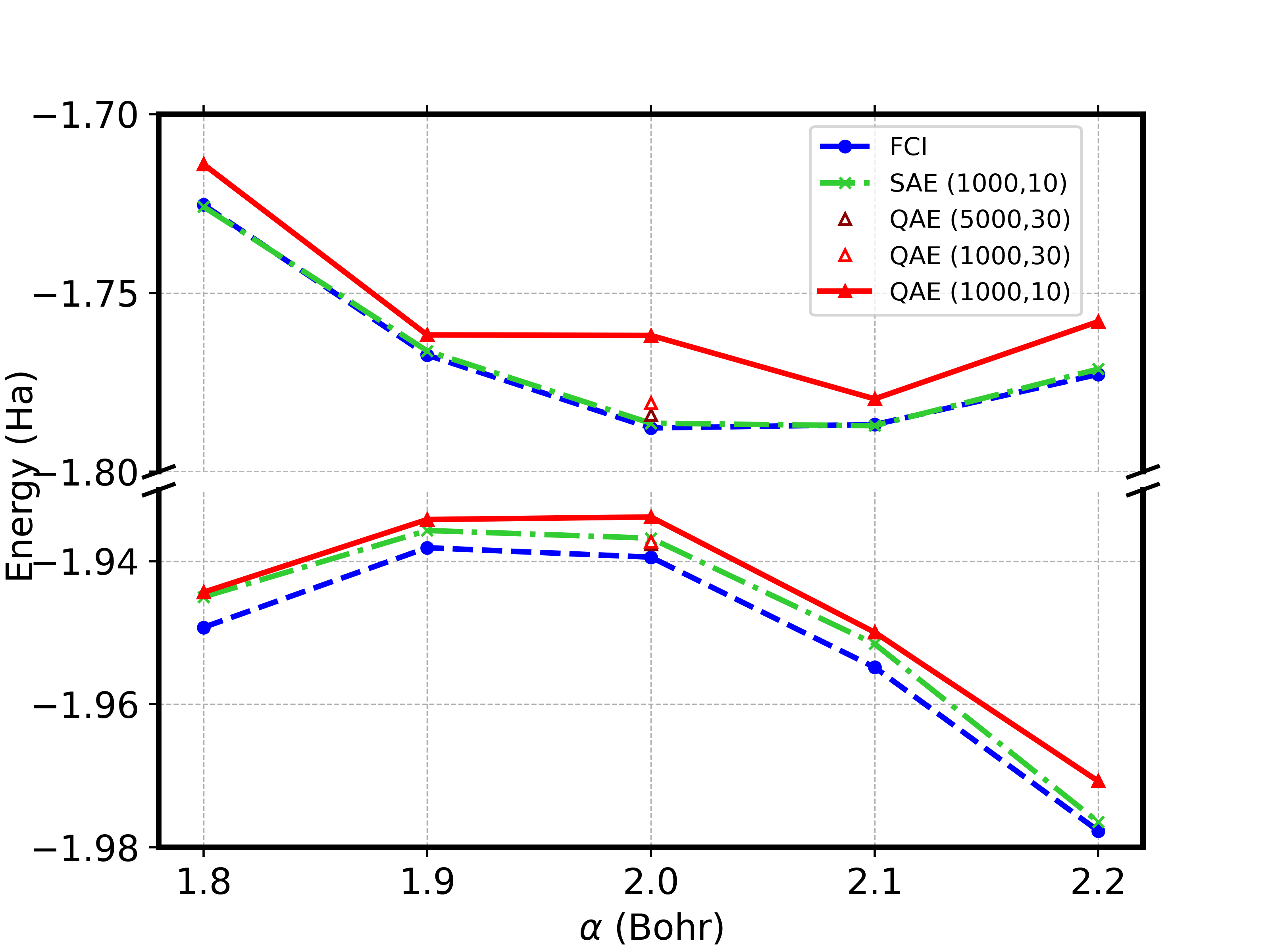}
	\centering
	\caption{Comparison of the potential energy curves of the ground electronic state (singlet) and the first excited electronic state (singlet) obtained for H$_4$ for different values of $\alpha$, using FCI, SAE, and QAE (on the D-Wave hardware). We calculate all of the data points with 1000 shots and 10 repetitions (denoted in the legend as (1000,10)), and setting an anneal time of 20 microseconds. The red and brown triangle markers at $\alpha = 2$ Bohr indicate our results obtained with 1000 shots and 30 repetitions (denoted in the legend as (1000,30)) and with 5000 shots and 30 repetitions (denoted in the legend as (5000,30)) respectively.}
    \label{H4_pec} 
\end{figure} 

\subsubsection{Results from computation}

We now discuss our results. Figure~\ref{H4_pec} presents our results from annealing (with accompanying data provided in Table \ref{tab:1}), where we compare the predicted avoided crossing from QAE with SAE and FCI, with 1000 shots and 10 repetitions. From the figure, we see that QAE, SAE, and FCI yield an energy difference (at the AC geometry, that is, 2 Bohr) of 0.17195, 0.15037, and 0.15169 Hartree (Ha) respectively. Thus, QAE is able to predict AC in the chosen system to within about 13 \% of the FCI value. The plot also shows that SAE is clearly in better agreement with the FCI values than QAE with both the ground and excited state energies at all of the geometries considered.
A major bottleneck in QAE which is otherwise absent in SAE is the requirement for embedding.
The QAE formulation results in a problem graph which is inherently fully connected regardless of the
specific problem instance. Embedding thus refers to the process of mapping this fully connected
graph onto sparsely connected qubits of the quantum annealer (here, Pegasus topology \cite{pegasus}). This process incurs a significant qubit overhead and thus more error due to the increase of the search space (given by the physical Hilbert space dimension) while the logical solution space remains constant. SAE on the other hand is less error prone since classical computers are more robust to errors. Furthermore, the quality of the predicted excited state energy is relatively poor for the AC geometry using QAE (as we shall discuss later, the quality of results from QAE is still substantially better than those from a VQE computation with the same parameters). We can backtrack the performance of QAE to the several input parameters that go into obtaining a QAE result, such as (but not limited to) the number of shots and number of repetitions, anneal time, and choice of $\lambda$ range. We devote the subsequent paragraphs to analyzing the effect of these three parameters on the AC results. 

\begin{table}[t]
  \centering
    \caption{Table presenting data (with 1000 shots and across 10 repetitions) on the percentage accuracies of the ground state ($\mathcal{A}_g$) and excited state energies ($\mathcal{A}_e$) as well as the avoided crossings ($\mathcal{A}_{AC}$) for the geometries considered for this work, all of them relative to their respective FCI values. The fourth row gives results with 1000 shots and 30 repetitions at the AC geometry, whereas the last row presents results with 5000 shots and across 30 repetitions (only at the AC geometry). The geometries are specified by $\alpha$ in units of Bohr. The entries shown in parentheses for the AC geometry indicates the difference between that energy and its FCI counterpart (in units of mHa). }
  \setlength{\tabcolsep}{6pt}
  \begin{tabular}{ccccc}
  \hline \hline 
    Shots &$\alpha$&$\mathcal{A}_g$&$\mathcal{A}_e$& $\mathcal{A}_{AC}$\\
    \hline
     & $1.8$& 99.75& 99.34& 102.87 \\
     & $1.9$& 99.80& 99.68& 100.98  \\
    1000 & $2.0$& 99.71 (5.6)
& 98.55 (25.8)
& 113.35 (20.2)
  \\
  & & 99.89 (2.17)
&  99.61 (7.0)
&  103.20 (2.6)
  \\
     & $2.1$& 99.75& 99.60& 101.36  \\
     & $2.2$& 99.65& 99.17& 103.81  \\
     \hline
     5000 & $2.0$& 99.91 (1.83)
& 99.79 (3.67)
& 101.22 (1.85)\\
    \hline \hline 
  \end{tabular}
  \label{tab:1}
\end{table} 
\begin{figure}[h]
\includegraphics[width=8cm]{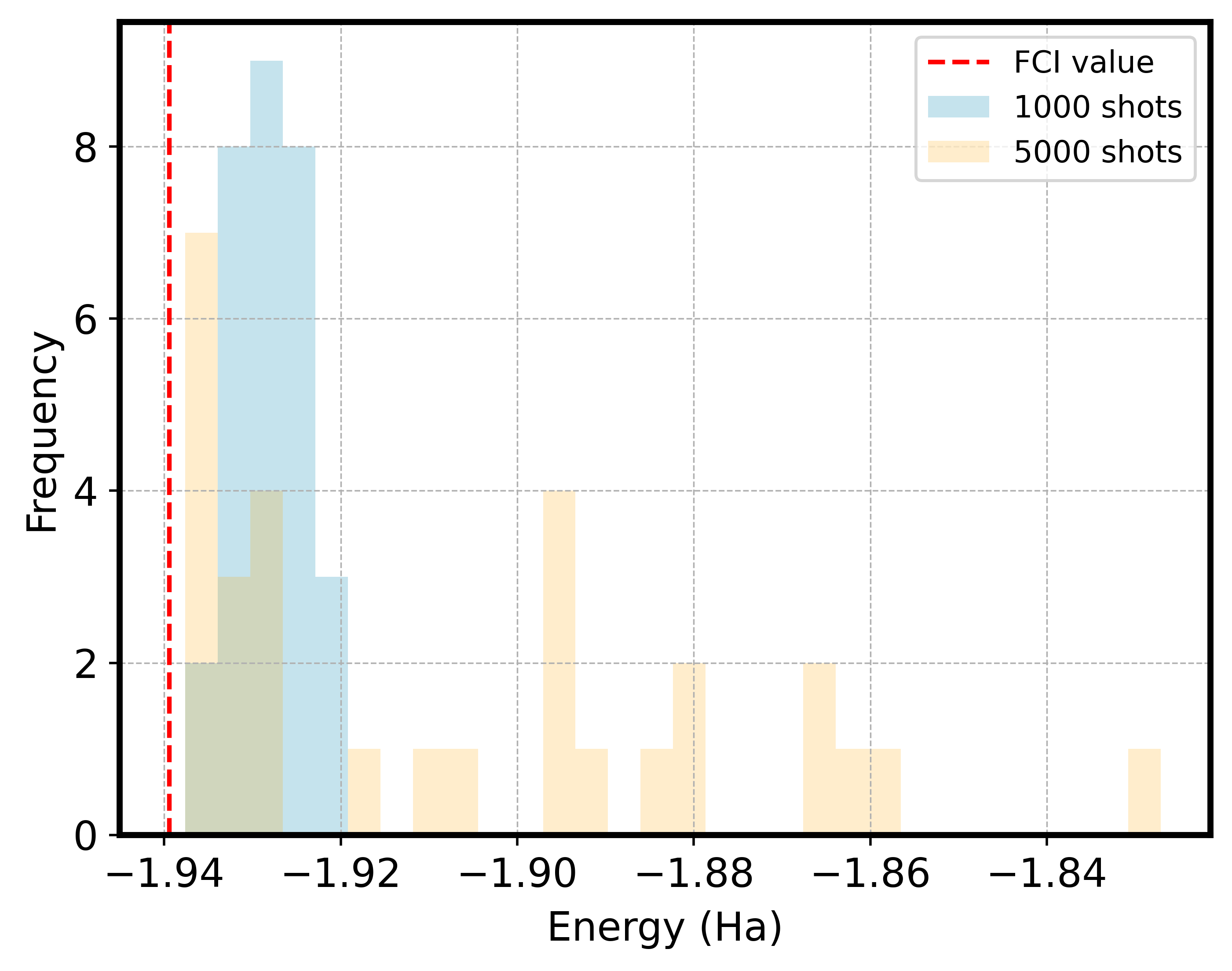}
 \caption{Comparison of the ground state energy at the AC geometry obtained by executing QAE (on the D-Wave hardware) with 1000 shots and 30 repetitions (marked in blue) and 5000 shots and 30 repetitions (marked in green). The reference FCI value is presented as a dashed line. }\label{energy_diff_shots}
\centering
\end{figure} 

\begin{figure}[t]
\includegraphics[width=8cm]{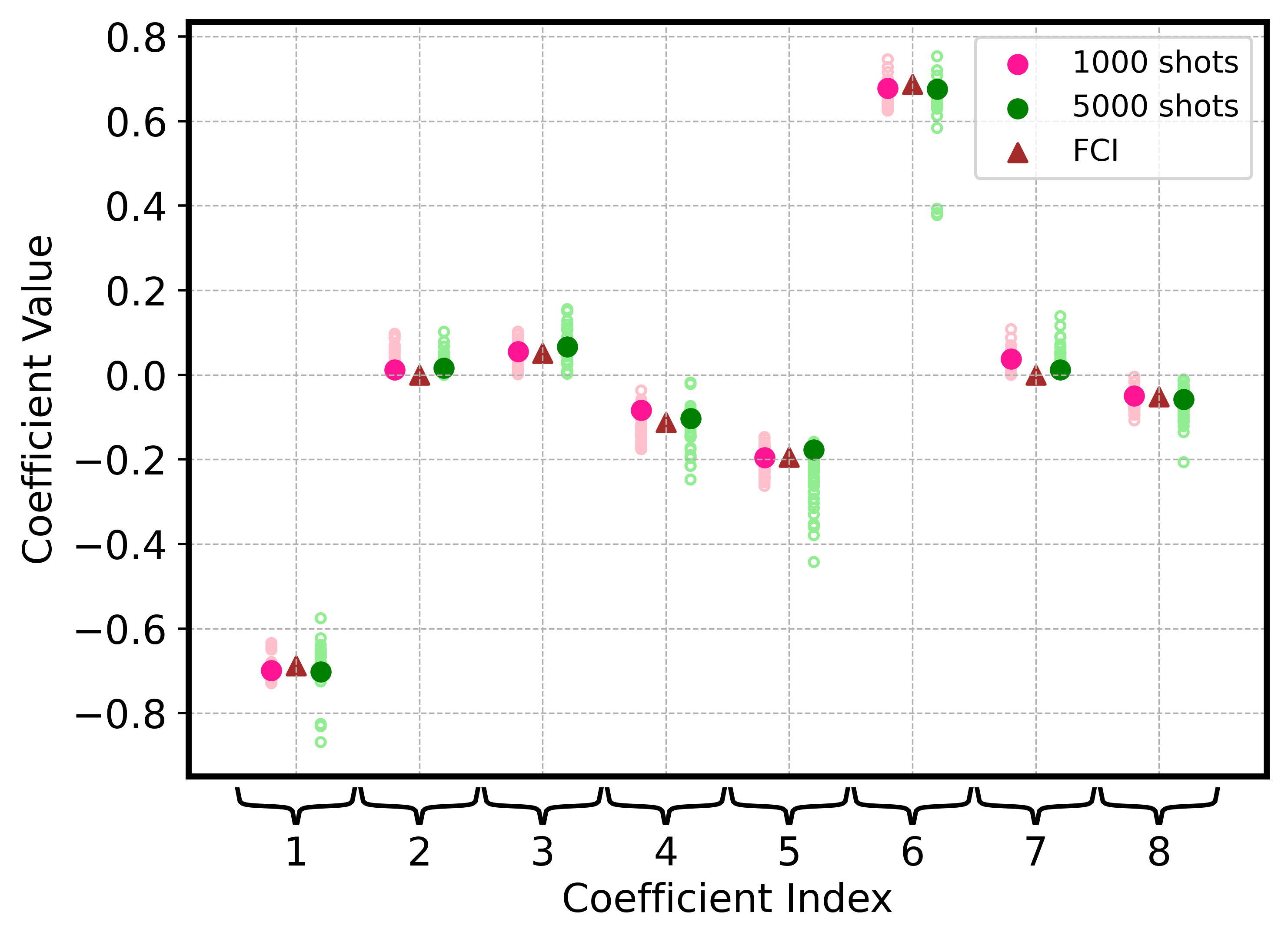}
\caption{Comparison of configuration interaction coefficients for the ground state obtained using QAE with exact values of the coefficients from FCI (denoted as a brown triangle) for 1000 and 5000 shots, both with 30 repetitions, at the AC geometry. Each of the pink (1000 shots) and green (5000 shots) circles denotes the result from a repetition. }\label{Coeff_vals_1000}
\centering
\end{figure} 

We begin by carrying out AC calculations with 1000 shots and 30 repetitions. As Figure~\ref{H4_pec} shows, increasing the number of repetitions drastically improves the quality of the excited state energy (the difference with respect to the FCI value) improves from 25.8 mHa to 7 mHa and improves the agreement of the ground state energy with the FCI value to 2.17 mHa (from 5.6 mHa). The AC value improves from 20.2 mHa for 10 repetitions to 2.6 mHa for 30 repetitions. We extend the analysis further by increasing the number of repetitions all the way to 100, but find that the AC value remains unchanged. Thus, we instead increase the number of shots to 5000 while keeping the number of repetitions fixed at 30. Figure \ref{energy_diff_shots} presents our findings, with the accompanying data given in Table \ref{tab:1}. The results indicate that as expected, one obtains better agreement with the FCI value for the AC with larger number of shots and repetitions. Furthermore, we see that the QAE result now is in much better agreement with the FCI value, differing only by 1.22 \% (5000 shots, 30 repetitions), as opposed to the 1000 shots 10 repetitions value of 13.35 \% and the 1000 shots 30 repetitions value of 3.2 \%. These percentages correspond to an energy difference ($E_{AC, FCI}-E_{AC, QAE}$) of 1.85 mHa,  20.2 mHa, and 2.6 mHa respectively. The values of AC obtained with 5000 shots and 30 repetitions, 1000 shots and 10 repetitions, and 1000 shots and 30 repetitions are 0.15354, 0.17195, and 0.15654 Ha respectively. 

\begin{figure}[t]
\begin{tabular}{c}
\includegraphics[width=8cm]{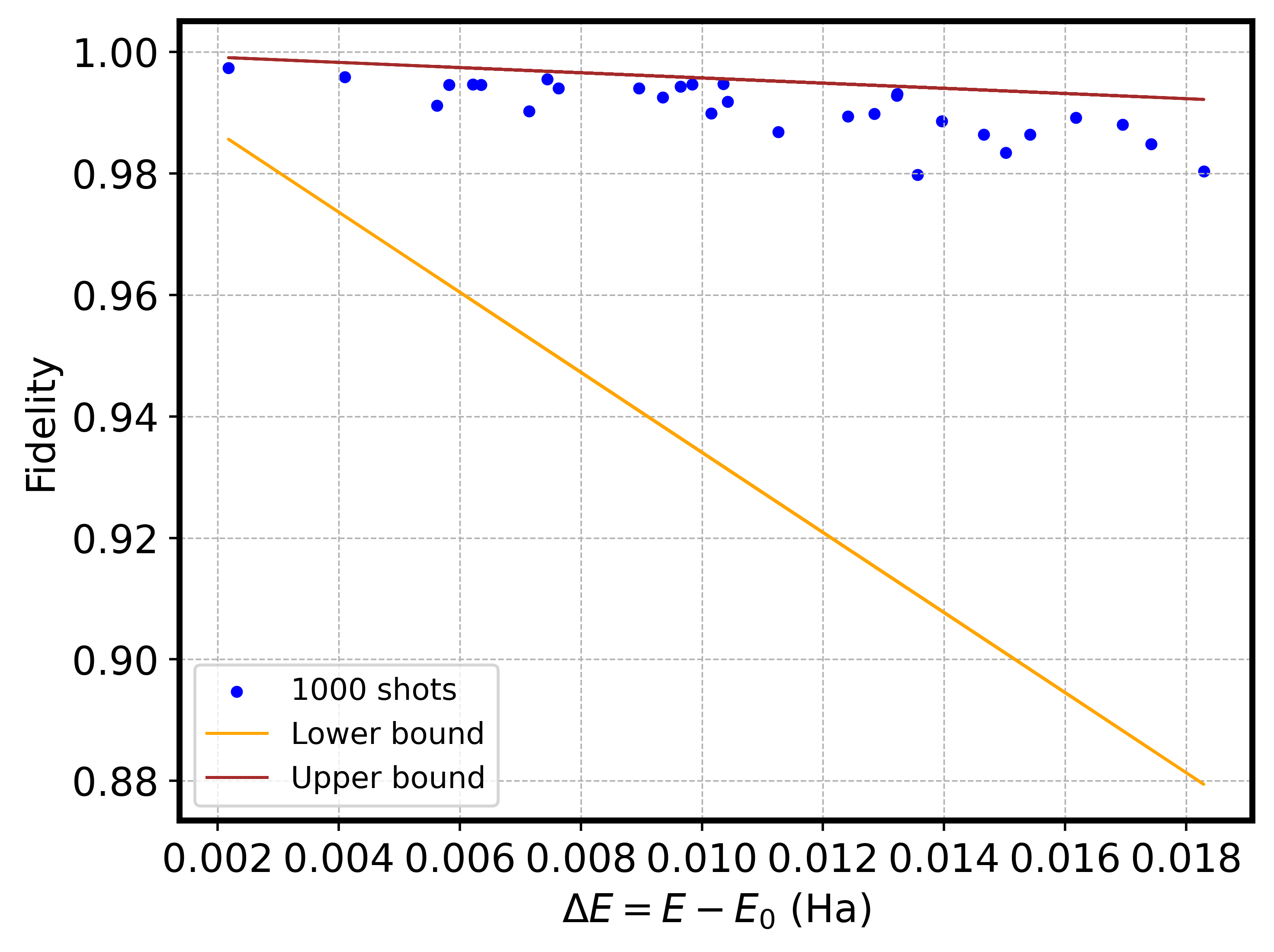}\\
(a)\\
\includegraphics[width=8cm]{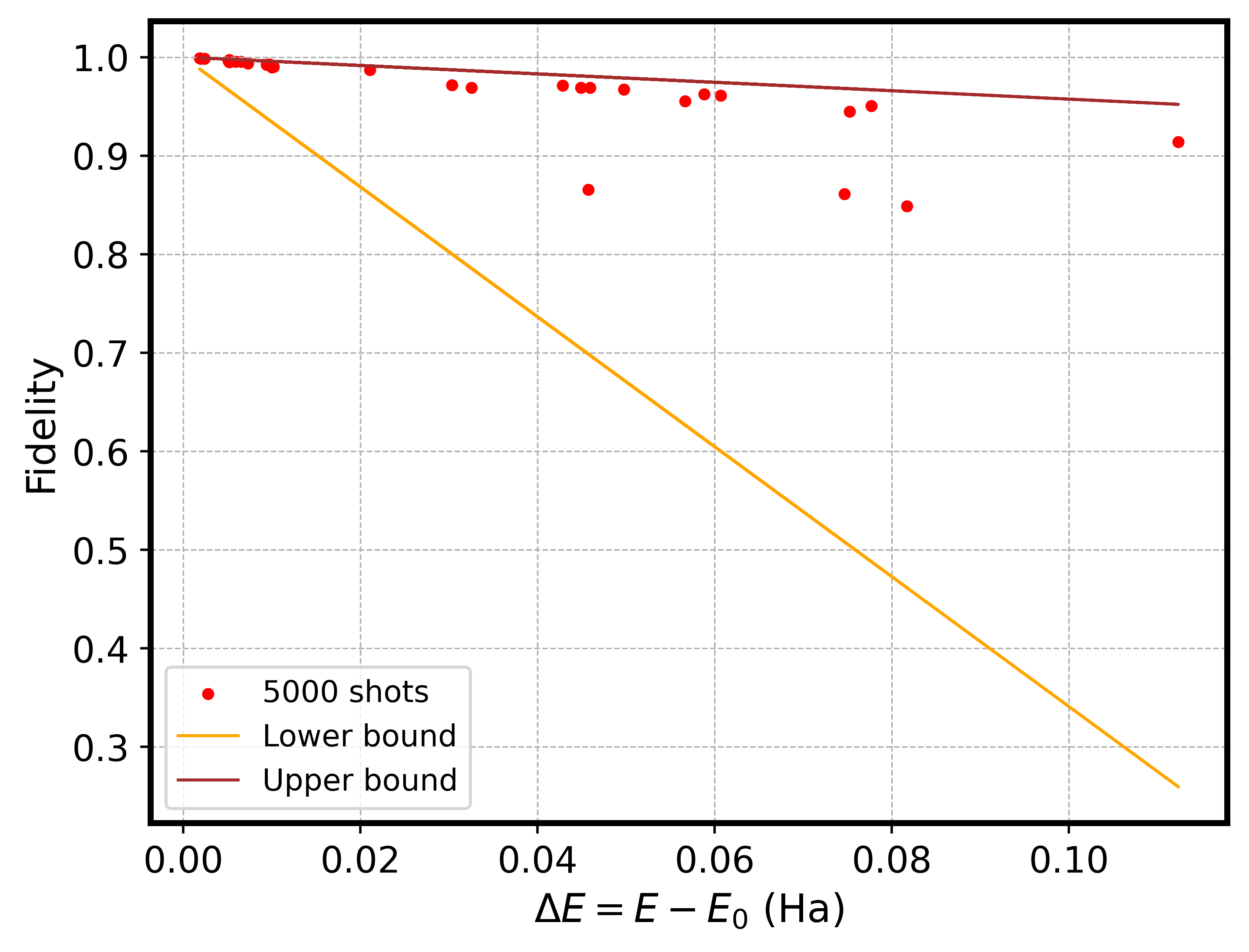}\\
(b)\\ 
\end{tabular}
 \caption{Scatter plots involving fidelity and $\Delta E$ values, for 1000 shots (sub-figure (a)) and 5000 shots (sub-figure (b)) cases. The former quantity refers to $|\langle \Psi_{QAE} | \Psi_{FCI}\rangle|^2$, where $|\Psi_{QAE}\rangle$ is the wave function constructed out of the CI coefficients obtained using QAE and $|\Psi_{FCI}\rangle$ is the FCI wave function. On the other hand, the latter is the difference between the QAE and the FCI ground state energy results. } \label{correl}
\centering 
\end{figure}

We now turn our attention to Figure \ref{Coeff_vals_1000}, where we plot the CI coefficients that we obtained from the ground state QAE calculation versus the coefficient index. We benchmark the computed CI coefficients employing QAE with those obtained from FCI. We carry out the exercise for both the 1000 and the 5000 shot cases, each with 30 repetitions. The purpose of the plot is to check the quality of the coefficients, which in turn indicates the quality of the predicted wave function. We see that the coefficients are all in reasonable agreement with their FCI counterparts. While the plot itself seems to yield almost similar set of coefficients by using 5000 shots over 1000 shots, this seemingly insignificant difference between their results is reflected in the subsequent ground state energy calculations as seen in the preceding paragraphs. Unsurprisingly, we also see the chemistry at play, where two coefficients are almost equally important (unlike the single reference scenario where only the HF state dominates). We also check the state fidelity as well as the difference between the AC gaps predicted using QAE and FCI, and find them to be 99.735\% and 2.6 mHa respectively for QAE calculation with 1000 shots and 30 repetitions, whereas they improve to 99.886\% and 1.85 mHa respectively when we employ 5000 shots and 30 repetitions. As a sanity check, we plot our data against the upper and lower bounds that can be obtained for the fidelity. Figures \ref{correl}(a) and (b) present scatter plots between fidelity ($\mathcal{F}$) and $\Delta E$, the difference between the QAE and the FCI ground state energy results. The fidelity is given by $|\langle \Psi_{QAE} | \Psi_{FCI}\rangle|^2$, where $|\Psi_{QAE}\rangle$ is the wave function constructed using the CI coefficients that QAE outputs, while $|\Psi_{FCI}\rangle$ refers to the FCI wave function. We observe from Fig.  \ref{correl}(a) (which has been plotted for the 1000 shots case) and (b) (the 5000 shots case) that our data respects the bounds for fidelity, given by $1-\frac{E_{g,QAE}-E_{g,FCI}}{E_{e,FCI}-E_{g,FCI}} \leq \mathcal{F} \leq 1-\frac{E_{g,QAE}-E_{g,FCI}}{E_{max,FCI}-E_{g,FCI}}$, where $E_{g,FCI}$ and $E_{g,QAE}$ are the ground state energies calculated using FCI and QAE respectively, whereas $E_{e,FCI}$ and $E_{max,FCI}$ refer to the excited state of interest to our AC problem and maximum eigenvalue of $H$ obtained from FCI. For the derivation of the bound, please see Appendix \ref{sec:bounds}. Our results also show that most of the data points are saturating the upper bound which means a majority of the population is contained in the ground state. Furthermore, we carry out linear regression analysis on our two data sets, and find that the Pearson coefficient, $R$, is anti-correlated at $-0.84$ for the 1000 shots case and $-1.04$ in the 5000 shots case. Ideally, we would expect that as we supply more shots, the strength of anti-correlation between the two quantities, $\mathcal{F}$ and $\Delta E$, would increase, based on the bounds seen earlier. 

\begin{figure}[t]
	\includegraphics[width=8cm]{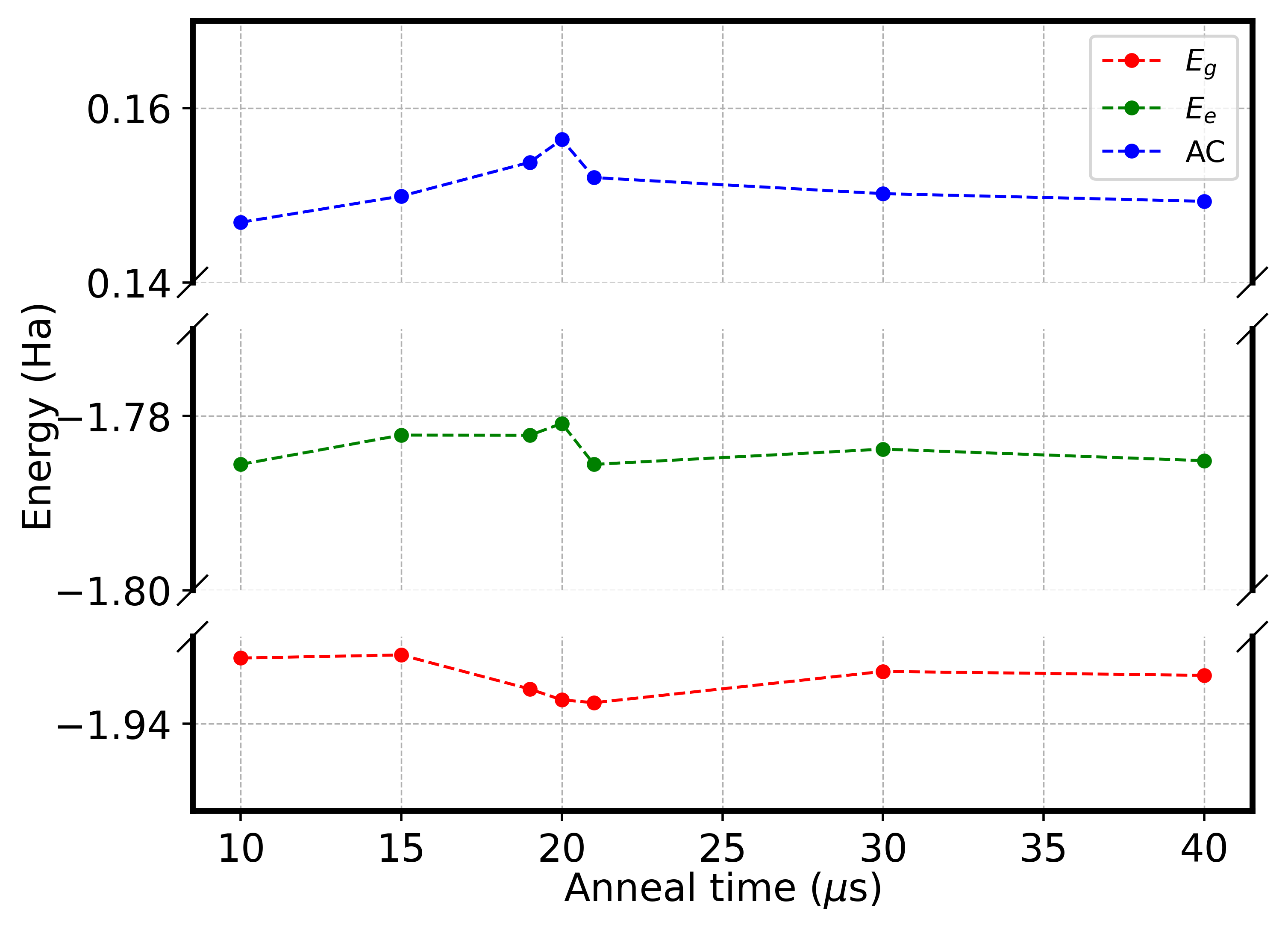}
 \caption{Plot showing the ground state energy ($E_g$), excited state energy ($E_e$), and avoided crossing (AC) values for different anneal times (in microseconds). The calculations are carried out with 1000 shots and 10 repetitions. } \label{anneal_times} 
	\centering
\end{figure}

We now comment on the effect of anneal time on our results. We begin by recalling that the Hamiltonian changes for the ground and excited states, and thus our assumption of setting the same anneal time for the ground and excited state computations leaves some room for further optimization of hyperparameters. We have studied this possibility in Figure \ref{anneal_times}, where we present our results for the variation of the ground state energy $E_g$, the excited state energy $E_e$, and AC  with anneal time, with the purpose of going beyond the assumption that we had made in our main results shown in Figure \ref{H4_pec}. We choose to carry out the study with anneal times chosen between 10 and 40 microseconds. We carry out a coarse grained analysis in the interest of computational cost, and with unequal time resolutions across this range so that we probe with smaller resolution in and around the default anneal time of 20 microseconds. We use 1000 shots and 10 repetitions for the analysis. The plot shows that the default time happens to yield particularly poor results for the excited state relative to the other values of anneal time. If we instead pick the anneal times that give the best results for the ground state and excited state energies, the AC value improves from 0.17195 Ha to 0.15201 Ha. In the interest of computational cost, we did not carry out the analysis for the case of 5000 shots and 30 repetitions, but we anticipate to obtain further improvement in the AC energy value with this approach.

\begin{figure}[t]
	\includegraphics[width=8cm]{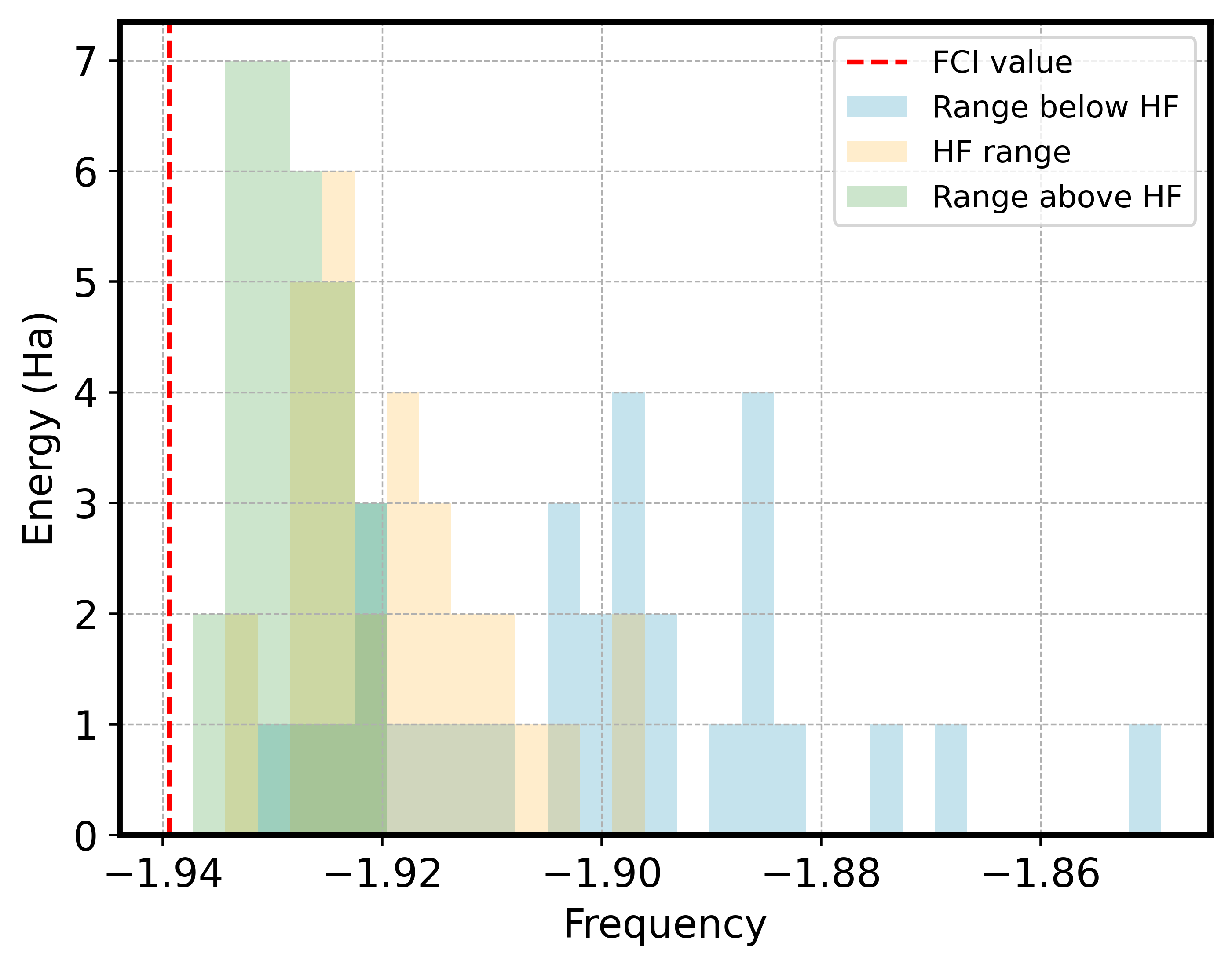}
 \caption{Comparison of the ground state energy across the three different $\lambda$ ranges considered. The FCI value of the ground state energy is provided for reference as a dashed line. } \label{lambda_ranges} 
	\centering
\end{figure}

Finally, we focus on the choice of $\lambda$ range for the problem. As mentioned in an earlier paragraph in Section \ref{sec:Theory}, we picked the $\lambda$ range that gave the best SAE result. For the purposes of this work, we pick a range of 0.1 Ha. However, since SAE and QAE are different approaches to search for a desired solution, we pick three such ranges of 0.1 Ha each: a range containing the HF value ($-1.77677$ Ha), which is $-1.8$ to $-1.7$ Ha, one that is immediately below and another that is immediately above it. We term the three ranges as A, B, and C respectively. We note that the range we picked for our main calculations (Figure~\ref{H4_pec}) was the one immediately \textit{above} the HF value, that is, range C. Figure \ref{lambda_ranges} presents our results justifying this choice. We observe that our choice of $\lambda$ range from SAE calculations were sufficiently good within a calculation involving 30 repetitions. 

\subsection{VQE results for ground state energies}\label{sec:hardware_results} 

\subsubsection{Input parameters} 

We carry out the VQE-UCCSD (unitary coupled cluster ansatz and limited to the singles and doubles approximation) calculation on the \texttt{ibm\_kiev} device. The choice of hardware was made by comparing the CLOPS for the three currently available quantum computers (Brisbane, Sherbrooke, and Kiev), and picking the one with the highest CLOPS. All of our calculations were performed with 4096 shots, and we quote our result by averaging over 10 repetitions. We begin by specifying the system parameters: processor type: Eagle R3, native gates: \{ECR, ID, RZ, SX, X\}, median ECR error: 1.086$\times 10^{-2}$, median readout error: 1.367$\times 10^{-2}$, median T1: 289.88 $\mu$s, median T2: 113.13 $\mu$s, and connectivity: heavy hexagonal lattice topology. We use the SLSQP optimizer for carrying out the classical optimization. We benchmark our results against statevector VQE excluding any noise model. For practical reasons, the  hardware results are reported under noisy conditions and employing error mitigation techniques could further improve their accuracy. The estimated cost of error mitigation is discussed in Point 8 of Appendix E. 

\subsubsection{Results from computation}

For our VQE calculations, we only consider the ground state energy, although it is possible in principle to obtain the excited state energy of interest, using an equation of motion based VQE approach. QAE naturally incorporates CI allowing it to accurately predict energies both near equilibrium and at far-from-equilibrium geometries, effectively capturing both weak and strong correlation effects. In contrast, VQE with the single reference UCCSD ansatz works best for near-equilibrium geometries and is unable to account for strong correlation effects. A fair comparison with QAE would require that we employ a multi-reference theory, for example, the icMRUCC (internally contracted multi-reference unitary coupled cluster) \cite{Andreas}, where the reference state is a linear combination of multiple determinants. However, pursuing a multi-reference treatment poses certain challenges. First, one must survey the class
of multi-reference theories and identify the one that is best suited for VQE in terms of cost of implementation. Furthermore, in multi-reference theories where redundancies are not naturally accounted for, the computational cost to remove redundancies (duplicate states that
need to be removed) is not necessarily easy. For example, in the icMRUCC theory, the process of obtaining the indices for the linearly independent excitation amplitudes involves computing square root of an overlap matrix, which is computationally hard for large matrices. 

Given the current limitations in two-qubit gate fidelities, we implement only a single-parameter VQE using the UCCSD ansatz, focusing on measuring only the dominant terms in the Hamiltonian. To capture correlation energy effectively in a multi-reference approach, we would need to measure at least two dominant terms from the Hamiltonian. Consequently, the multi-reference theory requires more resources. Thus, our UCCSD results serve as an upper
bound to the values obtained from a multi-reference UCCSD approach, which, if implemented on a quantum computer with improved gate fidelities, would yield better quality results for the ground state energy than UCCSD in strongly correlated regimes. 

Our 26-parameter VQE-UCCSD calculation incurred 1716 $ECR$ gates (the number of one-qubit gates are 16500, 5844, and 1444 for $RZ$, $SX$, and $X$ respectively). We note that even if the device had been fully connected, the number of two-qubit gates, $\mathcal{N}_{2qg}$, is still 1440, which would yield a rather low result fidelity. For example, assuming an optimistic estimate of $0.999$ for the two-qubit gate fidelity (for comparison, state-of-the-art commercial quantum computers such as the IonQ Forte-I have current fidelities hovering in and around $0.99$), an estimate of the result fidelity even with a fictitious all-to-all connected topology is a mere $\sim 0.999^{1440} \approx 0.2$. Thus, in order to obtain meaningful results, one needs to aggressively optimize the quantum circuit in addition to implementing additional resource reduction strategies to reduce the number of measurements, etc \cite{Palak2024VQE}. The authors of the cited work show that with a suite of such classically resource-intensive strategies, the number of gates can be reduced to the extent that makes computation on current-day quantum devices possible. We borrow their techniques, which we will briefly describe for completeness further below. We begin by carrying out a VQE calculation on a traditional computer. We then carry out optimization on two fronts: wave function (the quantum circuit itself) and Hamiltonian (reducing the number of terms in the Hamiltonian, so that the number of circuits evaluated is reduced). On the circuit optimization side, we first pass our quantum circuit with the optimal parameters into three optimization routines sequentially: Qiskit $(L=3)$ \cite{Qiskit2021}, Pytket \cite{Sivarajah2021tket}, and PyZX \cite{kissinger2020Pyzx}. This was followed by an additional layer of circuit optimization using reinforcement learning-based ZX calculus routine and a causal flow preserving ZX calculus module \cite{riu2024rlzx}, followed by another round of Qiskit $(L=3)$ optimization to yield a quantum circuit with the following gate count: $RZ$: 38, $SX$: 19, $X$=11, and $ECR$: 12. On the Hamiltonian front, we group the Hamiltonian into cliques, which are sets of mutually (qubit-wise) commuting terms, and then group cliques that yield common energies into supercliques, based on ideas from Ref. \cite{mruccvqe2025}. We then pick the  dominant superclique (in energy) for our hardware computation. 

We measured two circuits (that belong to the two dominant supercliques) on the optimized quantum circuit built with optimized parameters to obtain the ground state energy. We find that in spite of reducing $\mathcal{N}_{2qg}$ to 12, we obtain an energy of $-1.71197$ Ha (recalling that they were obtained by averaging over 10 repeats, each with 4096 shots), with a percentage fraction difference of 7.81 \% relative to the value obtained in the same setting on a simulator (noiseless). Relative to FCI, we obtain a percentage fraction difference of 11.72 \%, since the problem has now been reduced to accommodate it on current gate-based quantum hardware. It is also worth adding that the correlation energy itself in the noiseless setting is $-0.08015$ Ha, whereas the one predicted from the IBM execution is significantly different at 0.06480 Ha. Augmenting the computations with error mitigation techniques (outside the scope of the current work) can improve the quality of the result, but it is clear that in order to carry out more resource intensive MRUCC-VQE computations even in small active spaces such as the one we have chosen would require further advances on the hardware front. On the other hand, the QAE algorithm gives reasonable values for the ground state energy ($-1.93760$ Ha with 5000 shots and averaged over 30 repetitions) on the D-Wave hardware.

\section{Analysis of scaling behaviours: a comparison}\label{sec:Comparison_with_vqe} 

\begin{figure*}[t]
\begin{tabular}{c}
\includegraphics[width=18cm]{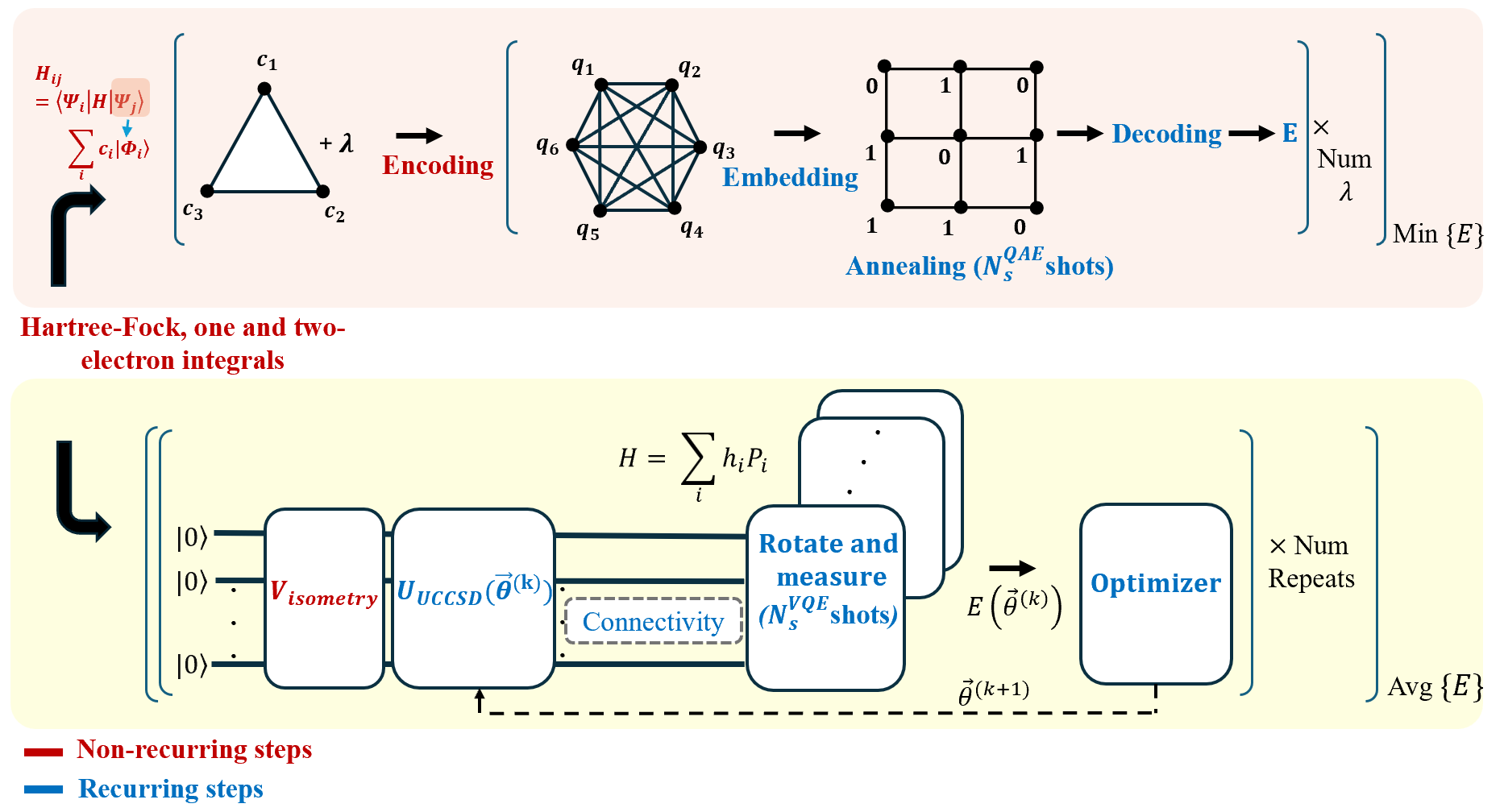}
\end{tabular}
 \caption{Schematic overview of the QAE (shaded light brown) and the VQE (shaded light yellow) algorithms with procedural steps categorized as non-recurring (red) or recurring (blue). Non-recurring steps common to both the algorithms include Hartree-Fock and generation of one- and two- electron integrals in the MO basis. Algorithm-specific non-recurring steps include generation of Hamiltonian matrix elements ($H_{ij}$) and QUBO encoding for QAE, while in VQE, we have input state preparation via $V_{isometry }$. Recurring steps in QAE encompasses embedding the fully connected problem graph onto the sparsely connected hardware graph, performing quantum annealing (with $N^{QAE}_s$ shots) and finally decoding to extract the real-valued coefficients $\{c_i\}$ which are combined with $H_{ij}$ to evaluate energy expectation value. In VQE, recurring steps include construction of the UCCSD ansatz, compensating for limited qubit connectivity by inserting SWAP gates (or CNOT gates), classical post-processing to interpret measurement histogram and finally passing the energy values to an optimizer. } \label{QAE_VQE}
\centering 
\end{figure*}

In the sections that follow, we comment on the overall cost of the QAE and the VQE algorithms by categorizing their procedural steps as non-recurring and recurring. Non-recurring steps refer to those performed only once during the execution of the algorithm, while 
recurring steps are repeated several times throughout the algorithm. Refer Figure~\ref{QAE_VQE}. For each algorithm, we identify the step with the highest scaling among both recurring and non-recurring steps, and report this as the net scaling for the algorithm. We then compare the net scaling expressions for both the algorithms to assess how QAE performs relative to VQE. 

\subsection{Assumptions} 
\begin{enumerate}[label=(\alph*)]
    \item We consider the traditional versions of both the QAE~\cite{Teplukhin2019CalculationAnnealer} and the VQE~\cite{Peruzzo2014VQE} algorithms. 
    \item We consider the chemistry-inspired UCCSD ansatz for VQE, and determinant-based configuration interaction in the singles and doubles approximation (CISD) for QAE (we use GUGA-based CI in our numerical calculations in the previous sections, but consider determinant-based CI for the scaling analysis on grounds of simplicity; as and when possible, we comment on GUGA-based CI). 
    \item We assume the existence of oracles, which supply the input state for VQE and a suitable $\lambda$ range for QAE (we recall that in QAE, we need to select a range of $\lambda$ values to scan, as well as the number of values in a selected range, the choice of which is not obvious, especially in light of the results shown in Figure \ref{lambda_ranges}). 
    \item On grounds of the problem being non-trivial, we assume that the number of shots for QAE is a function of the anneal time $t_a$, the precision $\epsilon$ and $N$, that is, $N^{QAE}_s = f(t_a, \epsilon, N)$. As $N$ increases, we have more possible bit strings that yield non-zero probability, and thus we expect that we need to supply more shots to capture the statistics reliably. It is also intuitive that if we seek better precision, we need to supply more shots in a computation to reduce statistical error. For VQE, the number of shots is $N_s^{VQE} = g(\epsilon, N)$.
\end{enumerate} 

\subsection{Non-recurring costs}

\begin{enumerate}[label=(\alph*)]
\item \textbf{One- and two- electron integrals: }Both QAE and VQE require us to supply one- and two- electron integrals. They are obtained from a classical computer, involving the HF (scales typically as $N^4$ for $N$ spin-orbitals~\cite{cramer}) and atomic orbital to molecular orbital integral transformation (scales as $N^5$~\cite{cramer}) steps (See Appendix \ref{cost_HF_AO_MO} for derivation). The number of Hamiltonian terms themselves, and thus the number of integrals, scale as $N^4$. This overhead can be reduced by employing techniques such as Hamiltonian factorization~\cite{sing_factor, double_factor, THC_factor} For example, the authors of Reference~\cite{exp_tens_hyp} suggest that one can ideally reach $N^2$ scaling with their explicit double factorization scheme. Reducing this overhead has the effect of reducing the number of circuit evaluations in VQE per iteration (since there are as many circuits per iteration as the number of Hamiltonian terms). On the other hand, in QAE, the `factorization' needs to be performed on the wave function in order to reduce the number of the quantum annealing runs. In fact, the GUGA-based approach that we employ in this work corresponds to this procedure. The determinant-based wave function expansion is inefficient with respect to the number of variables, because Slater determinants are not necessarily spin eigenfunctions. Qualitatively, GUGA generates spin symmetry-adapted CSFs by block-diagonalizing the electron spin $\mathbf{S}^2$ operator to reduce the number of variables to optimize.
\item \textbf{Evaluating Hamiltonian matrix elements: }Following the computation of integrals, the CI Hamiltonian matrix elements must be constructed classically prior to performing QAE. We therefore analyze the cost associated with such a construction. If there are $d_{CI}$ coefficients, then noting that the Hamiltonian is Hermitian, we need to evaluate $\frac{d_{CI}^2}{2} + \frac{d_{CI}}{2} \sim d_{CI}^2$ elements. We also introduce some standard notation: $n_o$ refers to the total number of occupied spin orbitals (i.e. total number of electrons), and $n_v$ the number of unoccupied spin orbitals. A cursory look shows that since the number of matrix elements to evaluate goes as $n_o^4n_v^4$ and the cost of constructing a CI matrix element, given the one- and two- electron integrals, by using the Slater--Condon rules goes at most as $n_o^2$ (corresponding to the case of the wavefunction in the bra and ket being the same), the cost of obtaining the CI Hamiltonian matrix elements is at most $n_o^6n_v^4$, that is, $N^{10}$. However, this naive evaluation does not account for sparsity of the Hamiltonian (via the Slater--Condon rules) and also assumes that every matrix element evaluation is as costly as the costliest one's evaluation. A more careful approach by accounting for the aforementioned considerations indicates a much cheaper scaling of $N^6$. The details of the derivation are presented in Appendix \ref{cisdHij}. The scaling behavior for evaluating matrix elements in GUGA-CI is very similar to that of the determinant-based approach. However, GUGA-CI incurs an extra cost which comes from evaluating additional coefficients associated with pathways of the Shavitt graph, which can introduce an extra scaling of $\mathcal{O}(N)$. Therefore the overall computational cost for GUGA scales as $\mathcal{O}(N^7)$~\cite{shavit1,shavit2}. While the number of Hamiltonian matrix elements are fewer in GUGA-based CI than for determinant-based CI, the time taken to evaluate them can be longer for the former, due to evaluation of the aforementioned coefficients. 

\begin{figure}[t]
\begin{tabular}{cc}
\includegraphics[width=4.5cm]{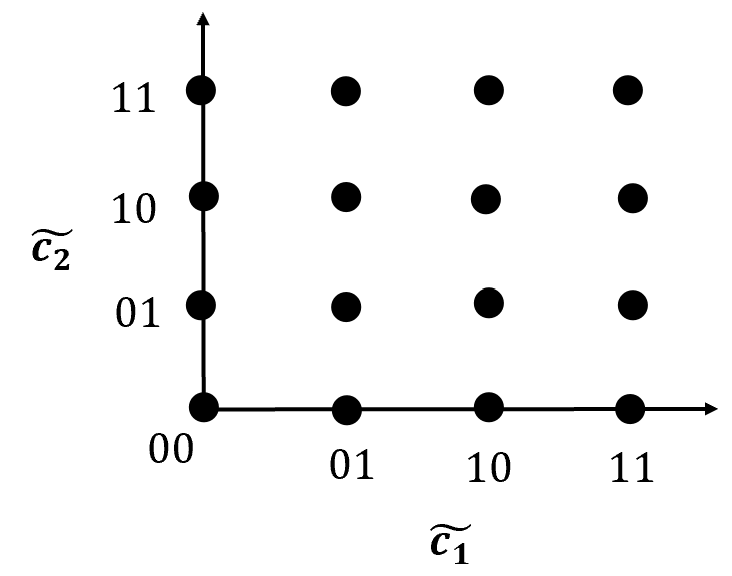}&
\includegraphics[width=4.5cm]{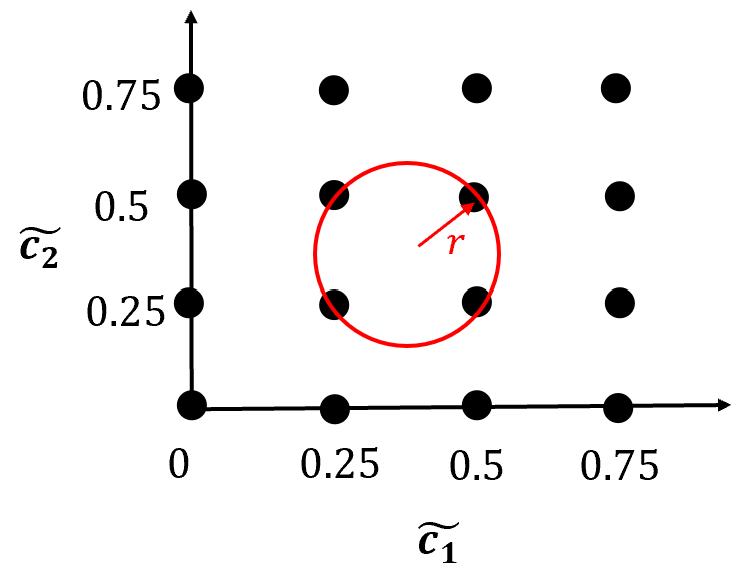}\\
(a)&(b)\\ 
\end{tabular}
 \caption{Optimization domain for $N=2$ and $K=2$ in (a) the space of binary variables and (b) the space of real variables. } 
\centering 
\label{binary_real_space}
\end{figure} 
\begin{table*}[t]
  \centering
    \caption{Table summarizing the worst case scaling for various steps of the QAE (for determinant-based CISD) and the VQE (for a multi-reference UCC) algorithms. The costs reported below correspond to an execution of QAE at a fixed $\lambda$ and a single iteration of VQE. The quantum computing overheads are marked in bold. In the table, $N$ is the total number of spin-orbitals. }
  \setlength{\tabcolsep}{10pt}
  \begin{tabular}{ccc}
  \hline \hline 
    Steps &QAE (end-to-end) &VQE (end-to-end)\\
    \hline 
    Hartree-Fock&$N^4$&$N^4$\\ 
     One- and two- electron integrals& $N^5$& $N^5$\\
     Evaluation of $H_{ij}$& $N^{6}$ & $-$ \\
     Number of CI coefficients& $N^4$ & $-$ \\
    \textbf{K scaling}& $\log_2(N)$& $-$\\
$\bm{\mathcal{N}^{\textbf{iso}}_{\text{\textbf{2qg}}}}$& $-$& $2^N$\footnote{Under the assumption that the number of determinants grow as $\sim log(N)$, for an $N$ qubit state for a representative example. Furthermore, our estimates are based on Qiskit's isometry routine. } \\
    $\bm{\mathcal{N}^{\textbf{uccsd}}_{\textbf{2qg}}}$& $-$& $N^9$\\
    Optimizer& $-$& $N^8$\footnote{Cost for the SLSQP optimizer. } \\
    \textbf{Embedding}& Num physical qubits (CISD) $\leq  N^8 log_2^2(N)$& Num additional SWAPs $\leq  N^3$\\ 
    Energy evaluation & $N^6$ & $N^8g(\epsilon, N)$\\ 
    \hline \hline 
  \end{tabular}
  \label{tab:2}
\end{table*} 
\item \textbf{CI coefficients and $K$:} We address the question of the scaling behaviour associated with the number of CI coefficients, $c_i$ as well as the required scaling for the discretization parameter, K which represents the number of qubits used to discretize each coefficient such that the optimization problem is over a set of binary variables. Although the number of coefficients in FCI scale as $d_{CI} \sim \binom{n_o}{m}\binom{n_v}{m} \sim \frac{n_o^m n^m_v}{(m!)^2}$, where $m$ is the excitation level, one performs truncated CI calculations where $m$ is fixed (for example, CISD, where $m=2$), and thus the number of coefficients, in practice, scales polynomially as $N^4$. 
\\
We qualitatively examine the scaling of K with the number of coefficients, $d_{CI}$. Each coefficient $c_i$ is approximated by a K bit binary representation defined as
\begin{eqnarray}\label{approx_coeff}
    c_i \sim \tilde{c}_i = 0.c_{i1} \ldots c_{ij}\ldots c_{iK},\ \nonumber \\  \forall j \text{ and each } i= 1,2,\ldots, d_{CI} \nonumber 
\end{eqnarray}
where $c_{ij} \in \{0,1\}$ implying that each coefficient can take $2^K$ distinct values. This allows us to visualize the binary optimization domain as an $d_{CI}$ dimensional hyper-cube with each side consisting of $2^K$ discrete points. For simplicity we show the case for $d_{CI}=2$ and $K=2$ in Figure \ref{binary_real_space} (a). The resulting space is uniformly discretized with neighbouring points spaced $\frac{1}{2^K}$ distance apart. Furthermore, we consider a unit cell of this space, which is a hyper-cube of volume $\frac{1}{2^{Kd_{CI}}}$ and circumscribe it by a hyper-sphere, as seen in Figure \ref{binary_real_space}(b). Thus, the discretization error is upper bounded by the radius of this hyper-sphere. Assuming equal discretization $\Delta x$ in each dimension, we have 
\begin{equation}
\begin{aligned}
    r^2 &= \sum^{d_{CI}}_{i=1} \bigg(\frac{\Delta x}{2}\bigg)^2\\
    r &= \sqrt{d_{CI}}\frac{\Delta x}{2}
\end{aligned}
\end{equation}
For constant error we demand $\sqrt{d_{CI}} \sim \frac{1}{\Delta x}$. Since $\Delta x = \frac{1}{2^K}$ we have
\begin{align}
    K \sim \log_2d_{CI}
\end{align}
Therefore K must grow logarithmically with the number of coefficients, that is, as $log_2{d_{CI}}$ which is equal to $log_2N^4$ for CISD. 
\item \textbf{Encoding in QAE : }The encoding procedure in QAE described in Equation~\eqref{encode_q_c} requires $\sim N^4\times K$ addition operations since each of the $N^4$ CI coefficients contribute $K$ summations.

\end{enumerate} 
\subsection{Recurring costs}
\subsubsection{QAE}
Recurring steps in the context of QAE involve those which are repeated $\mathcal{D}(N)$ times for each of the `$\mathcal{D}(N)$' $\lambda$ values in a given range. Below, we enumerate these steps along with their corresponding cost.
\begin{enumerate}[label=(\alph*)]

\item \textbf{Embedding:} The number of coefficients to optimize post the QAE encoding is $n_{coeff} = d_{CI} \times K$, which results in an $n_{coeff}$ qubit fully-connected problem graph. Minor embedding an $n_{coeff}$ qubit all-to-all connected problem requires a physical qubit overhead which in the worst case scales as $\mathcal{O}(n^2_{coeff})$ \cite{Ana}. Therefore, 
\begin{align}
    \text{Number of physical qubits} \sim \mathcal{O}(N^8 log^2(N)). 
\end{align}
\item \textbf{Shots: }We assume that a single execution of QAE (for a given $\lambda$ value in the range) requires $N^{Q}_s$ shots which exhibits a non-trivial dependence on $t_a, \epsilon,N $ i.e. $ N^{QAE}_s \sim f(t_a, \epsilon, N)$. 
\item \textbf{Decoding: }The decoding step, akin to the encoding step, incurs the same cost i.e. $\sim N^4\times K$ addition operations.
\item \textbf{Energy evaluation: }The expectation value of the Hamiltonian is computed classically as $\sum_{i,j} c_i c_jH_{ij}$. Although this involves two nested loops, each iterating over $N^4$ determinants, the overall cost scales as $\sim N^6$ corresponding to the asymptotic scaling of the number of non-zero Hamiltonian matrix elements. However, as each term in the summation is independent and can be distributed across multiple nodes in a high-performance computing cluster, the task can be efficiently parallelized. 
\item \textbf{Sorting: }Since the final energy is obtained by identifying the minimum, the cost of sorting $\mathcal{D}(N)$ energies must also be considered which goes as $\sim \mathcal{D}(N)$. 
\end{enumerate} 

\subsubsection{VQE} 

Recurring steps in the context of VQE refer to those which are repeated many times within a single VQE execution. The first two points below pertain to the recurring costs associated with the number of copies of the quantum circuit $U(\vec{\theta})\ket{\Phi_{inp}}$ to be prepared ($\ket{\Phi_{inp}}$ is the state constructed via $V_{isometry} \ket{0}^{\otimes N}$) and the last two are the recurring costs outside of the quantum circuit preparation, namely those involving expectation value calculation and the optimizer routine cost. Below, we list these steps along with their corresponding costs. 

\begin{enumerate}[label=(\alph*)]
\item \textbf{Number of two-qubit gates in the ansatz: }In the UCCSD ansatz, the number of two-qubit gates scale as $\sim N^5$. We can arrive at the estimate by considering that a Pauli gadget over $N$ qubits (recalling that the number of spin-orbitals, $N$, is the number of qubits in VQE) scales at most linearly in $N$ (assuming the Jordan--Wigner transformation), and that there are $n_o^2n_v^2$ double excitations and thus as many Pauli words. The analysis assumes order one step one Trotterization. Thus, for example, even for a calculation involving a small molecule, where $n_o$ and $n_v$ are about 20 and 40 respectively, $\mathcal{N}_{2qg}$ is already of the order of 100 million, and even with aggressive quantum circuit optimization strategies (for example, see Ref. \cite{Palak2024VQE}), it is unlikely to carry out such calculations at the physical qubit level on current and near-future quantum computers. 

\item \textbf{Embedding (Connectivity): }
In the gate model, the overhead of embedding can be approximately quantified in terms of the additional number of SWAP gates required to account for the limited connectivity between physical qubits. Consider the worst-case scenario of an $N$ qubit quantum circuit where every pair of qubits is connected by a CNOT. Suppose the hardware has only nearest neighbour 1D connectivity, that is, the qubits are arranged as nodes on a path graph (worst case scenario). In this case, the total number of CNOT gates which cannot be executed directly due to connectivity constraints are $\bigg(\frac{N(N-1)}{2} - N\bigg)$. Therefore the total number of SWAP gates incurred, $\mathcal{N}_{\mathrm{additional\ SWAPs}}$ are bounded by
\begin{equation}
\begin{aligned}
    \mathcal{N}_{\mathrm{additional\ SWAPs}}\\
    \leq \bigg(\frac{N(N-1)}{2} - N \bigg) \bigg(N-2\bigg)   
\end{aligned}
\end{equation}
which in the worst case scales as $\mathcal{O}(N^3)$. As a consequence, the overall result fidelity ($F_{\mathrm{res}}$) would degrade by a factor of $\sim F^{ \mathcal{N}_{\mathrm{additional\ SWAPs}}}$ where $F$ denotes the two qubit gate fidelity. This scaling arises from
\begin{eqnarray}
    F_{\mathrm{res}} \sim F^{\mathcal{N}_{\mathrm{2qg}}+ \mathcal{N}_{\mathrm{additional\ SWAPs}}}.
\end{eqnarray}
resulting in an exponential decrease of the result fidelity.

\item \textbf{Energy evaluation: }We recall that there are $N^4$ terms in the Hamiltonian, and thus in each of the $\mathbb{I}(N, \epsilon)$ iterations, one prepares $N^4$ copies of the UCCSD ansatz quantum circuit to obtain each of the $\langle H_i \rangle$ values. At the end of each quantum circuit, one needs to expend classical post-processing cost to extract $\langle H_i \rangle$ as given by Eq. \ref{eq:vqeexpec} (Appendix \ref{sec:additional_details}), by using the counts from the measurement outcomes from $N_s^{VQE}$ shots. We do not have the counts distributed across $2^N$ bitstrings with non-zero probabilities, but only $\sim N^4$ of them corresponding to those states which the UCCSD ansatz admits. Thus, obtaining the energy expectation value per iteration involves $N^4$ operations per $H_i$ for $N^4$ such Hamiltonian terms. The total cost thus comes out to be at most $N^8 \times \mathbb{I}(N, \epsilon)\times g(N, \epsilon)$. We assume a noiseless setting throughout for simplicity. 
\item \textbf{Optimizer: }For our numerical simulations whose results we had discussed in Section \ref{sec:hardware_results}, we use the SLSQP optimizer, which is widely employed in literature. For simplicity, we pick this as our optimizer of choice, although it is to be noted that the cost from optimization depends strongly on the chosen optimizer. For SLSQP, the cost function evaluation per iteration scales quadratically in the number of parameters (Refer section 2.2.4 in~\cite{slsqpreport}). This implies the cost scales roughly as $\sim N^8$. Therefore, the net cost from optimizer alone is $\sim N^8 \times \mathbb{I}(N, \epsilon)$. However, since the scaling of the number of iterations with system size is highly problem-specific and non-trivial in nature, we defer the analysis to a future study. 
\end{enumerate}

\subsection{Net scaling behaviours} 
We now identify the costliest term among both recurring and non-recurring steps for the QAE and VQE algorithms. For QAE, our empirical observations show that for small problem instances the cost incurred in executing the algorithm with $N^{QAE}_s$ shots is less than the cost of embedding evaluated per $\lambda$ division. Thus based on this simplifying assumption, we conclude that the net scaling for executing the QAE algorithm goes as $\sim N^8\log^2{N}\times \mathcal{D}(N)$. 

Likewise for VQE, the dominant contribution to the overall scaling comes from repeated ansatz preparation (assuming the existence of an oracle which supplies the input state). This is because the total number of two-qubit gates involved for the entire VQE execution with $g(N, \epsilon)$ shots and $\mathbb{I}(N, \epsilon)$ iterations is given by $\sim N^5$ per circuit multiplied by the total number of circuit evaluations $\sim N^4$ per iteration, thus resulting in the net scaling as $\sim N^9 \times g(N, \epsilon) \times \mathbb{I}(N,\epsilon)$. 

\begin{figure*}[t]
\begin{tabular}{c}
\includegraphics[width=15cm]{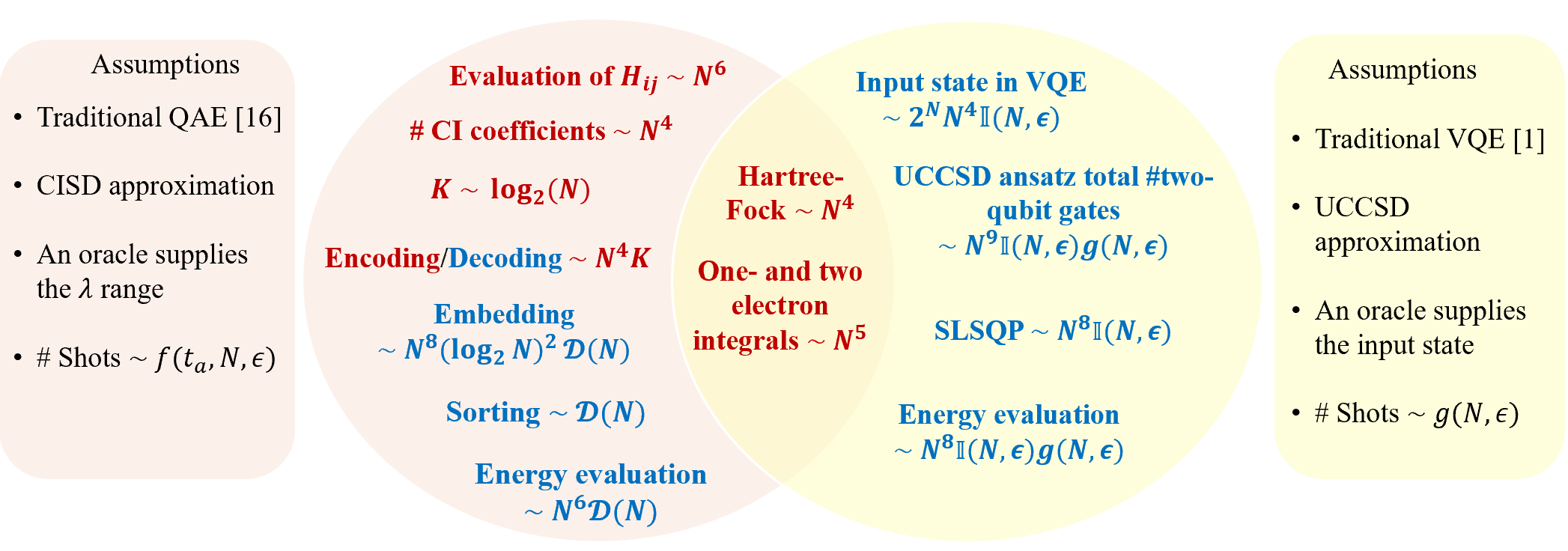} 
\end{tabular}
 \caption{Figure illustrating the non-recurring and recurring steps along with their associated costs for the entire QAE (shaded light brown) and VQE (shaded light yellow) algorithms.} \label{venn}
\centering 
\end{figure*}

Table \ref{tab:2} summarizes the upper bounds for the individual steps of the QAE and the VQE algorithms (end-to-end), while Figure \ref{venn} presents the information in a visual manner. Comparing the costliest terms in QAE ($N^8\log^2{N}\times \mathcal{D}(N)$ arising from embedding) and VQE ($N^9 \times g(N, \epsilon) \times \mathbb{I}(N, \epsilon)$ arising from circuit evaluations) we conclude that the question of which algorithm is costlier depends on the scaling of $g(N, \epsilon), \mathbb{I}(N, \epsilon)$ and $\mathcal{D}(N)$. 

For comparison with classical algorithms, a regular execution of a CCSD or a CISD routine on a traditional computer scales as $N^6$ and hence we do not expect to see any improvement in QAE or VQE relative to the classical methods. 

It is important to note that our finding does not definitively establish the scaling for QAE and VQE, and is to only be viewed as a first step towards a more rigorous scaling analysis for the future. Furthermore, VQE and QAE are both restricted by the current state-of-the-art quantum hardware; even with the current best NISQ computers, it is impractical to go beyond 5000 physical qubits and about 50 two-qubit gates for QAE and VQE executions respectively, and thus we do not expect any advantage anyway over traditional computers for such limited system sizes. 

For completeness, we provide additional details about both the algorithms that are not directly relevant to the scaling analysis but may be of interest in Appendix \ref{sec:additional_details}.

\section{Conclusion and future outlook} \label{sec:Conclusion}

In summary, we have explored the potential of the QAE algorithm as a NISQ tool for carrying out electronic structure calculations by (a) predicting an avoided crossing in the H$_4$ molecule on a D-Wave quantum computer, and (b) carrying out a preliminary yet detailed scaling analysis comparison between QAE and the celebrated gate-based VQE algorithm. We anticipate that our analysis would serve as a starting point to bridge a significant gap in literature with regard to the application of quantum annealers to NISQ-era chemistry, in contrast to the extensive body of research dedicated to the gate-based VQE algorithm in the same domain. 

We find from our calculations that using the QAE algorithm with 1000 shots and 10 repetitions yields AC with reasonable precision, to within about 20 \% of the reference full configuration interaction values. Furthermore, we find that the result can be substantially improved by increasing the number of shots to 5000, the number of repetitions to 30, and by choosing different anneal times for the ground and excited state energies. The choice of $\lambda$ range using SAE was found to be sufficient. The improved AC result with 5000 shots and 30 repetitions is to within about 1.2 \% of the FCI value. Among the factors considered, we find that the number of shots significantly influences the quality of our results. We do not carry out error mitigation in our computations for AC since our problem size is very small, although it may be required as we scale up in system size. 

Our hardware results indicate that in order to use the VQE algorithm to predict energies accurately within a considered active space, one requires quantum devices with gate fidelities well beyond the current best capabilities. On the other hand, the QAE algorithm is well-placed to obtain energies with reasonable precision on current-day D-Wave hardware. However, the next natural step of studying the effect of D-Wave hardware errors as we scale up the size of the molecular systems would require more physical qubits, in view of the embedding overheads. An alternative that has not been explored in this work is the use of sub-QUBOs to extend the calculations to larger system sizes in current D-Wave computers. 

It is also worth noting at this point that there are several other factors that could be tuned to improve the QAE results further, which we do not carry out in this pilot study. These include, but are not limited to: 

\begin{itemize}
\item Checking the effect of including more $\lambda$ values within a chosen $\lambda$ range. 
\item Studying the effect of the number of physical qubits that are expended for a computation via embedding. 
\item Our calculations are spread over a time period of about three months, and thus we expect that our results and analyses would have errors due to drift in machine parameters. It is, however, well beyond the scope of the current study to address the issue. 
\end{itemize} 

For the scaling analysis, we estimate non-recurring and recurring costs involved in executing the traditional versions of the QAE-CISD and the VQE-UCCSD algorithms, under the strong assumptions that oracles supply the input state for VQE and the bounds for $\lambda$ range choice for QAE. Furthermore, we assume that the number of shots in both of the algorithms is a non-trivial function of system size and precision sought, besides the anneal time in the case of QAE. We find that QAE scales as $\sim N^8\log^2{(N)}\times \mathcal{D}(N)$, while VQE scales as $\sim N^9 \times g(N, \epsilon) \times \mathbb{I}(N,\epsilon)$. Here, $\mathcal{D}(N)$ is the number of $\lambda$ divisions required for a QAE calculation, whereas $ g(N, \epsilon)$ is the required number of shots for VQE, and $\mathbb{I}(N,\epsilon)$ is the number of iterations for VQE. Thus, under our assumptions that these functions scale polynomially, both the algorithms consist of at most polynomially scaling sub-routines. However, they do not scale better than the classical approaches, CISD and CCSD, which both scale as $N^6$. 

We anticipate that our study paves way for more comprehensive and detailed studies on the underexplored QAE algorithm, and also for algorithmic advances in NISQ approaches for quantum chemistry in the quantum annealing framework. 

\begin{acknowledgments}
This work was supported by the European Commission EIC-Transition project RoCCQeT (GA 101112839). VSP acknowledges support from CRG grant (CRG/2023/002558) for some of the computations. AAZ and VSP sincerely acknowledge Palak Chawla for her tremendous support and patience in helping us with the VQE computations, and Disha Shetty for patiently reading the initial drafts and giving her feedback, which in turn helped improve the quality of the manuscript. VSP and AAZ acknowledge Prof. Bhanu Pratap Das for discussions on QAE and support with D-Wave time, and Prof. Debashis Mukherjee for discussions on the general nature of AC. The authors acknowledge Mr. Jose Miralles for fruitful discussions on Sub-QUBOs. KS acknowledges support from Quantum Leap Flagship Program (Grant No. JPMXS0120319794) from the MEXT, Japan, Center of Innovations for Sustainable Quantum AI (JPMJPF2221) from JST, Japan, and Grants-in-Aid for Scientific Research C (21K03407) and for Transformative Research Area B (23H03819) from JSPS, Japan. J.N and J.R thankfully acknowledge RES resources provided by Barcelona Supercomputing Center in Marenostrum 5 to FI-2025-1-0043.
\end{acknowledgments} 

\section{Data availability} 

All the data is included in the article. The details of our code will be available on reasonable request. 

\bibliography{apssamp} 

\begin{thebibliography}{73}%
\makeatletter
\providecommand \@ifxundefined [1]{%
 \@ifx{#1\undefined}
}%
\providecommand \@ifnum [1]{%
 \ifnum #1\expandafter \@firstoftwo
 \else \expandafter \@secondoftwo
 \fi
}%
\providecommand \@ifx [1]{%
 \ifx #1\expandafter \@firstoftwo
 \else \expandafter \@secondoftwo
 \fi
}%
\providecommand \natexlab [1]{#1}%
\providecommand \enquote  [1]{``#1''}%
\providecommand \bibnamefont  [1]{#1}%
\providecommand \bibfnamefont [1]{#1}%
\providecommand \citenamefont [1]{#1}%
\providecommand \href@noop [0]{\@secondoftwo}%
\providecommand \href [0]{\begingroup \@sanitize@url \@href}%
\providecommand \@href[1]{\@@startlink{#1}\@@href}%
\providecommand \@@href[1]{\endgroup#1\@@endlink}%
\providecommand \@sanitize@url [0]{\catcode `\\12\catcode `\$12\catcode `\&12\catcode `\#12\catcode `\^12\catcode `\_12\catcode `\%12\relax}%
\providecommand \@@startlink[1]{}%
\providecommand \@@endlink[0]{}%
\providecommand \url  [0]{\begingroup\@sanitize@url \@url }%
\providecommand \@url [1]{\endgroup\@href {#1}{\urlprefix }}%
\providecommand \urlprefix  [0]{URL }%
\providecommand \Eprint [0]{\href }%
\providecommand \doibase [0]{https://doi.org/}%
\providecommand \selectlanguage [0]{\@gobble}%
\providecommand \bibinfo  [0]{\@secondoftwo}%
\providecommand \bibfield  [0]{\@secondoftwo}%
\providecommand \translation [1]{[#1]}%
\providecommand \BibitemOpen [0]{}%
\providecommand \bibitemStop [0]{}%
\providecommand \bibitemNoStop [0]{.\EOS\space}%
\providecommand \EOS [0]{\spacefactor3000\relax}%
\providecommand \BibitemShut  [1]{\csname bibitem#1\endcsname}%
\let\auto@bib@innerbib\@empty
\bibitem [{\citenamefont {Peruzzo}\ \emph {et~al.}(2014)\citenamefont {Peruzzo} \emph {et~al.}}]{Peruzzo2014VQE}%
  \BibitemOpen
  \bibfield  {author} {\bibinfo {author} {\bibfnamefont {A.}~\bibnamefont {Peruzzo}} \emph {et~al.},\ }\href {https://www.nature.com/articles/ncomms5213} {\bibfield  {journal} {\bibinfo  {journal} {Nat. Commun.}\ }\textbf {\bibinfo {volume} {5}},\ \bibinfo {pages} {4213} (\bibinfo {year} {2014})}\BibitemShut {NoStop}%
\bibitem [{\citenamefont {Kandala}\ \emph {et~al.}(2017)\citenamefont {Kandala}, \citenamefont {Mezzacapo}, \citenamefont {Temme}, \citenamefont {Takita}, \citenamefont {Brink}, \citenamefont {Chow},\ and\ \citenamefont {Gambetta}}]{kandala2017hardware}%
  \BibitemOpen
  \bibfield  {author} {\bibinfo {author} {\bibfnamefont {A.}~\bibnamefont {Kandala}}, \bibinfo {author} {\bibfnamefont {A.}~\bibnamefont {Mezzacapo}}, \bibinfo {author} {\bibfnamefont {K.}~\bibnamefont {Temme}}, \bibinfo {author} {\bibfnamefont {M.}~\bibnamefont {Takita}}, \bibinfo {author} {\bibfnamefont {M.}~\bibnamefont {Brink}}, \bibinfo {author} {\bibfnamefont {J.~M.}\ \bibnamefont {Chow}},\ and\ \bibinfo {author} {\bibfnamefont {J.~M.}\ \bibnamefont {Gambetta}},\ }\href {https://www.nature.com/articles/nature23879} {\bibfield  {journal} {\bibinfo  {journal} {Nature}\ }\textbf {\bibinfo {volume} {549}},\ \bibinfo {pages} {242} (\bibinfo {year} {2017})}\BibitemShut {NoStop}%
\bibitem [{\citenamefont {Arute}\ \emph {et~al.}(2020)\citenamefont {Arute} \emph {et~al.}}]{google2020hartree}%
  \BibitemOpen
  \bibfield  {author} {\bibinfo {author} {\bibfnamefont {F.}~\bibnamefont {Arute}} \emph {et~al.},\ }\href {https://www.science.org/doi/10.1126/science.abb9811} {\bibfield  {journal} {\bibinfo  {journal} {Science}\ }\textbf {\bibinfo {volume} {369}},\ \bibinfo {pages} {1084} (\bibinfo {year} {2020})}\BibitemShut {NoStop}%
\bibitem [{\citenamefont {Nam}\ \emph {et~al.}(2020)\citenamefont {Nam} \emph {et~al.}}]{nam2020ground}%
  \BibitemOpen
  \bibfield  {author} {\bibinfo {author} {\bibfnamefont {Y.}~\bibnamefont {Nam}} \emph {et~al.},\ }\href {https://www.nature.com/articles/s41534-020-0259-3} {\bibfield  {journal} {\bibinfo  {journal} {npj Quantum Inf.}\ }\textbf {\bibinfo {volume} {6}},\ \bibinfo {pages} {33} (\bibinfo {year} {2020})}\BibitemShut {NoStop}%
\bibitem [{\citenamefont {Ying}\ \emph {et~al.}(2023)\citenamefont {Ying} \emph {et~al.}}]{Pan2023vqe12q}%
  \BibitemOpen
  \bibfield  {author} {\bibinfo {author} {\bibfnamefont {C.}~\bibnamefont {Ying}} \emph {et~al.},\ }\href {https://doi.org/10.1103/PhysRevLett.130.110601} {\bibfield  {journal} {\bibinfo  {journal} {Phys. Rev. Lett.}\ }\textbf {\bibinfo {volume} {130}},\ \bibinfo {pages} {110601} (\bibinfo {year} {2023})}\BibitemShut {NoStop}%
\bibitem [{\citenamefont {Sugisaki}\ \emph {et~al.}(2022{\natexlab{a}})\citenamefont {Sugisaki}, \citenamefont {Kato}, \citenamefont {Minato}, \citenamefont {Okuwaki},\ and\ \citenamefont {Mochizuki}}]{C2vpathway}%
  \BibitemOpen
  \bibfield  {author} {\bibinfo {author} {\bibfnamefont {K.}~\bibnamefont {Sugisaki}}, \bibinfo {author} {\bibfnamefont {T.}~\bibnamefont {Kato}}, \bibinfo {author} {\bibfnamefont {Y.}~\bibnamefont {Minato}}, \bibinfo {author} {\bibfnamefont {K.}~\bibnamefont {Okuwaki}},\ and\ \bibinfo {author} {\bibfnamefont {Y.}~\bibnamefont {Mochizuki}},\ }\href {https://pubs.rsc.org/en/content/articlelanding/2022/cp/d1cp04318h} {\bibfield  {journal} {\bibinfo  {journal} {Phys. Chem. Chem. Phys.}\ }\textbf {\bibinfo {volume} {24}},\ \bibinfo {pages} {8439} (\bibinfo {year} {2022}{\natexlab{a}})}\BibitemShut {NoStop}%
\bibitem [{\citenamefont {Guo}\ \emph {et~al.}(2024)\citenamefont {Guo} \emph {et~al.}}]{guo2024experimental}%
  \BibitemOpen
  \bibfield  {author} {\bibinfo {author} {\bibfnamefont {S.}~\bibnamefont {Guo}} \emph {et~al.},\ }\href {https://www.nature.com/articles/s41567-024-02530-z} {\bibfield  {journal} {\bibinfo  {journal} {Nat. Phys.}\ }\textbf {\bibinfo {volume} {20}},\ \bibinfo {pages} {1240} (\bibinfo {year} {2024})}\BibitemShut {NoStop}%
\bibitem [{\citenamefont {Ostaszewski}\ \emph {et~al.}(2021)\citenamefont {Ostaszewski}, \citenamefont {Grant},\ and\ \citenamefont {Benedetti}}]{Ostaszewski2021optimisation_qc}%
  \BibitemOpen
  \bibfield  {author} {\bibinfo {author} {\bibfnamefont {M.}~\bibnamefont {Ostaszewski}}, \bibinfo {author} {\bibfnamefont {E.}~\bibnamefont {Grant}},\ and\ \bibinfo {author} {\bibfnamefont {M.}~\bibnamefont {Benedetti}},\ }\href {https://quantum-journal.org/papers/q-2021-01-28-391/} {\bibfield  {journal} {\bibinfo  {journal} {{Quantum}}\ }\textbf {\bibinfo {volume} {5}},\ \bibinfo {pages} {391} (\bibinfo {year} {2021})}\BibitemShut {NoStop}%
\bibitem [{\citenamefont {Tang}\ \emph {et~al.}(2021)\citenamefont {Tang}, \citenamefont {Shkolnikov}, \citenamefont {Barron}, \citenamefont {Grimsley}, \citenamefont {Mayhall}, \citenamefont {Barnes},\ and\ \citenamefont {Economou}}]{Tang2021AdaptVqe}%
  \BibitemOpen
  \bibfield  {author} {\bibinfo {author} {\bibfnamefont {H.~L.}\ \bibnamefont {Tang}}, \bibinfo {author} {\bibfnamefont {V.}~\bibnamefont {Shkolnikov}}, \bibinfo {author} {\bibfnamefont {G.~S.}\ \bibnamefont {Barron}}, \bibinfo {author} {\bibfnamefont {H.~R.}\ \bibnamefont {Grimsley}}, \bibinfo {author} {\bibfnamefont {N.~J.}\ \bibnamefont {Mayhall}}, \bibinfo {author} {\bibfnamefont {E.}~\bibnamefont {Barnes}},\ and\ \bibinfo {author} {\bibfnamefont {S.~E.}\ \bibnamefont {Economou}},\ }\href {https://journals.aps.org/prxquantum/abstract/10.1103/PRXQuantum.2.020310} {\bibfield  {journal} {\bibinfo  {journal} {PRX Quantum}\ }\textbf {\bibinfo {volume} {2}},\ \bibinfo {pages} {020310} (\bibinfo {year} {2021})}\BibitemShut {NoStop}%
\bibitem [{\citenamefont {Villela}\ \emph {et~al.}(2022)\citenamefont {Villela}, \citenamefont {Prasannaa},\ and\ \citenamefont {Das}}]{Villela2022sim_ionizationenergy}%
  \BibitemOpen
  \bibfield  {author} {\bibinfo {author} {\bibfnamefont {R.}~\bibnamefont {Villela}}, \bibinfo {author} {\bibfnamefont {V.~S.}\ \bibnamefont {Prasannaa}},\ and\ \bibinfo {author} {\bibfnamefont {B.~P.}\ \bibnamefont {Das}},\ }\href {https://link.springer.com/article/10.1140/epjp/s13360-022-03198-1} {\bibfield  {journal} {\bibinfo  {journal} {Eur. Phys. J. Plus}\ }\textbf {\bibinfo {volume} {137}},\ \bibinfo {pages} {1017} (\bibinfo {year} {2022})}\BibitemShut {NoStop}%
\bibitem [{\citenamefont {Sugisaki}\ \emph {et~al.}(2021)\citenamefont {Sugisaki}, \citenamefont {Toyota}, \citenamefont {Sato}, \citenamefont {Shiomi},\ and\ \citenamefont {Takui}}]{sugisaki2021quantum}%
  \BibitemOpen
  \bibfield  {author} {\bibinfo {author} {\bibfnamefont {K.}~\bibnamefont {Sugisaki}}, \bibinfo {author} {\bibfnamefont {K.}~\bibnamefont {Toyota}}, \bibinfo {author} {\bibfnamefont {K.}~\bibnamefont {Sato}}, \bibinfo {author} {\bibfnamefont {D.}~\bibnamefont {Shiomi}},\ and\ \bibinfo {author} {\bibfnamefont {T.}~\bibnamefont {Takui}},\ }\href {https://pubs.acs.org/doi/10.1021/acs.jpclett.1c00283} {\bibfield  {journal} {\bibinfo  {journal} {J. Phys. Chem. Lett.}\ }\textbf {\bibinfo {volume} {12}},\ \bibinfo {pages} {2880} (\bibinfo {year} {2021})}\BibitemShut {NoStop}%
\bibitem [{\citenamefont {Sumeet}\ \emph {et~al.}(2022)\citenamefont {Sumeet}, \citenamefont {Prasannaa~V}, \citenamefont {Das},\ and\ \citenamefont {Sahoo}}]{Sumeet2022precision_vqe}%
  \BibitemOpen
  \bibfield  {author} {\bibinfo {author} {\bibnamefont {Sumeet}}, \bibinfo {author} {\bibfnamefont {S.}~\bibnamefont {Prasannaa~V}}, \bibinfo {author} {\bibfnamefont {B.~P.}\ \bibnamefont {Das}},\ and\ \bibinfo {author} {\bibfnamefont {B.~K.}\ \bibnamefont {Sahoo}},\ }\href {https://www.mdpi.com/2624-960X/4/2/12} {\bibfield  {journal} {\bibinfo  {journal} {Quantum Rep.}\ }\textbf {\bibinfo {volume} {4}},\ \bibinfo {pages} {173} (\bibinfo {year} {2022})}\BibitemShut {NoStop}%
\bibitem [{\citenamefont {Bharti}\ \emph {et~al.}(2022)\citenamefont {Bharti} \emph {et~al.}}]{Bharti2022NISQalgo}%
  \BibitemOpen
  \bibfield  {author} {\bibinfo {author} {\bibfnamefont {K.}~\bibnamefont {Bharti}} \emph {et~al.},\ }\href {https://doi.org/10.1103/RevModPhys.94.015004} {\bibfield  {journal} {\bibinfo  {journal} {Rev. Mod. Phys.}\ }\textbf {\bibinfo {volume} {94}},\ \bibinfo {pages} {015004} (\bibinfo {year} {2022})}\BibitemShut {NoStop}%
\bibitem [{\citenamefont {Motta}\ and\ \citenamefont {Rice}(2022)}]{Mario2021QCalgos_chemistry}%
  \BibitemOpen
  \bibfield  {author} {\bibinfo {author} {\bibfnamefont {M.}~\bibnamefont {Motta}}\ and\ \bibinfo {author} {\bibfnamefont {J.~E.}\ \bibnamefont {Rice}},\ }\href {https://wires.onlinelibrary.wiley.com/doi/abs/10.1002/wcms.1580} {\bibfield  {journal} {\bibinfo  {journal} {Wiley Interdiscip. Rev. Comput. Mol. Sci.}\ }\textbf {\bibinfo {volume} {12}},\ \bibinfo {pages} {e1580} (\bibinfo {year} {2022})}\BibitemShut {NoStop}%
\bibitem [{\citenamefont {Chawla}\ \emph {et~al.}(2025{\natexlab{a}})\citenamefont {Chawla} \emph {et~al.}}]{Palak2024VQE}%
  \BibitemOpen
  \bibfield  {author} {\bibinfo {author} {\bibfnamefont {P.}~\bibnamefont {Chawla}} \emph {et~al.},\ }\href {https://journals.aps.org/pra/abstract/10.1103/PhysRevA.111.022817} {\bibfield  {journal} {\bibinfo  {journal} {Phys. Rev. A}\ }\textbf {\bibinfo {volume} {111}},\ \bibinfo {pages} {022817} (\bibinfo {year} {2025}{\natexlab{a}})}\BibitemShut {NoStop}%
\bibitem [{\citenamefont {Teplukhin}\ \emph {et~al.}(2019)\citenamefont {Teplukhin}, \citenamefont {Kendrick},\ and\ \citenamefont {Babikov}}]{Teplukhin2019CalculationAnnealer}%
  \BibitemOpen
  \bibfield  {author} {\bibinfo {author} {\bibfnamefont {A.}~\bibnamefont {Teplukhin}}, \bibinfo {author} {\bibfnamefont {B.~K.}\ \bibnamefont {Kendrick}},\ and\ \bibinfo {author} {\bibfnamefont {D.}~\bibnamefont {Babikov}},\ }\href {https://pubs.acs.org/doi/10.1021/acs.jctc.9b00402} {\bibfield  {journal} {\bibinfo  {journal} {J. Chem. Theory Comput.}\ }\textbf {\bibinfo {volume} {15}},\ \bibinfo {pages} {4555} (\bibinfo {year} {2019})}\BibitemShut {NoStop}%
\bibitem [{\citenamefont {Kumar}\ \emph {et~al.}(2024)\citenamefont {Kumar}, \citenamefont {Baskaran}, \citenamefont {Prasannaa}, \citenamefont {Sugisaki}, \citenamefont {Mukherjee}, \citenamefont {Dyall},\ and\ \citenamefont {Das}}]{Vikrant2024}%
  \BibitemOpen
  \bibfield  {author} {\bibinfo {author} {\bibfnamefont {V.}~\bibnamefont {Kumar}}, \bibinfo {author} {\bibfnamefont {N.}~\bibnamefont {Baskaran}}, \bibinfo {author} {\bibfnamefont {V.~S.}\ \bibnamefont {Prasannaa}}, \bibinfo {author} {\bibfnamefont {K.}~\bibnamefont {Sugisaki}}, \bibinfo {author} {\bibfnamefont {D.}~\bibnamefont {Mukherjee}}, \bibinfo {author} {\bibfnamefont {K.~G.}\ \bibnamefont {Dyall}},\ and\ \bibinfo {author} {\bibfnamefont {B.~P.}\ \bibnamefont {Das}},\ }\href {https://doi.org/https://doi.org/10.1103/PhysRevA.109.042808} {\bibfield  {journal} {\bibinfo  {journal} {Phys. Rev. A}\ }\textbf {\bibinfo {volume} {109}},\ \bibinfo {pages} {042808} (\bibinfo {year} {2024})}\BibitemShut {NoStop}%
\bibitem [{\citenamefont {Teplukhin}\ \emph {et~al.}(2020{\natexlab{a}})\citenamefont {Teplukhin}, \citenamefont {Kendrick},\ and\ \citenamefont {Babikov}}]{Teplukhin2020complex}%
  \BibitemOpen
  \bibfield  {author} {\bibinfo {author} {\bibfnamefont {A.}~\bibnamefont {Teplukhin}}, \bibinfo {author} {\bibfnamefont {B.~K.}\ \bibnamefont {Kendrick}},\ and\ \bibinfo {author} {\bibfnamefont {D.}~\bibnamefont {Babikov}},\ }\href {https://pubs.rsc.org/en/content/articlelanding/2020/cp/d0cp04272b} {\bibfield  {journal} {\bibinfo  {journal} {Phys. Chem. Chem. Phys.}\ }\textbf {\bibinfo {volume} {22}},\ \bibinfo {pages} {26136} (\bibinfo {year} {2020}{\natexlab{a}})}\BibitemShut {NoStop}%
\bibitem [{\citenamefont {Teplukhin}\ \emph {et~al.}(2020{\natexlab{b}})\citenamefont {Teplukhin}, \citenamefont {Kendrick}, \citenamefont {Tretiak},\ and\ \citenamefont {Dub}}]{Teplukhin2020ElectronicAnnealer}%
  \BibitemOpen
  \bibfield  {author} {\bibinfo {author} {\bibfnamefont {A.}~\bibnamefont {Teplukhin}}, \bibinfo {author} {\bibfnamefont {B.~K.}\ \bibnamefont {Kendrick}}, \bibinfo {author} {\bibfnamefont {S.}~\bibnamefont {Tretiak}},\ and\ \bibinfo {author} {\bibfnamefont {P.~A.}\ \bibnamefont {Dub}},\ }\href {https://doi.org/10.1038/s41598-020-77315-4} {\bibfield  {journal} {\bibinfo  {journal} {Sci. Rep.}\ }\textbf {\bibinfo {volume} {10}},\ \bibinfo {pages} {20753} (\bibinfo {year} {2020}{\natexlab{b}})}\BibitemShut {NoStop}%
\bibitem [{\citenamefont {Teplukhin}\ \emph {et~al.}(2021)\citenamefont {Teplukhin} \emph {et~al.}}]{Teplukhin2021ComputingAnnealer}%
  \BibitemOpen
  \bibfield  {author} {\bibinfo {author} {\bibfnamefont {A.}~\bibnamefont {Teplukhin}} \emph {et~al.},\ }\href {https://doi.org/10.1038/s41598-021-98331-y} {\bibfield  {journal} {\bibinfo  {journal} {Sci. Rep.}\ }\textbf {\bibinfo {volume} {11}},\ \bibinfo {pages} {18796} (\bibinfo {year} {2021})}\BibitemShut {NoStop}%
\bibitem [{\citenamefont {Illa}\ and\ \citenamefont {Savage}(2022)}]{Illaqae}%
  \BibitemOpen
  \bibfield  {author} {\bibinfo {author} {\bibfnamefont {M.}~\bibnamefont {Illa}}\ and\ \bibinfo {author} {\bibfnamefont {M.~J.}\ \bibnamefont {Savage}},\ }\href {https://doi.org/https://doi.org/10.1103/PhysRevA.106.052605} {\bibfield  {journal} {\bibinfo  {journal} {Phys. Rev. A}\ }\textbf {\bibinfo {volume} {106}},\ \bibinfo {pages} {052605} (\bibinfo {year} {2022})}\BibitemShut {NoStop}%
\bibitem [{\citenamefont {Rahman}\ \emph {et~al.}(2020)\citenamefont {Rahman}, \citenamefont {Lewis}, \citenamefont {Mendicelli},\ and\ \citenamefont {Powell}}]{Rahmanqae}%
  \BibitemOpen
  \bibfield  {author} {\bibinfo {author} {\bibfnamefont {S.~A.}\ \bibnamefont {Rahman}}, \bibinfo {author} {\bibfnamefont {R.}~\bibnamefont {Lewis}}, \bibinfo {author} {\bibfnamefont {E.}~\bibnamefont {Mendicelli}},\ and\ \bibinfo {author} {\bibfnamefont {S.}~\bibnamefont {Powell}},\ }\href {https://doi.org/https://doi.org/10.1103/PhysRevD.104.034501} {\bibfield  {journal} {\bibinfo  {journal} {Phys. Rev. D}\ }\textbf {\bibinfo {volume} {104}},\ \bibinfo {pages} {034501} (\bibinfo {year} {2020})}\BibitemShut {NoStop}%
\bibitem [{\citenamefont {Sugisaki}\ \emph {et~al.}(2022{\natexlab{b}})\citenamefont {Sugisaki}, \citenamefont {Kato}, \citenamefont {Minato}, \citenamefont {Okuwaki},\ and\ \citenamefont {Mochizuki}}]{scorrel000}%
  \BibitemOpen
  \bibfield  {author} {\bibinfo {author} {\bibfnamefont {K.}~\bibnamefont {Sugisaki}}, \bibinfo {author} {\bibfnamefont {T.}~\bibnamefont {Kato}}, \bibinfo {author} {\bibfnamefont {Y.}~\bibnamefont {Minato}}, \bibinfo {author} {\bibfnamefont {K.}~\bibnamefont {Okuwaki}},\ and\ \bibinfo {author} {\bibfnamefont {Y.}~\bibnamefont {Mochizuki}},\ }\href {https://pubs.rsc.org/en/content/articlelanding/2022/cp/d1cp04318h} {\bibfield  {journal} {\bibinfo  {journal} {Phys. Chem. Chem. Phys.}\ }\textbf {\bibinfo {volume} {24}},\ \bibinfo {pages} {8439} (\bibinfo {year} {2022}{\natexlab{b}})}\BibitemShut {NoStop}%
\bibitem [{\citenamefont {Halder}\ \emph {et~al.}(2022)\citenamefont {Halder}, \citenamefont {Prasannaa},\ and\ \citenamefont {Maitra}}]{scorrel00}%
  \BibitemOpen
  \bibfield  {author} {\bibinfo {author} {\bibfnamefont {D.}~\bibnamefont {Halder}}, \bibinfo {author} {\bibfnamefont {V.~S.}\ \bibnamefont {Prasannaa}},\ and\ \bibinfo {author} {\bibfnamefont {R.}~\bibnamefont {Maitra}},\ }\href {https://pubs.aip.org/aip/jcp/article-abstract/157/17/174117/2842086/Dual-exponential-coupled-cluster-theory-Unitary?redirectedFrom=fulltext} {\bibfield  {journal} {\bibinfo  {journal} {J. Chem. Phys.}\ }\textbf {\bibinfo {volume} {157}},\ \bibinfo {pages} {174117} (\bibinfo {year} {2022})}\BibitemShut {NoStop}%
\bibitem [{\citenamefont {Halder}\ \emph {et~al.}(2023)\citenamefont {Halder}, \citenamefont {Prasannaa}, \citenamefont {Agarawal},\ and\ \citenamefont {Maitra}}]{scorrel01}%
  \BibitemOpen
  \bibfield  {author} {\bibinfo {author} {\bibfnamefont {D.}~\bibnamefont {Halder}}, \bibinfo {author} {\bibfnamefont {V.~S.}\ \bibnamefont {Prasannaa}}, \bibinfo {author} {\bibfnamefont {V.}~\bibnamefont {Agarawal}},\ and\ \bibinfo {author} {\bibfnamefont {R.}~\bibnamefont {Maitra}},\ }\href {https://onlinelibrary.wiley.com/doi/abs/10.1002/qua.27021} {\bibfield  {journal} {\bibinfo  {journal} {Int. J. Quantum Chem.}\ }\textbf {\bibinfo {volume} {123}},\ \bibinfo {pages} {e27021} (\bibinfo {year} {2023})}\BibitemShut {NoStop}%
\bibitem [{\citenamefont {Sokolov}\ \emph {et~al.}(2020)\citenamefont {Sokolov}, \citenamefont {Barkoutsos}, \citenamefont {Ollitrault}, \citenamefont {Greenberg}, \citenamefont {Rice}, \citenamefont {Pistoia},\ and\ \citenamefont {Tavernelli}}]{scorrel02}%
  \BibitemOpen
  \bibfield  {author} {\bibinfo {author} {\bibfnamefont {I.~O.}\ \bibnamefont {Sokolov}}, \bibinfo {author} {\bibfnamefont {P.~K.}\ \bibnamefont {Barkoutsos}}, \bibinfo {author} {\bibfnamefont {P.~J.}\ \bibnamefont {Ollitrault}}, \bibinfo {author} {\bibfnamefont {D.}~\bibnamefont {Greenberg}}, \bibinfo {author} {\bibfnamefont {J.}~\bibnamefont {Rice}}, \bibinfo {author} {\bibfnamefont {M.}~\bibnamefont {Pistoia}},\ and\ \bibinfo {author} {\bibfnamefont {I.}~\bibnamefont {Tavernelli}},\ }\href {https://pubs.aip.org/aip/jcp/article-abstract/152/12/124107/953761/Quantum-orbital-optimized-unitary-coupled-cluster?redirectedFrom=fulltext} {\bibfield  {journal} {\bibinfo  {journal} {J. Chem. Phys.}\ }\textbf {\bibinfo {volume} {152}},\ \bibinfo {pages} {124107} (\bibinfo {year} {2020})}\BibitemShut {NoStop}%
\bibitem [{\citenamefont {Pavosevic}\ \emph {et~al.}(2023)\citenamefont {Pavosevic}, \citenamefont {Tavernelli},\ and\ \citenamefont {Rubio}}]{scorrel0}%
  \BibitemOpen
  \bibfield  {author} {\bibinfo {author} {\bibfnamefont {F.}~\bibnamefont {Pavosevic}}, \bibinfo {author} {\bibfnamefont {I.}~\bibnamefont {Tavernelli}},\ and\ \bibinfo {author} {\bibfnamefont {A.}~\bibnamefont {Rubio}},\ }\href {https://pubs.acs.org/doi/10.1021/acs.jpclett.3c01935} {\bibfield  {journal} {\bibinfo  {journal} {J. Phys. Chem. Lett.}\ }\textbf {\bibinfo {volume} {14}},\ \bibinfo {pages} {7876} (\bibinfo {year} {2023})}\BibitemShut {NoStop}%
\bibitem [{\citenamefont {Rossmannek}\ \emph {et~al.}(2023)\citenamefont {Rossmannek}, \citenamefont {Pavosevic}, \citenamefont {Rubio},\ and\ \citenamefont {Tavernelli}}]{scorrel1}%
  \BibitemOpen
  \bibfield  {author} {\bibinfo {author} {\bibfnamefont {M.}~\bibnamefont {Rossmannek}}, \bibinfo {author} {\bibfnamefont {F.}~\bibnamefont {Pavosevic}}, \bibinfo {author} {\bibfnamefont {A.}~\bibnamefont {Rubio}},\ and\ \bibinfo {author} {\bibfnamefont {I.}~\bibnamefont {Tavernelli}},\ }\href {https://pubs.acs.org/doi/10.1021/acs.jpclett.3c00330} {\bibfield  {journal} {\bibinfo  {journal} {J. Phys. Chem. Lett.}\ }\textbf {\bibinfo {volume} {14}},\ \bibinfo {pages} {3491} (\bibinfo {year} {2023})}\BibitemShut {NoStop}%
\bibitem [{\citenamefont {Xia}\ \emph {et~al.}(2018)\citenamefont {Xia}, \citenamefont {Bian},\ and\ \citenamefont {Kais}}]{Xia2018ElectronicHamiltonian}%
  \BibitemOpen
  \bibfield  {author} {\bibinfo {author} {\bibfnamefont {R.}~\bibnamefont {Xia}}, \bibinfo {author} {\bibfnamefont {T.}~\bibnamefont {Bian}},\ and\ \bibinfo {author} {\bibfnamefont {S.}~\bibnamefont {Kais}},\ }\href {https://doi.org/10.1021/ACS.JPCB.7B10371/ASSET/IMAGES/MEDIUM/JP-2017-10371H{\_}0008.GIF} {\bibfield  {journal} {\bibinfo  {journal} {J. Phys. Chem. B}\ }\textbf {\bibinfo {volume} {122}},\ \bibinfo {pages} {3384} (\bibinfo {year} {2018})}\BibitemShut {NoStop}%
\bibitem [{\citenamefont {Streif}\ \emph {et~al.}(2019)\citenamefont {Streif}, \citenamefont {Neukart},\ and\ \citenamefont {Leib}}]{XBKhardware}%
  \BibitemOpen
  \bibfield  {author} {\bibinfo {author} {\bibfnamefont {M.}~\bibnamefont {Streif}}, \bibinfo {author} {\bibfnamefont {F.}~\bibnamefont {Neukart}},\ and\ \bibinfo {author} {\bibfnamefont {M.}~\bibnamefont {Leib}},\ }\href@noop {} {\bibfield  {journal} {\bibinfo  {journal} {Quantum Technology and Optimization Problems QTOP 2019}\ }\textbf {\bibinfo {volume} {11413}} (\bibinfo {year} {2019})}\BibitemShut {NoStop}%
\bibitem [{\citenamefont {Ryabinkin}\ \emph {et~al.}(2018)\citenamefont {Ryabinkin}, \citenamefont {Yen}, \citenamefont {Genin},\ and\ \citenamefont {Izmaylov}}]{QCC}%
  \BibitemOpen
  \bibfield  {author} {\bibinfo {author} {\bibfnamefont {I.~G.}\ \bibnamefont {Ryabinkin}}, \bibinfo {author} {\bibfnamefont {T.~C.}\ \bibnamefont {Yen}}, \bibinfo {author} {\bibfnamefont {S.~N.}\ \bibnamefont {Genin}},\ and\ \bibinfo {author} {\bibfnamefont {A.}~\bibnamefont {Izmaylov}},\ }\href {https://pubs.acs.org/doi/pdf/10.1021/acs.jctc.8b00932?casa_token=uie_w1hepjkAAAAA:yCRmMnCOVKQ0bsomWWjSauGVNRou7pUMgqYI8AqhFgMZQElopCuAWzsYBV80W5topNgUulJT2nwNT1o} {\bibfield  {journal} {\bibinfo  {journal} {J. Chem. Theory Comput.}\ }\textbf {\bibinfo {volume} {14}},\ \bibinfo {pages} {6317} (\bibinfo {year} {2018})}\BibitemShut {NoStop}%
\bibitem [{\citenamefont {Genin}\ \emph {et~al.}(2019)\citenamefont {Genin}, \citenamefont {Ryabinkin},\ and\ \citenamefont {Izmaylov}}]{QuantChemonQA}%
  \BibitemOpen
  \bibfield  {author} {\bibinfo {author} {\bibfnamefont {S.~N.}\ \bibnamefont {Genin}}, \bibinfo {author} {\bibfnamefont {I.~G.}\ \bibnamefont {Ryabinkin}},\ and\ \bibinfo {author} {\bibfnamefont {A.~F.}\ \bibnamefont {Izmaylov}},\ }\href {https://arxiv.org/pdf/1901.04715} {\bibfield  {journal} {\bibinfo  {journal} {arXiv:1901.04715}\ } (\bibinfo {year} {2019})}\BibitemShut {NoStop}%
\bibitem [{\citenamefont {Copenhaver}\ \emph {et~al.}(2021)\citenamefont {Copenhaver}, \citenamefont {Wasserman},\ and\ \citenamefont {Kaufmann}}]{QCC2}%
  \BibitemOpen
  \bibfield  {author} {\bibinfo {author} {\bibfnamefont {J.}~\bibnamefont {Copenhaver}}, \bibinfo {author} {\bibfnamefont {A.}~\bibnamefont {Wasserman}},\ and\ \bibinfo {author} {\bibfnamefont {B.~W.}\ \bibnamefont {Kaufmann}},\ }\href {https://pubs.aip.org/aip/jcp/article-abstract/154/3/034105/199868/Using-quantum-annealers-to-calculate-ground-state?redirectedFrom=PDF} {\bibfield  {journal} {\bibinfo  {journal} {J. Chem. Phys.}\ }\textbf {\bibinfo {volume} {154}},\ \bibinfo {pages} {034105} (\bibinfo {year} {2021})}\BibitemShut {NoStop}%
\bibitem [{\citenamefont {Paldus}\ \emph {et~al.}(1993)\citenamefont {Paldus}, \citenamefont {Piecuch}, \citenamefont {Pylypow},\ and\ \citenamefont {Jeziorski}}]{H4paper}%
  \BibitemOpen
  \bibfield  {author} {\bibinfo {author} {\bibfnamefont {J.}~\bibnamefont {Paldus}}, \bibinfo {author} {\bibfnamefont {P.}~\bibnamefont {Piecuch}}, \bibinfo {author} {\bibfnamefont {L.}~\bibnamefont {Pylypow}},\ and\ \bibinfo {author} {\bibfnamefont {B.}~\bibnamefont {Jeziorski}},\ }\href {https://doi.org/10.1103/PhysRevA.47.2738} {\bibfield  {journal} {\bibinfo  {journal} {Phys. Rev. A}\ }\textbf {\bibinfo {volume} {47}},\ \bibinfo {pages} {2738} (\bibinfo {year} {1993})}\BibitemShut {NoStop}%
\bibitem [{\citenamefont {Barca}\ \emph {et~al.}(2022)\citenamefont {Barca} \emph {et~al.}}]{GAMESS}%
  \BibitemOpen
  \bibfield  {author} {\bibinfo {author} {\bibfnamefont {G.~M.~J.}\ \bibnamefont {Barca}} \emph {et~al.},\ }\href {https://doi.org/10.1063/5.0005188} {\bibfield  {journal} {\bibinfo  {journal} {J. Chem. Phys.}\ }\textbf {\bibinfo {volume} {152}},\ \bibinfo {pages} {154102} (\bibinfo {year} {2022})}\BibitemShut {NoStop}%
\bibitem [{\citenamefont {Kadowaki}\ and\ \citenamefont {Nishimori}(1998)}]{Kadowaki1998quantum_annealing}%
  \BibitemOpen
  \bibfield  {author} {\bibinfo {author} {\bibfnamefont {T.}~\bibnamefont {Kadowaki}}\ and\ \bibinfo {author} {\bibfnamefont {H.}~\bibnamefont {Nishimori}},\ }\href {https://journals.aps.org/pre/abstract/10.1103/PhysRevE.58.5355} {\bibfield  {journal} {\bibinfo  {journal} {Phys. Rev. E}\ }\textbf {\bibinfo {volume} {58}},\ \bibinfo {pages} {5355} (\bibinfo {year} {1998})}\BibitemShut {NoStop}%
\bibitem [{\citenamefont {Kadowaki}(2002)}]{Kadowaki2002quantum_annealing}%
  \BibitemOpen
  \bibfield  {author} {\bibinfo {author} {\bibfnamefont {T.}~\bibnamefont {Kadowaki}},\ }\href {https://arxiv.org/abs/quant-ph/0205020} {\bibfield  {journal} {\bibinfo  {journal} {arXiv preprint quant-ph/0205020}\ } (\bibinfo {year} {2002})}\BibitemShut {NoStop}%
\bibitem [{\citenamefont {Farhi}\ \emph {et~al.}(2000)\citenamefont {Farhi}, \citenamefont {Goldstone}, \citenamefont {Gutmann},\ and\ \citenamefont {Sipser}}]{Farhi2000adiabatic_computation}%
  \BibitemOpen
  \bibfield  {author} {\bibinfo {author} {\bibfnamefont {E.}~\bibnamefont {Farhi}}, \bibinfo {author} {\bibfnamefont {J.}~\bibnamefont {Goldstone}}, \bibinfo {author} {\bibfnamefont {S.}~\bibnamefont {Gutmann}},\ and\ \bibinfo {author} {\bibfnamefont {M.}~\bibnamefont {Sipser}},\ }\href {https://arxiv.org/abs/quant-ph/0001106} {\bibfield  {journal} {\bibinfo  {journal} {arXiv preprint quant-ph/0001106}\ } (\bibinfo {year} {2000})}\BibitemShut {NoStop}%
\bibitem [{\citenamefont {Fan}\ \emph {et~al.}(2015)\citenamefont {Fan}, \citenamefont {Dong},\ and\ \citenamefont {Piecuch}}]{Piecuch}%
  \BibitemOpen
  \bibfield  {author} {\bibinfo {author} {\bibnamefont {Fan}}, \bibinfo {author} {\bibfnamefont {P.}~\bibnamefont {Dong}},\ and\ \bibinfo {author} {\bibfnamefont {P.}~\bibnamefont {Piecuch}},\ }\href {https://www.sciencedirect.com/science/article/abs/pii/S0065327606510019} {\bibfield  {journal} {\bibinfo  {journal} {Adv. Quantum Chem.}\ }\textbf {\bibinfo {volume} {51}},\ \bibinfo {pages} {1} (\bibinfo {year} {2015})}\BibitemShut {NoStop}%
\bibitem [{\citenamefont {Hubbard}(1963)}]{Hubbard}%
  \BibitemOpen
  \bibfield  {author} {\bibinfo {author} {\bibfnamefont {J.}~\bibnamefont {Hubbard}},\ }\href {https://royalsocietypublishing.org/doi/10.1098/rspa.1963.0204#} {\bibfield  {journal} {\bibinfo  {journal} {Math. Phys. Sci.}\ }\textbf {\bibinfo {volume} {276}},\ \bibinfo {pages} {238} (\bibinfo {year} {1963})}\BibitemShut {NoStop}%
\bibitem [{\citenamefont {Dukelsky}\ \emph {et~al.}(2003)\citenamefont {Dukelsky}, \citenamefont {Dussel}, \citenamefont {Hirsch},\ and\ \citenamefont {Schuck}}]{Dukelsky}%
  \BibitemOpen
  \bibfield  {author} {\bibinfo {author} {\bibfnamefont {J.}~\bibnamefont {Dukelsky}}, \bibinfo {author} {\bibfnamefont {G.}~\bibnamefont {Dussel}}, \bibinfo {author} {\bibfnamefont {J.~G.}\ \bibnamefont {Hirsch}},\ and\ \bibinfo {author} {\bibfnamefont {P.}~\bibnamefont {Schuck}},\ }\href {https://www.sciencedirect.com/science/article/pii/S0375947402013611} {\bibfield  {journal} {\bibinfo  {journal} {Nucl. Phys. A}\ }\textbf {\bibinfo {volume} {714}},\ \bibinfo {pages} {63} (\bibinfo {year} {2003})}\BibitemShut {NoStop}%
\bibitem [{\citenamefont {Henderson}\ \emph {et~al.}(2014)\citenamefont {Henderson}, \citenamefont {Dukelsky}, \citenamefont {Gustavo}, \citenamefont {Signoracci},\ and\ \citenamefont {Duguet}}]{Henderson}%
  \BibitemOpen
  \bibfield  {author} {\bibinfo {author} {\bibfnamefont {T.}~\bibnamefont {Henderson}}, \bibinfo {author} {\bibfnamefont {J.}~\bibnamefont {Dukelsky}}, \bibinfo {author} {\bibfnamefont {E.}~\bibnamefont {Gustavo}}, \bibinfo {author} {\bibfnamefont {A.}~\bibnamefont {Signoracci}},\ and\ \bibinfo {author} {\bibfnamefont {T.}~\bibnamefont {Duguet}},\ }\href {https://journals.aps.org/prc/abstract/10.1103/PhysRevC.89.054305} {\bibfield  {journal} {\bibinfo  {journal} {Phys. Rev. C}\ }\textbf {\bibinfo {volume} {89}},\ \bibinfo {pages} {63} (\bibinfo {year} {2014})}\BibitemShut {NoStop}%
\bibitem [{\citenamefont {von Neumann}\ and\ \citenamefont {Wigner}(1929)}]{Neumann2000ONTB}%
  \BibitemOpen
  \bibfield  {author} {\bibinfo {author} {\bibfnamefont {J.}~\bibnamefont {von Neumann}}\ and\ \bibinfo {author} {\bibfnamefont {E.~P.}\ \bibnamefont {Wigner}},\ }\href {https://www.worldscientific.com/doi/abs/10.1142/9789812795762_0002} {\bibfield  {journal} {\bibinfo  {journal} {Phys. Z}\ }\textbf {\bibinfo {volume} {30}} (\bibinfo {year} {1929})}\BibitemShut {NoStop}%
\bibitem [{\citenamefont {dwavesystems}(2024)}]{Ocean}%
  \BibitemOpen
  \bibfield  {author} {\bibinfo {author} {\bibnamefont {dwavesystems}},\ }\href {https://github.com/dwavesystems/dwave-samplers/tree/main/dwave/samplers/sa} {\bibinfo {title} {dwave-samplers}} (\bibinfo {year} {2024}),\ \bibinfo {note} {accessed: 2024-12-12}\BibitemShut {NoStop}%
\bibitem [{\citenamefont {Boothby}\ \emph {et~al.}(2020)\citenamefont {Boothby}, \citenamefont {Bunyk}, \citenamefont {Raymond},\ and\ \citenamefont {Roy}}]{pegasus}%
  \BibitemOpen
  \bibfield  {author} {\bibinfo {author} {\bibfnamefont {K.}~\bibnamefont {Boothby}}, \bibinfo {author} {\bibfnamefont {P.}~\bibnamefont {Bunyk}}, \bibinfo {author} {\bibfnamefont {J.}~\bibnamefont {Raymond}},\ and\ \bibinfo {author} {\bibfnamefont {A.}~\bibnamefont {Roy}},\ }\href@noop {} {\bibinfo {title} {Next-generation topology of {D}-wave quantum processors}} (\bibinfo {year} {2020}),\ \Eprint {https://arxiv.org/abs/2003.00133} {arXiv:2003.00133} \BibitemShut {NoStop}%
\bibitem [{\citenamefont {Hanauer}\ and\ \citenamefont {Kohn}(2011)}]{Andreas}%
  \BibitemOpen
  \bibfield  {author} {\bibinfo {author} {\bibfnamefont {M.}~\bibnamefont {Hanauer}}\ and\ \bibinfo {author} {\bibfnamefont {A.}~\bibnamefont {Kohn}},\ }\href {https://doi.org/https://doi.org/10.1063/1.3592786} {\bibfield  {journal} {\bibinfo  {journal} {{J. Chem. Phys.}}\ }\textbf {\bibinfo {volume} {134}},\ \bibinfo {pages} {204111} (\bibinfo {year} {2011})}\BibitemShut {NoStop}%
\bibitem [{\citenamefont {Treinish}\ \emph {et~al.}(2022)\citenamefont {Treinish} \emph {et~al.}}]{Qiskit2021}%
  \BibitemOpen
  \bibfield  {author} {\bibinfo {author} {\bibfnamefont {M.}~\bibnamefont {Treinish}} \emph {et~al.},\ }\href@noop {} {\bibinfo {title} {Qiskit: An open-source framework for quantum computing}} (\bibinfo {year} {2022})\BibitemShut {NoStop}%
\bibitem [{\citenamefont {Sivarajah}\ \emph {et~al.}(2020)\citenamefont {Sivarajah}, \citenamefont {Dilkes}, \citenamefont {Cowtan}, \citenamefont {Simmons}, \citenamefont {Edgington},\ and\ \citenamefont {Duncan}}]{Sivarajah2021tket}%
  \BibitemOpen
  \bibfield  {author} {\bibinfo {author} {\bibfnamefont {S.}~\bibnamefont {Sivarajah}}, \bibinfo {author} {\bibfnamefont {S.}~\bibnamefont {Dilkes}}, \bibinfo {author} {\bibfnamefont {A.}~\bibnamefont {Cowtan}}, \bibinfo {author} {\bibfnamefont {W.}~\bibnamefont {Simmons}}, \bibinfo {author} {\bibfnamefont {A.}~\bibnamefont {Edgington}},\ and\ \bibinfo {author} {\bibfnamefont {R.}~\bibnamefont {Duncan}},\ }\href {https://doi.org/10.1088/2058-9565/ab8e92} {\bibfield  {journal} {\bibinfo  {journal} {{Quantum Sci. Technol.}}\ }\textbf {\bibinfo {volume} {6}},\ \bibinfo {pages} {014003} (\bibinfo {year} {2020})}\BibitemShut {NoStop}%
\bibitem [{\citenamefont {Kissinger}\ and\ \citenamefont {van~de Wetering}(2020)}]{kissinger2020Pyzx}%
  \BibitemOpen
  \bibfield  {author} {\bibinfo {author} {\bibfnamefont {A.}~\bibnamefont {Kissinger}}\ and\ \bibinfo {author} {\bibfnamefont {J.}~\bibnamefont {van~de Wetering}},\ }in\ \href {https://doi.org/10.4204/EPTCS.318.14} {\emph {\bibinfo {booktitle} {{\rm Proceedings 16th International Conference on} Quantum Physics and Logic, {\rm Chapman University, Orange, CA, USA., 10-14 June 2019}}}},\ \bibinfo {series} {Electronic Proceedings in Theoretical Computer Science}, Vol.\ \bibinfo {volume} {318},\ \bibinfo {editor} {edited by\ \bibinfo {editor} {\bibfnamefont {B.}~\bibnamefont {Coecke}}\ and\ \bibinfo {editor} {\bibfnamefont {M.}~\bibnamefont {Leifer}}}\ (\bibinfo  {publisher} {Open Publishing Association},\ \bibinfo {year} {2020})\ pp.\ \bibinfo {pages} {229--241}\BibitemShut {NoStop}%
\bibitem [{\citenamefont {Riu}\ \emph {et~al.}(2024)\citenamefont {Riu}, \citenamefont {Nogué}, \citenamefont {Vilaplana}, \citenamefont {Garcia-Saez},\ and\ \citenamefont {Estarellas}}]{riu2024rlzx}%
  \BibitemOpen
  \bibfield  {author} {\bibinfo {author} {\bibfnamefont {J.}~\bibnamefont {Riu}}, \bibinfo {author} {\bibfnamefont {J.}~\bibnamefont {Nogué}}, \bibinfo {author} {\bibfnamefont {G.}~\bibnamefont {Vilaplana}}, \bibinfo {author} {\bibfnamefont {A.}~\bibnamefont {Garcia-Saez}},\ and\ \bibinfo {author} {\bibfnamefont {M.~P.}\ \bibnamefont {Estarellas}},\ }\href {https://arxiv.org/abs/2312.11597} {\bibinfo {title} {Reinforcement learning based quantum circuit optimization via {ZX}-calculus}} (\bibinfo {year} {2024}),\ \Eprint {https://arxiv.org/abs/2312.11597} {arXiv:2312.11597 [quant-ph]} \BibitemShut {NoStop}%
\bibitem [{\citenamefont {Chawla}\ \emph {et~al.}(2025{\natexlab{b}})\citenamefont {Chawla}, \citenamefont {Shetty}, \citenamefont {Tsemo}, \citenamefont {Sugisaki}, \citenamefont {Riu}, \citenamefont {Nogué}, \citenamefont {Mukherjee},\ and\ \citenamefont {Prasannaa}}]{mruccvqe2025}%
  \BibitemOpen
  \bibfield  {author} {\bibinfo {author} {\bibfnamefont {P.}~\bibnamefont {Chawla}}, \bibinfo {author} {\bibfnamefont {D.}~\bibnamefont {Shetty}}, \bibinfo {author} {\bibfnamefont {P.~B.}\ \bibnamefont {Tsemo}}, \bibinfo {author} {\bibfnamefont {K.}~\bibnamefont {Sugisaki}}, \bibinfo {author} {\bibfnamefont {J.}~\bibnamefont {Riu}}, \bibinfo {author} {\bibfnamefont {J.}~\bibnamefont {Nogué}}, \bibinfo {author} {\bibfnamefont {D.}~\bibnamefont {Mukherjee}},\ and\ \bibinfo {author} {\bibfnamefont {V.~S.}\ \bibnamefont {Prasannaa}},\ }\href {https://doi.org/10.48550/arXiv.2504.07037} {\bibinfo {title} {{VQE} calculations on a {NISQ} era trapped ion quantum computer using a multireference unitary coupled cluster ansatz: application to the {B}e{H}$_2$ insertion problem}} (\bibinfo {year} {2025}{\natexlab{b}}),\ \Eprint {https://arxiv.org/abs/2504.07037} {arXiv:2504.07037} \BibitemShut {NoStop}%
\bibitem [{\citenamefont {Shavitt}(1961)}]{cramer}%
  \BibitemOpen
  \bibfield  {author} {\bibinfo {author} {\bibfnamefont {I.}~\bibnamefont {Shavitt}},\ }\href@noop {} {\emph {\bibinfo {title} {Essentials of computational chemistry pp. 237}}}\ (\bibinfo  {publisher} {Wiley},\ \bibinfo {year} {1961})\BibitemShut {NoStop}%
\bibitem [{\citenamefont {Berry}\ \emph {et~al.}(2019)\citenamefont {Berry}, \citenamefont {Gidney}, \citenamefont {Motta}, \citenamefont {McClean},\ and\ \citenamefont {Babbush}}]{sing_factor}%
  \BibitemOpen
  \bibfield  {author} {\bibinfo {author} {\bibfnamefont {D.~W.}\ \bibnamefont {Berry}}, \bibinfo {author} {\bibfnamefont {C.}~\bibnamefont {Gidney}}, \bibinfo {author} {\bibfnamefont {M.}~\bibnamefont {Motta}}, \bibinfo {author} {\bibfnamefont {J.~R.}\ \bibnamefont {McClean}},\ and\ \bibinfo {author} {\bibfnamefont {R.}~\bibnamefont {Babbush}},\ }\href {https://quantum-journal.org/papers/q-2019-12-02-208/} {\bibfield  {journal} {\bibinfo  {journal} {Quantum}\ }\textbf {\bibinfo {volume} {3}},\ \bibinfo {pages} {208} (\bibinfo {year} {2019})}\BibitemShut {NoStop}%
\bibitem [{\citenamefont {von Burg}\ \emph {et~al.}(2021)\citenamefont {von Burg}, \citenamefont {Low}, \citenamefont {Haner}, \citenamefont {Steiger}, \citenamefont {Reiher}, \citenamefont {Roetteler},\ and\ \citenamefont {Troyer}}]{double_factor}%
  \BibitemOpen
  \bibfield  {author} {\bibinfo {author} {\bibfnamefont {V.}~\bibnamefont {von Burg}}, \bibinfo {author} {\bibfnamefont {G.~H.}\ \bibnamefont {Low}}, \bibinfo {author} {\bibfnamefont {T.}~\bibnamefont {Haner}}, \bibinfo {author} {\bibfnamefont {D.~S.}\ \bibnamefont {Steiger}}, \bibinfo {author} {\bibfnamefont {M.}~\bibnamefont {Reiher}}, \bibinfo {author} {\bibfnamefont {M.}~\bibnamefont {Roetteler}},\ and\ \bibinfo {author} {\bibfnamefont {M.}~\bibnamefont {Troyer}},\ }\href {https://journals.aps.org/prresearch/abstract/10.1103/PhysRevResearch.3.033055} {\bibfield  {journal} {\bibinfo  {journal} {Phys. Rev. Res.}\ }\textbf {\bibinfo {volume} {3}},\ \bibinfo {pages} {033055} (\bibinfo {year} {2021})}\BibitemShut {NoStop}%
\bibitem [{\citenamefont {Lee}\ \emph {et~al.}(2021)\citenamefont {Lee}, \citenamefont {Berry}, \citenamefont {Gidney}, \citenamefont {Huggins}, \citenamefont {McClean}, \citenamefont {Wiebe},\ and\ \citenamefont {Babbush}}]{THC_factor}%
  \BibitemOpen
  \bibfield  {author} {\bibinfo {author} {\bibfnamefont {J.}~\bibnamefont {Lee}}, \bibinfo {author} {\bibfnamefont {D.~W.}\ \bibnamefont {Berry}}, \bibinfo {author} {\bibfnamefont {C.}~\bibnamefont {Gidney}}, \bibinfo {author} {\bibfnamefont {W.~J.}\ \bibnamefont {Huggins}}, \bibinfo {author} {\bibfnamefont {J.~R.}\ \bibnamefont {McClean}}, \bibinfo {author} {\bibfnamefont {N.}~\bibnamefont {Wiebe}},\ and\ \bibinfo {author} {\bibfnamefont {R.}~\bibnamefont {Babbush}},\ }\href {https://journals.aps.org/prxquantum/abstract/10.1103/PRXQuantum.2.030305} {\bibfield  {journal} {\bibinfo  {journal} {PRX Quantum}\ }\textbf {\bibinfo {volume} {2}},\ \bibinfo {pages} {030305} (\bibinfo {year} {2021})}\BibitemShut {NoStop}%
\bibitem [{\citenamefont {Jeffrey~Cohn}\ and\ \citenamefont {Parrish}(2021)}]{exp_tens_hyp}%
  \BibitemOpen
  \bibfield  {author} {\bibinfo {author} {\bibfnamefont {M.~M.}\ \bibnamefont {Jeffrey~Cohn}}\ and\ \bibinfo {author} {\bibfnamefont {R.~M.}\ \bibnamefont {Parrish}},\ }\href {https://journals.aps.org/prxquantum/abstract/10.1103/PRXQuantum.2.040352} {\bibfield  {journal} {\bibinfo  {journal} {PRX Quantum}\ }\textbf {\bibinfo {volume} {2}},\ \bibinfo {pages} {040352} (\bibinfo {year} {2021})}\BibitemShut {NoStop}%
\bibitem [{\citenamefont {Shavitt}(1988{\natexlab{a}})}]{shavit1}%
  \BibitemOpen
  \bibfield  {author} {\bibinfo {author} {\bibfnamefont {I.}~\bibnamefont {Shavitt}},\ }\href@noop {} {\emph {\bibinfo {title} {Mathematical Frontiers in Computational Chemical Physics in D. G. Truhlar ed. pp. 300-349}}}\ (\bibinfo  {publisher} {Springer, New York},\ \bibinfo {year} {1988})\BibitemShut {NoStop}%
\bibitem [{\citenamefont {Shavitt}(1988{\natexlab{b}})}]{shavit2}%
  \BibitemOpen
  \bibfield  {author} {\bibinfo {author} {\bibfnamefont {I.}~\bibnamefont {Shavitt}},\ }\href@noop {} {\emph {\bibinfo {title} {The Unitary Group (Lecture Notes in Chemistry No. 22 in J. Hinze, ed. pp. 51-99}}}\ (\bibinfo  {publisher} {Springer, Berlin},\ \bibinfo {year} {1988})\BibitemShut {NoStop}%
\bibitem [{\citenamefont {Palacios}\ \emph {et~al.}(2025)\citenamefont {Palacios}, \citenamefont {Garcia-Saez}, \citenamefont {Julia-Diaz},\ and\ \citenamefont {Estarellas}}]{Ana}%
  \BibitemOpen
  \bibfield  {author} {\bibinfo {author} {\bibfnamefont {A.}~\bibnamefont {Palacios}}, \bibinfo {author} {\bibfnamefont {A.}~\bibnamefont {Garcia-Saez}}, \bibinfo {author} {\bibfnamefont {B.}~\bibnamefont {Julia-Diaz}},\ and\ \bibinfo {author} {\bibfnamefont {M.~P.}\ \bibnamefont {Estarellas}},\ }\href {https://doi.org/10.1103/PhysRevApplied.23.054070} {\bibfield  {journal} {\bibinfo  {journal} {Phys. Rev. Appl.}\ }\textbf {\bibinfo {volume} {23}},\ \bibinfo {pages} {054070} (\bibinfo {year} {2025})}\BibitemShut {NoStop}%
\bibitem [{\citenamefont {Kraft}(1988)}]{slsqpreport}%
  \BibitemOpen
  \bibfield  {author} {\bibinfo {author} {\bibfnamefont {D.}~\bibnamefont {Kraft}},\ }\href@noop {} {\emph {\bibinfo {title} {A software package for sequential quadratic programming}}},\ \bibinfo {type} {Technical Report}\ \bibinfo {number} {DFVLR-FR 88–28}\ (\bibinfo  {institution} {Deutsche Forschungs- und Versuchsanstalt für Luft- und Raumfahrt (DFVLR)},\ \bibinfo {year} {1988})\BibitemShut {NoStop}%
\bibitem [{\citenamefont {Iten}\ \emph {et~al.}(2016)\citenamefont {Iten}, \citenamefont {Colbeck}, \citenamefont {Kukuljan}, \citenamefont {Home},\ and\ \citenamefont {Christandl}}]{Isometry}%
  \BibitemOpen
  \bibfield  {author} {\bibinfo {author} {\bibfnamefont {R.}~\bibnamefont {Iten}}, \bibinfo {author} {\bibfnamefont {R.}~\bibnamefont {Colbeck}}, \bibinfo {author} {\bibfnamefont {I.}~\bibnamefont {Kukuljan}}, \bibinfo {author} {\bibfnamefont {J.}~\bibnamefont {Home}},\ and\ \bibinfo {author} {\bibfnamefont {M.}~\bibnamefont {Christandl}},\ }\href {https://journals.aps.org/pra/abstract/10.1103/PhysRevA.93.032318} {\bibfield  {journal} {\bibinfo  {journal} {Phys. Rev. A}\ }\textbf {\bibinfo {volume} {93}},\ \bibinfo {pages} {032318} (\bibinfo {year} {2016})}\BibitemShut {NoStop}%
\bibitem [{\citenamefont {Shende}\ \emph {et~al.}(2006)\citenamefont {Shende}, \citenamefont {Bullock},\ and\ \citenamefont {Markov}}]{Shende}%
  \BibitemOpen
  \bibfield  {author} {\bibinfo {author} {\bibfnamefont {V.~V.}\ \bibnamefont {Shende}}, \bibinfo {author} {\bibfnamefont {S.~S.}\ \bibnamefont {Bullock}},\ and\ \bibinfo {author} {\bibfnamefont {I.~L.}\ \bibnamefont {Markov}},\ }\href@noop {} {\bibfield  {journal} {\bibinfo  {journal} {IEEE Trans. on Computer-Aided Design}\ }\textbf {\bibinfo {volume} {25(6)}},\ \bibinfo {pages} {1000} (\bibinfo {year} {2006})}\BibitemShut {NoStop}%
\bibitem [{\citenamefont {Sugisaki}\ \emph {et~al.}(2019)\citenamefont {Sugisaki}, \citenamefont {Nakazawa}, \citenamefont {Toyota}, \citenamefont {Sato}, \citenamefont {Shiomi},\ and\ \citenamefont {Takui}}]{multiref_1}%
  \BibitemOpen
  \bibfield  {author} {\bibinfo {author} {\bibfnamefont {K.}~\bibnamefont {Sugisaki}}, \bibinfo {author} {\bibfnamefont {S.}~\bibnamefont {Nakazawa}}, \bibinfo {author} {\bibfnamefont {K.}~\bibnamefont {Toyota}}, \bibinfo {author} {\bibfnamefont {K.}~\bibnamefont {Sato}}, \bibinfo {author} {\bibfnamefont {D.}~\bibnamefont {Shiomi}},\ and\ \bibinfo {author} {\bibfnamefont {T.}~\bibnamefont {Takui}},\ }\href {https://pubs.acs.org/doi/10.1021/acscentsci.8b00788} {\bibfield  {journal} {\bibinfo  {journal} {ACS. Cent. Sci.}\ }\textbf {\bibinfo {volume} {5}},\ \bibinfo {pages} {167} (\bibinfo {year} {2019})}\BibitemShut {NoStop}%
\bibitem [{\citenamefont {Ino}\ \emph {et~al.}(2024)\citenamefont {Ino}, \citenamefont {Yonekawa}, \citenamefont {Yuzawa}, \citenamefont {Minato},\ and\ \citenamefont {Sugisaki}}]{multiref_2}%
  \BibitemOpen
  \bibfield  {author} {\bibinfo {author} {\bibfnamefont {Y.}~\bibnamefont {Ino}}, \bibinfo {author} {\bibfnamefont {M.}~\bibnamefont {Yonekawa}}, \bibinfo {author} {\bibfnamefont {H.}~\bibnamefont {Yuzawa}}, \bibinfo {author} {\bibfnamefont {Y.}~\bibnamefont {Minato}},\ and\ \bibinfo {author} {\bibfnamefont {K.}~\bibnamefont {Sugisaki}},\ }\href {https://pubs.rsc.org/en/content/articlelanding/2024/cp/d4cp03454f} {\bibfield  {journal} {\bibinfo  {journal} {Phys. Chem. Chem. Phys.}\ }\textbf {\bibinfo {volume} {26}},\ \bibinfo {pages} {30044} (\bibinfo {year} {2024})}\BibitemShut {NoStop}%
\bibitem [{\citenamefont {Larocca}\ \emph {et~al.}(2025)\citenamefont {Larocca}, \citenamefont {Thanasilp}, \citenamefont {Wang}, \citenamefont {Sharma}, \citenamefont {Biamonte}, \citenamefont {Coles}, \citenamefont {Cincio}, \citenamefont {McClean}, \citenamefont {Holmes},\ and\ \citenamefont {Cerezo}}]{bp}%
  \BibitemOpen
  \bibfield  {author} {\bibinfo {author} {\bibfnamefont {M.}~\bibnamefont {Larocca}}, \bibinfo {author} {\bibfnamefont {S.}~\bibnamefont {Thanasilp}}, \bibinfo {author} {\bibfnamefont {S.}~\bibnamefont {Wang}}, \bibinfo {author} {\bibfnamefont {K.}~\bibnamefont {Sharma}}, \bibinfo {author} {\bibfnamefont {J.}~\bibnamefont {Biamonte}}, \bibinfo {author} {\bibfnamefont {P.~J.}\ \bibnamefont {Coles}}, \bibinfo {author} {\bibfnamefont {L.}~\bibnamefont {Cincio}}, \bibinfo {author} {\bibfnamefont {J.~R.}\ \bibnamefont {McClean}}, \bibinfo {author} {\bibfnamefont {Z.}~\bibnamefont {Holmes}},\ and\ \bibinfo {author} {\bibfnamefont {M.}~\bibnamefont {Cerezo}},\ }\href {https://doi.org/10.1038/s42254-025-00813-9} {\bibfield  {journal} {\bibinfo  {journal} {Nat. Rev. Phys.}\ }\textbf {\bibinfo {volume} {7}},\ \bibinfo {pages} {174} (\bibinfo {year} {2025})}\BibitemShut {NoStop}%
\bibitem [{\citenamefont {Tilly}\ \emph {et~al.}(2022)\citenamefont {Tilly}, \citenamefont {Chen}, \citenamefont {Cao}, \citenamefont {Picozzi}, \citenamefont {Setia}, \citenamefont {Li}, \citenamefont {Grant}, \citenamefont {Wossnig}, \citenamefont {Rungger}, \citenamefont {Booth},\ and\ \citenamefont {Tennyson}}]{Tilly}%
  \BibitemOpen
  \bibfield  {author} {\bibinfo {author} {\bibfnamefont {J.}~\bibnamefont {Tilly}}, \bibinfo {author} {\bibfnamefont {H.}~\bibnamefont {Chen}}, \bibinfo {author} {\bibfnamefont {S.}~\bibnamefont {Cao}}, \bibinfo {author} {\bibfnamefont {D.}~\bibnamefont {Picozzi}}, \bibinfo {author} {\bibfnamefont {K.}~\bibnamefont {Setia}}, \bibinfo {author} {\bibfnamefont {Y.}~\bibnamefont {Li}}, \bibinfo {author} {\bibfnamefont {E.}~\bibnamefont {Grant}}, \bibinfo {author} {\bibfnamefont {L.}~\bibnamefont {Wossnig}}, \bibinfo {author} {\bibfnamefont {I.}~\bibnamefont {Rungger}}, \bibinfo {author} {\bibfnamefont {G.~H.}\ \bibnamefont {Booth}},\ and\ \bibinfo {author} {\bibfnamefont {J.}~\bibnamefont {Tennyson}},\ }\href {https://doi.org/10.1016/j.physrep.2022.08.003} {\bibfield  {journal} {\bibinfo  {journal} {Phys. Rep.}\ }\textbf {\bibinfo {volume} {986}},\ \bibinfo {pages} {1} (\bibinfo {year} {2022})}\BibitemShut {NoStop}%
\bibitem [{\citenamefont {Atobe}\ \emph {et~al.}(2022)\citenamefont {Atobe}, \citenamefont {Tawada},\ and\ \citenamefont {Togawa}}]{Atobe}%
  \BibitemOpen
  \bibfield  {author} {\bibinfo {author} {\bibfnamefont {Y.}~\bibnamefont {Atobe}}, \bibinfo {author} {\bibfnamefont {M.}~\bibnamefont {Tawada}},\ and\ \bibinfo {author} {\bibfnamefont {N.}~\bibnamefont {Togawa}},\ }\href {https://doi.org/10.1109/TC.2021.3138629} {\bibfield  {journal} {\bibinfo  {journal} {IEEE Transactions on Computers}\ }\textbf {\bibinfo {volume} {71}},\ \bibinfo {pages} {2606} (\bibinfo {year} {2022})}\BibitemShut {NoStop}%
\bibitem [{\citenamefont {Temme}\ and\ \citenamefont {Gambetta}(2017)}]{Brav_ZNE}%
  \BibitemOpen
  \bibfield  {author} {\bibinfo {author} {\bibfnamefont {K.}~\bibnamefont {Temme}}\ and\ \bibinfo {author} {\bibfnamefont {S.~B. J.~M.}\ \bibnamefont {Gambetta}},\ }\href {https://journals.aps.org/prl/abstract/10.1103/PhysRevLett.119.180509} {\bibfield  {journal} {\bibinfo  {journal} {Phys. Rev. L}\ }\textbf {\bibinfo {volume} {119}},\ \bibinfo {pages} {180509} (\bibinfo {year} {2017})}\BibitemShut {NoStop}%
\bibitem [{\citenamefont {Majumdar}\ \emph {et~al.}(2023)\citenamefont {Majumdar}, \citenamefont {Rivero}, \citenamefont {Metz}, \citenamefont {Hasan},\ and\ \citenamefont {Wang}}]{best_pract_qem}%
  \BibitemOpen
  \bibfield  {author} {\bibinfo {author} {\bibfnamefont {R.}~\bibnamefont {Majumdar}}, \bibinfo {author} {\bibfnamefont {P.}~\bibnamefont {Rivero}}, \bibinfo {author} {\bibfnamefont {F.}~\bibnamefont {Metz}}, \bibinfo {author} {\bibfnamefont {A.}~\bibnamefont {Hasan}},\ and\ \bibinfo {author} {\bibfnamefont {D.~S.}\ \bibnamefont {Wang}},\ }in\ \href {https://doi.org/10.1109/QCE57702.2023.00102} {\emph {\bibinfo {booktitle} {2023 IEEE International Conference on Quantum Computing and Engineering (QCE)}}},\ Vol.~\bibinfo {volume} {01}\ (\bibinfo {year} {2023})\ pp.\ \bibinfo {pages} {881--887}\BibitemShut {NoStop}%
\bibitem [{\citenamefont {Bravyi}\ \emph {et~al.}(2021)\citenamefont {Bravyi}, \citenamefont {Sheldon}, \citenamefont {Kandala}, \citenamefont {Mckay},\ and\ \citenamefont {Gambetta}}]{Brav_inv_matrix}%
  \BibitemOpen
  \bibfield  {author} {\bibinfo {author} {\bibfnamefont {S.}~\bibnamefont {Bravyi}}, \bibinfo {author} {\bibfnamefont {S.}~\bibnamefont {Sheldon}}, \bibinfo {author} {\bibfnamefont {A.}~\bibnamefont {Kandala}}, \bibinfo {author} {\bibfnamefont {D.~C.}\ \bibnamefont {Mckay}},\ and\ \bibinfo {author} {\bibfnamefont {J.~M.}\ \bibnamefont {Gambetta}},\ }\href {https://journals.aps.org/pra/abstract/10.1103/PhysRevA.103.042605} {\bibfield  {journal} {\bibinfo  {journal} {Phys. Rev. A}\ }\textbf {\bibinfo {volume} {103}},\ \bibinfo {pages} {042605} (\bibinfo {year} {2021})}\BibitemShut {NoStop}%
\bibitem [{\citenamefont {et~al}(2025)}]{QEM_QA}%
  \BibitemOpen
  \bibfield  {author} {\bibinfo {author} {\bibfnamefont {J.~R.}\ \bibnamefont {et~al}},\ }\href {https://www.nature.com/articles/s41534-025-00977-3} {\bibfield  {journal} {\bibinfo  {journal} {npj Quantum Inf}\ }\textbf {\bibinfo {volume} {11}},\ \bibinfo {pages} {38} (\bibinfo {year} {2025})}\BibitemShut {NoStop}%
\bibitem [{\citenamefont {Koczor}(2021)}]{Exp_err_sup}%
  \BibitemOpen
  \bibfield  {author} {\bibinfo {author} {\bibfnamefont {B.}~\bibnamefont {Koczor}},\ }\href {https://journals.aps.org/prx/abstract/10.1103/PhysRevX.11.031057} {\bibfield  {journal} {\bibinfo  {journal} {Phys Rev. X}\ }\textbf {\bibinfo {volume} {11}},\ \bibinfo {pages} {031057} (\bibinfo {year} {2021})}\BibitemShut {NoStop}%
\bibitem [{\citenamefont {Shingu}\ \emph {et~al.}(2024)\citenamefont {Shingu}, \citenamefont {Nikuni}, \citenamefont {Kawabata},\ and\ \citenamefont {Matsuzaki}}]{QA_with_EM}%
  \BibitemOpen
  \bibfield  {author} {\bibinfo {author} {\bibfnamefont {Y.}~\bibnamefont {Shingu}}, \bibinfo {author} {\bibfnamefont {T.}~\bibnamefont {Nikuni}}, \bibinfo {author} {\bibfnamefont {S.}~\bibnamefont {Kawabata}},\ and\ \bibinfo {author} {\bibfnamefont {Y.}~\bibnamefont {Matsuzaki}},\ }\href {https://journals.aps.org/pra/abstract/10.1103/PhysRevA.109.042606} {\bibfield  {journal} {\bibinfo  {journal} {Phys Rev. A}\ }\textbf {\bibinfo {volume} {11}},\ \bibinfo {pages} {031057} (\bibinfo {year} {2024})}\BibitemShut {NoStop}%
\end{thebibliography}%

\appendix 

\section{Derivation of fidelity bounds}\label{sec:bounds} 

The energy of a quantum state $\ket{\Psi}$ is given by 

\begin{eqnarray}
E = \langle \Psi|H|\Psi\rangle. 
\end{eqnarray}

Substituting the Hamiltonian, $H$, with its spectral decomposition, we obtain 

\begin{eqnarray}
E &=& \sum_{n=0}^{d-1}E_np_n \nonumber \\ 
&=& E_0p_0 + \sum_{n=1}^{d-1}E_np_n, 
\end{eqnarray}

where $E_n$ is the energy of the eigenstate $\ket{E_n}$, and $p_n = |\langle E_n|\Psi \rangle|^2$ represents its population. 

Since the eigenvalues, $E_n$, are ordered in increasing magnitude, any energy, $E_n$, with $n \in {1,\cdots,d-1}$, satisfies the bounds: 

\begin{eqnarray}
E_1 \leq E_n \leq E_{max}, 
\end{eqnarray}

where $E_{max} = E_{d-1}$ denotes the largest eigenvalue of $H$. 

\begin{table*}[t]
\caption{Table showing worst-case scaling for the cost of computing different CISD Hamiltonian matrix elements. In the table, ($p, q, r, \ldots$  ) refer to general indices, ($i, j, k, \ldots$  ) refer to indices of occupied spin orbitals, and ($a, b, c, \ldots$) refer to indices of unoccupied spin orbitals. } 
  \setlength{\tabcolsep}{10pt}
  \hskip-1.5cm
  \begin{tabular}{cccccccc}
  \hline \hline 
     S. No. & $\ket{\Phi_m}$ &$\ket{\Phi_n}$&$\#$ same indices& Examples of $\ket{\Phi_n}$ & Pattern & Scaling \\
    \hline 
    1. & $\ket{0}$ & $\ket{0}$& 0 & $\ket{0}$ &1 &$\mathcal{O}(1) \times \mathcal{O}(1) \times \mathcal{O}(N^2) = \mathcal{O}(N^2)$\\
    2. & $\ket{0}$ & $\ket{S}$& 0 & $\ket{\Phi_i^a}$ &2 &$\mathcal{O}(1) \times \mathcal{O}(N^2) \times \mathcal{O}(N) = \mathcal{O}(N^3)$\\
    3. & $\ket{0}$ & $\ket{D}$& 0 & $\ket{\Phi_{ij}^{ab}}$ &3 &$\mathcal{O}(1) \times \mathcal{O}(N^4) \times \mathcal{O}(1) = \mathcal{O}(N^4)$\\
    4. & $\ket{\Phi_i^a}$ & $\ket{S}$& 2 & $\ket{\Phi_i^a}$ &1 &$\mathcal{O}(N^2) \times \mathcal{O}(1) \times \mathcal{O}(N^2) = \mathcal{O}(N^4)$\\
    5. & $\ket{\Phi_i^a}$ & $\ket{S}$& 1 & $\ket{\Phi_i^b}$, $\ket{\Phi_j^a}$ &2 &$\mathcal{O}(N^2) \times \mathcal{O}(N) \times \mathcal{O}(N) = \mathcal{O}(N^5)$\\
    6. & $\ket{\Phi_i^a}$ & $\ket{S}$& 0 & $\ket{\Phi_j^b}$ &3 &$\mathcal{O}(N^2) \times \mathcal{O}(N^2) \times \mathcal{O}(1) = \mathcal{O}(N^4)$\\
    7. & $\ket{\Phi_i^a}$ & $\ket{D}$& 2 & $\ket{\Phi_{ij}^{ab}}$, $\ket{\Phi_{ij}^{ba}}$ &2 &$\mathcal{O}(N^2) \times \mathcal{O}(N^2) \times \mathcal{O}(N) = \mathcal{O}(N^5)$\\
    8. & $\ket{\Phi_i^a}$ & $\ket{D}$& 1 & $\ket{\Phi_{ij}^{bc}}$, $\ket{\Phi_{jk}^{ab}}$ &3 &$\mathcal{O}(N^2) \times \mathcal{O}(N^3) \times \mathcal{O}(1) = \mathcal{O}(N^5)$\\
    9. & $\ket{\Phi_i^a}$ & $\ket{D}$& 0 & $\ket{\Phi_{jk}^{bc}}$ &4 &-\\
    10. & $\ket{\Phi_{ij}^{ab}}$ & $\ket{D}$& 4 & $\ket{\Phi_{ij}^{ab}}$ &1 &$\mathcal{O}(N^4) \times \mathcal{O}(1) \times \mathcal{O}(N^2) = \mathcal{O}(N^6)$\\
    11. & $\ket{\Phi_{ij}^{ab}}$ & $\ket{D}$& 3 & $\ket{\Phi_{ik}^{ab}}$, $\ket{\Phi_{kj}^{ab}}$, $\ket{\Phi_{ij}^{ac}}$, $\ket{\Phi_{ij}^{cb}}$, $\cdots$ &2 &$\mathcal{O}(N^4) \times \mathcal{O}(N) \times \mathcal{O}(N) = \mathcal{O}(N^6)$\\
    12. & $\ket{\Phi_{ij}^{ab}}$ & $\ket{D}$& 2 & $\ket{\Phi_{ij}^{cd}}$, $\ket{\Phi_{ik}^{ac}}$, $\ket{\Phi_{kj}^{ac}}$, $\ket{\Phi_{kj}^{cb}}$, $\cdots$ &3 &$\mathcal{O}(N^4) \times \mathcal{O}(N^2) \times \mathcal{O}(1) = \mathcal{O}(N^6)$\\
    13. & $\ket{\Phi_{ij}^{ab}}$ & $\ket{D}$& 1 & $\ket{\Phi_{ik}^{ab}}$, $\ket{\Phi_{ik}^{cd}}$, $\ket{\Phi_{kl}^{cb}}$, $\ket{\Phi_{kl}^{ac}}$, $\cdots$ &4 &-\\
    14. & $\ket{\Phi_{ij}^{ab}}$ & $\ket{D}$& 0 & $\ket{\Phi_{kl}^{cd}}$ &4 &-\\
    \hline \hline 
  \end{tabular}
  \label{tab:3}
\end{table*} 

Thus, the second term in the right hand side of Eq. (A2) is bounded as 

\begin{eqnarray}
E_1(1-p_0) \leq \sum_{n=1}^{d-1}E_np_n \leq E_{max}(1-p_0), 
\end{eqnarray}

where we have used the normalization condition $\sum_{n=1}^{d-1} p_n = 1-p_0$. Substituting
this bound into Eq. (A2), the energy of the state $\ket{\Psi}$ is bounded by 

\begin{eqnarray}
E_0p_0 + E_1(1-p_0) \leq E \leq E_0p_0 + E_{max}(1-p_0). 
\end{eqnarray}

After some algebra and identifying $p_0 = |\langle E_0|\Psi \rangle|^2 = \mathcal{F}$ as the fidelity of $\ket{\Psi}$ with the ground state, we derive the following bounds for the fidelity: 

\begin{eqnarray}
1-\frac{E-E_0}{E_1-E_0} \leq \mathcal{F} \leq 1-\frac{E-E_0}{E_{max}-E_0}. 
\end{eqnarray} 

In our case, we check if the bounds are satisfied for the case of our avoided crossing problem for H$_4$ in square geometry where $E_0 = E_g$, $E_1 = E_e$ and $E = E_{g, QAE}$.

The lower bound
represents the case where the population outside the ground state is entirely in
the first excited state, the one with the lowest energy. On the other hand, the
upper bound corresponds to the scenario where all the population outside the
ground state is in the state with the highest energy. In any realistic situation,
the true population distribution will lie somewhere between these two extremes. 

\section{Cost of Hartree-Fock and one- and two-electron integrals in the MO basis}\label{cost_HF_AO_MO}

The computational cost of the Hartree-Fock method scales as $\mathcal{O}(N^4)$ due to the number of two electron integrals involved in construction of the Fock matrix. A typical two-electron integral in the atomic orbital basis (an electron repulsion integral) takes the form  $\langle\chi_{\mu}\chi_{\nu} |\chi_{\rho} \chi_{\sigma}\rangle$ (this is the shorthand for $\iint dr_1dr_2 \chi^*_{\mu}(r_1)\chi^*_{\nu}(r_1)\frac{1}{r_{12}}\chi_{\rho}(r_2)\chi_{\sigma}(r_2)$), where the indices $\mu, \nu, \rho, \sigma$ range from $1$ to $N$. Here $\chi_{i}$s are the atomic orbitals.
Also, note that a spatial molecular orbital $\phi_p(\vec{r})$ can be expressed as a linear combination of atomic orbitals via  $\phi_p(\vec{r}) = \sum^{N}_{\mu=1}c_{p\mu}\chi_{\mu}$. Assuming that all the coefficients are real, a typical two-electron integral in the MO basis takes the form
\begin{align}
    \langle\phi_{p}\phi_{q}|\frac{1}{r} |\phi_{r} \phi_{s}\rangle = \sum^{N}_{\mu, \nu, \rho, \sigma = 1}c_{\mu p}c_{\nu q}c_{\rho r}c_{\sigma s}\langle\chi_{\mu}\chi_{\nu} |\frac{1}{r} |\chi_{\rho} \chi_{\sigma}\rangle, 
\end{align}
where the right hand side is evaluated by a sequence of four quarter-transformations, each scaling as $\mathcal{O}(N^5)$ as shown below
\begin{align*}
    \langle\phi_{p}\chi_{\nu} |\frac{1}{r} |\chi_{\rho} \chi_{\sigma}\rangle &= \sum^{N}_{\mu = 1}c_{\mu p}\langle\chi_{\mu}\chi_{\nu} |\frac{1}{r} |\chi_{\rho} \chi_{\sigma}\rangle\\
    \langle\phi_{p}\phi_{q} |\frac{1}{r} |\chi_{\rho} \chi_{\sigma}\rangle &= \sum^{N}_{\nu = 1}c_{\nu q}\langle\phi_{p}\chi_{\nu} |\frac{1}{r} |\chi_{\rho} \chi_{\sigma}\rangle\\
    \langle\phi_{p}\phi_{q} |\frac{1}{r} |\phi_{r} \chi_{\sigma}\rangle &= \sum^{N}_{\rho = 1}c_{\rho r}\langle\phi_{p}\phi_{q} |\frac{1}{r} |\chi_{\rho} \chi_{\sigma}\rangle\\
    \langle\phi_{p}\phi_{q} |\frac{1}{r} |\phi_{r} \phi_{s}\rangle &= \sum^{N}_{\rho = 1}c_{\sigma s}\langle\phi_{p}\phi_{q} |\frac{1}{r} |\phi_{r}\chi_{\sigma}\rangle.
\end{align*} 

\section{Cost of evaluating CISD Hamiltonian matrix elements}\label{cisdHij} 

Here, we present the cost of generating the CISD Hamiltonian matrix elements, $\langle \Phi_m | H | \Phi_n \rangle$, which are fed as inputs to QAE. The cost is analyzed by categorizing the matrix elements into 4 patterns and applying the Slater--Condon rules. We begin by classifying $\langle \Phi_m | H | \Phi_n \rangle$: 

\begin{itemize}
\item Pattern 1: $\ket{\Psi_m} = \ket{\Psi_n}$ (diagonal element of the CI matrix), for which the cost of evaluating the matrix element goes as $\mathcal{O}(N^2)$. 
\item Pattern 2: $\ket{\Phi_m}$ and $\ket{\Phi_n}$ have two different occupation indices (for example, $\ket{\Phi_m}=\ket{\Phi_i^a}$ and $\ket{\Phi_n}=\ket{\Phi_i^b}$ . The cost of evaluating $\langle \Phi_m | H | \Phi_n \rangle$ then scales as $\mathcal{O}(N)$. 
\item Pattern 3: $\ket{\Psi_m}$ and $\ket{\Psi_n}$ have four different occupation indices (for instance, $\ket{\Phi_m}=\ket{\Phi_i^a}$ and $\ket{\Phi_{ij}^{bc}}$). The cost of evaluating $\langle \Phi_m | H | \Phi_n \rangle$ then scales as $\mathcal{O}(1)$. 
\item Pattern 4: $\ket{\Psi_m}$ and $\ket{\Psi_n}$ have more than four differing occupation indices (for instance, $\ket{\Phi_m}=\ket{\Phi_i^a}$ and $\ket{\Phi_{jk}^{bc}}$). These matrix elements are zero due to the Slater--Condon rules. 
\end{itemize} 

From the table, we see that the cost of evaluating the matrix elements for determinant-based CISD is at most $N^6$. We comment on why the scaling of serial number 5 is not $\mathcal{O}(N^2) \times \mathcal{O}(N^2) \times \mathcal{O}(N) = \mathcal{O}(N^5)$. This is because one of the indices of $\ket{\Phi_m}=\ket{\Phi_j^b}$ is fixed to either $j=i$ or $a=b$. Thus, the number of possible $\ket{\Phi_n}$ scales as $^2C_1 \times \mathcal{O}(N) = \mathcal{O}(N)$. Similarly, for serial number 11, the scaling is not $\mathcal{O}(N^4) \times \mathcal{O}(N^4) \times \mathcal{O}(N) = \mathcal{O}(N^9)$, because three out of four indices of $\ket{\Phi_n}$ should be fixed, and the number of possible $\ket{\Phi_n}$ scales as $^4C_3 \times \mathcal{O}(N) = \mathcal{O}(N)$. 

In the case of GUGA-based CISD, we want to know the cost scaling for evaluating $\langle \Phi_m | H | \Phi_n \rangle$, where $\ket{\Phi_m}$ and $\ket{\Phi_n}$ are symmetry-adapted CSFs. The scaling behavior of the matrix elements evaluation is very similar to that of determinant-based one. However, in the GUGA-CI approach, we incur extra cost for evaluating an additional coefficient depending on the pathways of Shavitt graph, with the cost for this step possibly scaling as $\mathcal{O}(N)$. 

\section{Details of input state preparation: a representative example}\label{sec:isometry} 
In VQE, one needs to construct the input reference state via an isometry, which in the case of a multi-reference UCC method, can be non-trivial. We note that not only does one need to employ classical many-body methods to find the coefficients of the entangled state, but also the cost of an input state preparation isometry in general scales exponentially in the number of two-qubit gates, $\mathcal{N}_{2qg}$. For instance, Table II in Ref.~\cite{Isometry} outlines the methods used and presents the CNOT gate bounds for general $m$ to $n$ qubit isometries. Using chemical intuition to guess a small number of determinants (for instance, assuming a log log growth of the number of determinants in the number of all possible determinants) reduces the scaling ($\lceil log(log(2^N)) \rceil \sim \lceil log(N) \rceil$), but the resulting circuits are still very deep for the NISQ era. We consider a toy example where we go from $N$ of 4 to 16 in steps of 2, and thus the number of determinants for the input state would go from 2 to 4. We restrict the number of occupied spin-orbitals to 2, and find that $\mathcal{N}_{2qg}$ increases as $2^N$ (the choice for coefficients is explained in Appendix \ref{sec:isometry}, while Figure \ref{2q_scaling} presents the result from our numerical analysis as a plot), in spite of a log log reduction in the number of determinants (the worst case scaling for $2^N$ basis states is $4^N$ two-qubit gates \cite{Shende, Isometry}). This toy example already underscores the hardness of one of the biggest open problems in gate-based algorithms: efficient input state preparation. Having said this, we also stress that our conclusions are drawn from numerical simulations that employ Qiskit's isometry routines, and there are works in literature that attempt to address the topic of manual construction of polynomial-depth quantum circuits to prepare multi-configurational wave
functions (for example, see Refs.~\cite{multiref_1} and ~\cite{multiref_2}). 
Here, we present the details of the conditions under which our representative example yielded the $2^N$ scaling for the number of two-qubit gates for the isometry associated with input state preparation for VQE. We consider a simplified model of a two-electron system where the number of determinants in the wavefunction grow as $\log(\log(2^N))$. Below we explicitly report the dominant excitations considered along with their respective coefficients (chosen to be $\leq 0.721$). The choice of determinants is based on excitations that remain close to the valence (occupied) orbitals. The determinants are reported using interleaved notation where occupancies are represented as $(\alpha, \beta, \alpha, \beta, \ldots)$: 
\begin{figure}[t]
	\includegraphics[width=8cm]{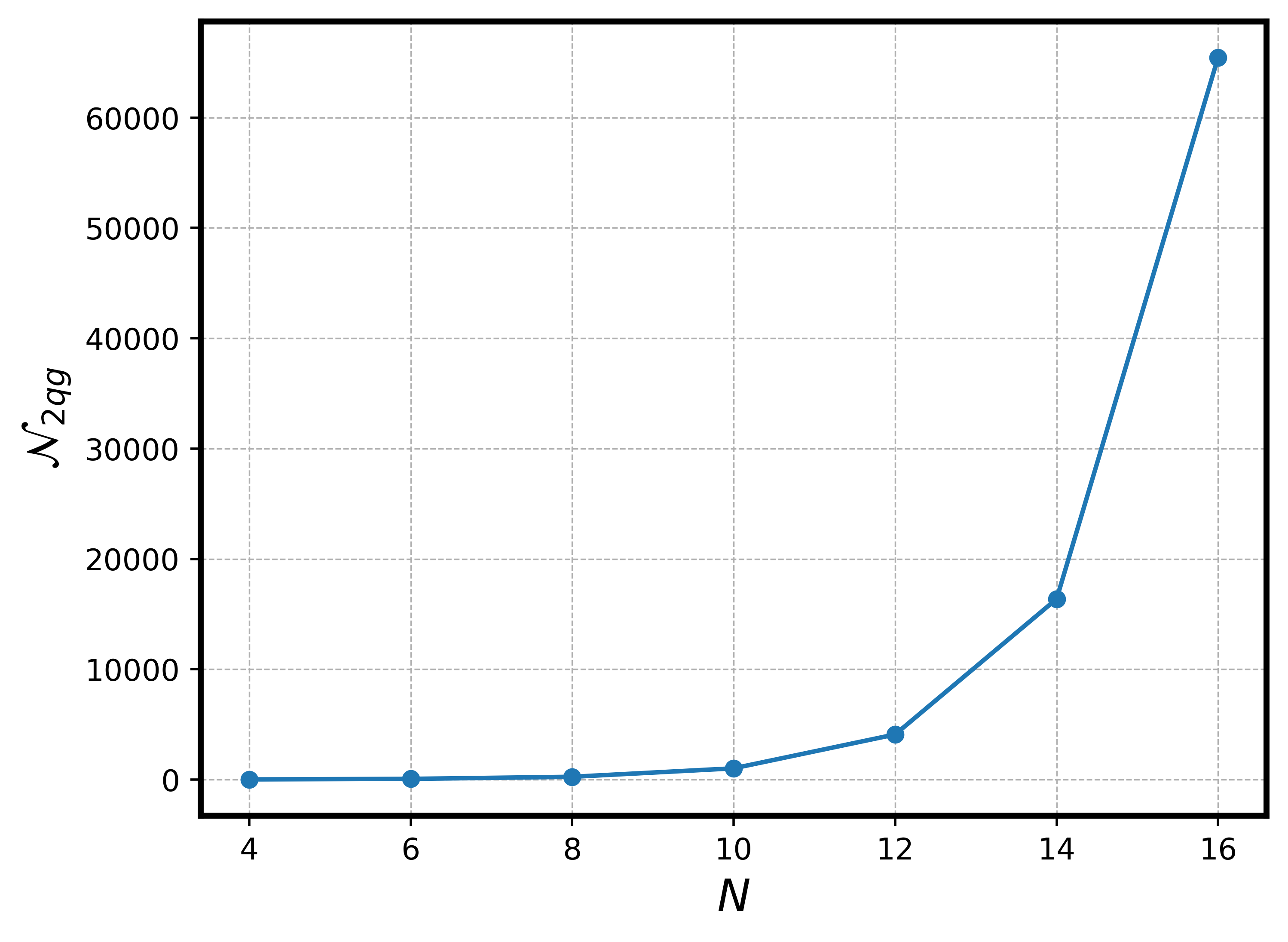}
 \caption{Figure illustrating the growth of the number of two qubit gates, $\mathcal{N}_{2qg}$, with number of qubits (written as $N$ in the figure), for input state preparation in VQE. }\label{2q_scaling} 
	\centering
\end{figure} 
\begin{itemize}
   \item For $N=4$: 
   \begin{eqnarray}
       \ket{\Psi} = 0.70\ket{1100} + 0.71\ket{0011}. \nonumber
   \end{eqnarray}
   \item For $N=6$: 
   \begin{eqnarray}
       \ket{\Psi} &=& 0.60\ket{110000} - 0.6\ket{001100} \nonumber \\ &+& 0.529\ket{000011}. \nonumber
   \end{eqnarray}
   \item For $N=8$: 
   \begin{eqnarray}
       \ket{\Psi} &=& 0.60\ket{11000000} - 0.6\ket{00110000} \nonumber \\ &+& 0.529\ket{00001100}. \nonumber
   \end{eqnarray}
   \item For $N=10$: 
   \begin{eqnarray}
       \ket{\Psi} &=& 0.72\ket{1100000000} - 0.6\ket{0011000000} \nonumber \\ &+& 0.245\ket{0000110000}- 0.245\ket{1001000000}. \nonumber
   \end{eqnarray}
   \item For $N=12$: 
   \begin{eqnarray}
       \ket{\Psi} &=& 0.72\ket{110000000000} - 0.6\ket{001100000000} \nonumber \\ &+& 0.245\ket{100100000000}- 0.245\ket{011000000000}. \nonumber
   \end{eqnarray} 
   \item For $N=14$: 
   \begin{eqnarray}
       \ket{\Psi} &=& 0.72\ket{11000000000000} - 0.6\ket{00110000000000} \nonumber \\ &+& 0.245\ket{10010000000000}- 0.245\ket{01100000000000}. \nonumber
   \end{eqnarray}
   \item For $N=16$: 
   \begin{eqnarray}
       \ket{\Psi} &=& 0.72\ket{1100000000000000} - 0.6\ket{0011000000000000} \nonumber \\ &+& 0.245\ket{1001000000000000}- 0.245\ket{0110000000000000}. \nonumber
   \end{eqnarray}
\end{itemize}

\section{Additional details : QAE and VQE algorithms}\label{sec:additional_details}

In this section, we provide some additional details about VQE and QAE outside of the discussions in Section \ref{sec:Comparison_with_vqe} of the main manuscript. 

\begin{enumerate}
\item \textbf{Choice of ansatz: }In VQE, the choice of ansatz is paramount to convergence. While physics-inspired/chemistry-based ansatze such as the UCC based ones yield excellent results in noiseless simulations, the entire class of hardware efficient ansatze often suffer from the notorious barren plateau problem \cite{bp}. On the other hand, we are unlikely to run into barren plateau issues in QAE, since by construction, we employ a CI wave function, which is physics-inspired/chemistry-based, and the algorithm involves an explore-exploit metaheuristic strategy, due to which we do not have a notion of iterations and convergence. It is worth adding that modified QAE workflows could have a convergence aspect to it, for example, see Ref. \cite{Vikrant2024}, but in our current study, we use the originally proposed QAE workflow, where we simply execute many shots of quantum annealing for each pre-selected $\lambda$ value within a range. 

\item \textbf{Initial guess for VQE parameters: }In VQE, we need to choose the initial guess for the parameters. Although we choose zeroes as initial guess values for our parameters, one can accelerate convergence by feeding in better-informed guesses such as MP2 (M{\o}ller--Plesset perturbation theory to second order in energy) initial guess (for example, see Ref. \cite{C2vpathway}). This, however, is accompanied by appropriate classical pre-processing steps. 

\item \textbf{Number of parameters to optimize: } This grows identically for both UCC-based VQE and QAE (which relies on CI). For example, in a UCCSD ansatz-based VQE computation, the number of parameters to be 
optimized grow as $n_o^2 n_v^2 \sim N^4$. However, it is important to also note that in the VQE-UCC framework, some higher-order excitations are accounted through the non-linear expansion of the UCC state, and the number of Slater determinants (or CSFs as in our case) included in the wave function expansion is larger for VQE-UCC than for QAE, even if the same number of parameters are included. For instance, even though UCCSD explicitly considers only $T_1$ and $T_2$ terms, it inherently includes correlation contributions from terms like $T_3, T_1^2, T_1^3, T_1^4, T_1 T_2$ etc. Therefore, while the number of parameters remain unchanged, the number of determinants involved is significantly larger in UCCSD compared to CISD.
\item \textbf{Convergence: }This also leads us to the next point: since VQE is iterative in nature, the convergence behaviour heavily relies on the choice of optimizer (with the correct choice not being necessarily easy), the precision sought, as well as the nature of the specific system of interest. On the other hand, one can think of quantum annealing itself as performing the optimization in the case of QAE. 
\item We note that QAE possesses another inherent advantage in giving better results, since we pick the lowest energies across repetitions by construction. This is in contrast to VQE, where the mean expectation value across repetitions is typically chosen as the final result. This is because evaluating the expectation value of an observable $A$ using VQE involves computing the weighted sum, with the equation given by 
\begin{eqnarray}\label{eq:vqeexpec}
\langle A \rangle = \sum_{i}\lambda_i p_i, 
\end{eqnarray}
where $\{\lambda_i\}$ are the eigenvalues of $A$.  Here $p_i = \tilde{N}_i/N^{VQE}_s$ represents the probability of the state being in the $i^{th}$ eigenstate, $\tilde{N}_i$ denotes the number of shots yielding that eigenstate, and $N^{VQE}_s$ is the total number of shots. Assuming the exact distribution has probabilities given by $p^{0}_i = \tilde{N_i^{0}}/N^{VQE}_s$, achieving an accurate estimate of the expectation value requires generating sufficient statistics such that $\tilde{N}_i \approx \tilde{N}^{0}_i$, $\forall i$. In contrast, in QAE, a single occurrence of the minimum energy is sufficient in order for it to qualify as the optimal solution. 
\item \textbf{Number of shots: }The previous point leads us to the question of the number of shots required for a calculation. If we seek a precision $\epsilon$ in our calculation, VQE requires that we supply $\sim N^4/\epsilon^2$ number of shots (see Ref. \cite{Tilly}, Section 5.1.1). The cost function evaluation in each iteration and for each Hamiltonian term involves cost from number of shots and the number of two-qubit gates for that circuit, which as we saw scales as $\sim N^5$. On the other hand, for QAE, determining the scaling behaviour of the number of shots is highly non-trivial, although it is central to the completion of the scaling analysis. Furthermore, noting that the anneal time itself scales as the inverse of minimum gap squared (where gap refers to the energy gap between the ground state and the first excited state of the total Hamiltonian, whose final Hamiltonian is the QUBO described in Eq. \ref{qubo}), we require that for annealing to be efficient, the gap must scale polynomially in $N$. We defer these two important considerations of the anneal time and the number of shots scaling behaviours with system size for a future work, but note that they may favourably or adversely impact the overall scaling of QAE. 
\begin{figure}[t]
	\includegraphics[width=8cm]{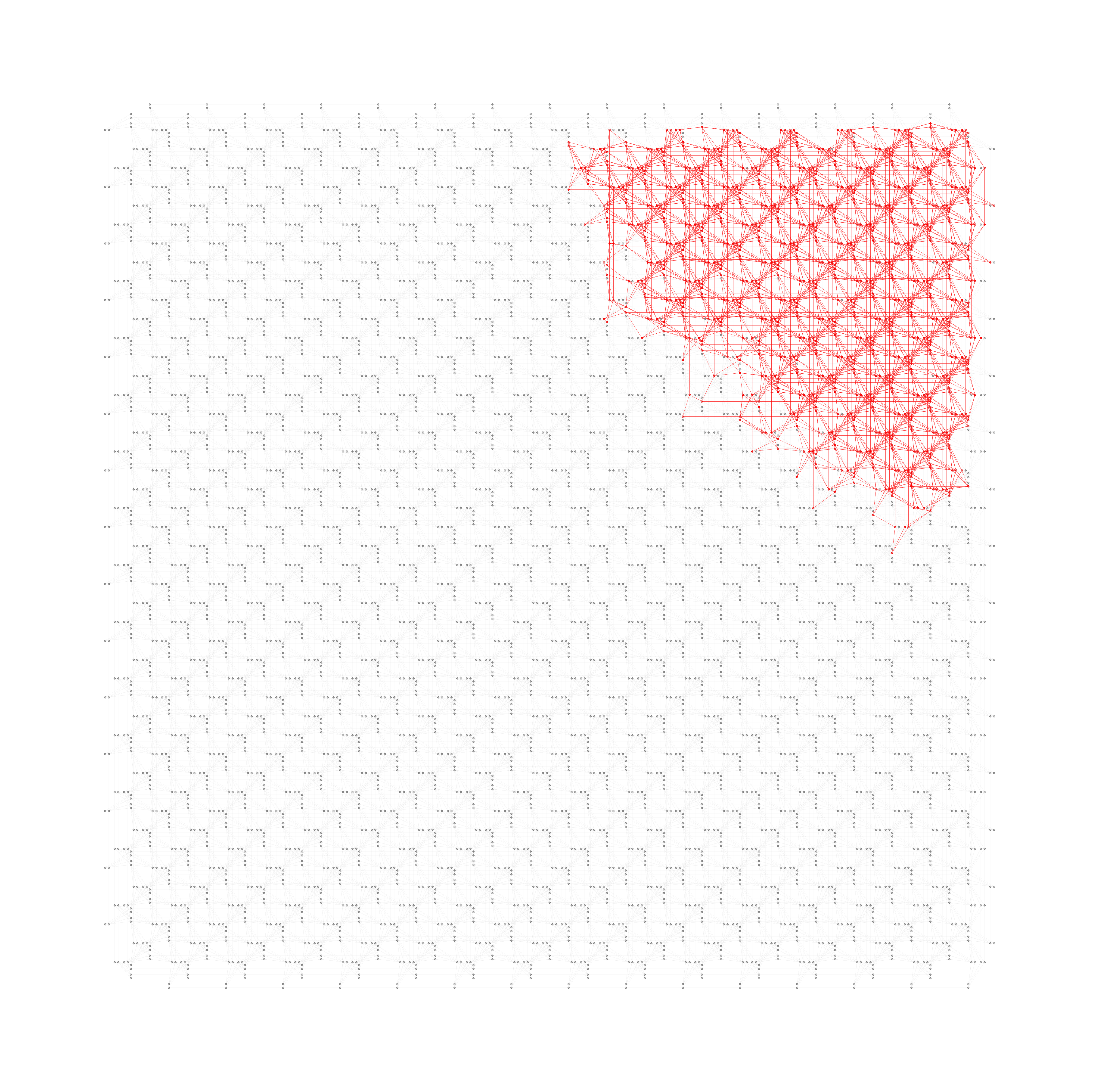}
 \caption{Figure illustrating an instance of the exact embedding used for mapping an 80 qubit fully connected QUBO onto the Pegasus topology of the D-Wave hardware. }\label{embedding} 
	\centering
\end{figure} 
\item \textbf{Embedding additional details: }In VQE, although there are gate-based quantum computers to execute the algorithm with over 100 qubits, we are limited by the number of gate operations even for small numbers of physical qubits. On the other hand, in QAE, we are limited by the number of physical qubits. For example, we find that for a given repetition, that is, across hundred $\lambda$ values that are scanned for our chosen $\lambda$ range, the number of physical qubits can vary between about 700 to about 900 (Figure \ref{embedding} depicts an embedding for a single instance) to construct our QUBO with 80 logical qubits.  This also reflects when we wish to increase K in a computation. For example, were we to carry out our computations on H$_4$ with a split valence basis set such as the 6-31G (4 electron- 5 orbital active space), a QUBO size of 170 (assuming $K=10$) requires on an average about 4000 physical qubits (over those scanned $\lambda$ values for which an embedding was found). Finding an embedding with any further increase in K is not possible, since the D-Wave annealers have only 5000 physical qubits. In such cases, one needs to opt for Sub-QUBOs (for example, see Ref. \cite{Atobe}), and additional errors incurred in approximations involved in choice of sub-QUBOs also add to the error budget. It is worth adding that the embedding issue we consider is specific to current state-of-the-art D-Wave hardware and not to the QAE algorithm itself; future advances in this direction, for example, machines with substantially better connectivity, could alleviate embedding-related issues. 

\item In a noisy setting, error mitigation techniques such as zero noise extrapolation (ZNE)~\cite{Brav_ZNE, best_pract_qem} could further improve the accuracy of our results. However, the improvement in accuracy is accompanied by additional costs associated with executing more circuits and the  overheads related to an extrapolation model fit. Measurement errors can be mitigated by techniques that rely on constructing an $N$-qubit noise matrix~\cite{Brav_inv_matrix}, whose cost scales as $\sim 2^N$. This is because the column vectors of this matrix are populated by probabilities obtained via preparing $2^N$ circuits, each initialized in a given basis state and measured. The noiseless state is then retrieved via $\vec{c}_{\text{noiseless}} = M^{-1}\vec{c}_{\text{noisy}}$. Therefore the net cost not only includes the cost of measuring the $2^N$ circuits which is $N\times 2^{N}$ but also the the cost of inverting $M$, that is, $\sim 2^{3N}$. 
There has been an effort in literature for adapting ZNE to quantum annealing~\cite{QEM_QA}, and it mainly improves the quality of the expectation value and not the sample set itself. Hence one could extrapolate probabilities to obtain the underlying error-free probability distribution, which however demands extensive sampling. Other methods like virtual distillation~\cite{Exp_err_sup}
offer exponential suppression in error at the cost of requiring $MN$ ($M$ refers to the number of copies of the noisy quantum states composed of $N$ qubits) qubits of the quantum state~\cite{QA_with_EM}, which is costly for NISQ era computers. 
\end{enumerate}
\end{document}